In Mar. '21, Submitted to the Publisher (Springer) – Transfer of Copyright Yet to Be Requested (as of Mar. 26 '21)

# The Serrated-Flow Behavior in High-Entropy Alloys

Jamieson Brechtl[1,*] and Peter K. Liaw[2,3]

[1] *Oak Ridge National Laboratory, 1 Bethel Valley Rd, Oak Ridge, TN, 37831*
[2] *Materials Science and Engineering, University of Tennessee, Knoxville, TN, 37996*

* Correspondence: pliaw@utk.edu
Tel.: 865-974-6356

This chapter presents a literature review of the serrated flow phenomenon in high-entropy alloys (HEAs). The serrated flow is important as it can result in permanent macroscopic and microstructural changes in HEAs. The literature reveals several important findings regarding the effect of strain rate and test temperature on the serrated flow. Furthermore, this chapter explores the relationship among the composition, microstructure, testing condition, and serration behavior. Towards the end of the chapter, a concise summary is presented for the temperature, strain rate, mechanical-testing type (compression/tension/nanoindentation), and serration type for HEAs. This chapter also provides an overview of the different types of analytical methods that have been successfully implemented to model and analyze the serration behavior in HEAs. Such techniques include the mean-field theory (MFT) formalism, complexity-analysis method, and multifractal technique. Finally, future research topics in this area are presented, such as the effects of twinning and irradiation on the serration behavior.

*Keywords: High-entropy alloys; Mechanical testing; Nanoindentation; Plastic deformation; Serrated flow; Dislocation Pinning; Twinning; Temperature effects; Strain-rate effects; Compositional Effects; Microstructural characterization; Mean-field theory; Algorithms; Data Analysis; Complexity*

# Table of Contents

# 1. Introduction

Typically, the serrated-flow (i.e., serrations) behavior in a material is characterized by fluctuations in the stress-strain curve during plastic deformation [1]. Such a phenomenon typically occurs during dynamic strain aging (DSA) and is known in this case as the Portevin–Le Chatelier (PLC) effect [2]. This type of unstable plastic deformation behavior is important as it can result in significant changes in the microstructure [1, 3-6] and can help elucidate the fundamental mechanisms that govern microscale-plastic deformation [6, 7]. Typically, the serrated flow is observed during tension and compression tests, but has also been seen during nanoindentation testing in the form of "pop-ins" [8]. This type of behavior is ubiquitous in materials as it has been observed in a variety of systems, including granular systems [9], single crystals and polycrystalline metals [10-13], steels [1, 14-28], Cu alloys [29], V alloys [30, 31], Al alloys [1, 32-43], bulk amorphous alloys [1, 9, 44-63], and medium-entropy and high-entropy alloys [1, 3, 5, 6, 8, 64-92]. Besides materials undergoing plastic deformations, serrations have been observed in other phenomena, including economic indices [93-97], neuronal avalanches [98-106], Barkhausen noise [1, 107-110], and crackling noise that accompanies earthquakes [111-114].

Figure 1 displays the five main types of serrations that have been exhibited by both traditional alloys and HEAs during plastic deformation, as based on the literature [67]. These serrations have been designated as Types-A, B, C, D, and E [1, 3, 65, 115, 116], which is based on the nomenclature from the traditional classification for the PLC effect, as proposed by Rodriguez [115]. The serration type can depend on the experimental conditions, such as test temperature and strain rate during mechanical testing [1, 3, 44, 65]. For example, the serrated flow has been observed to transition from Type-A to C serrations with an increase in temperature [6, 91]. Each type has different characteristics. Type-A serrations consist of periodic fluctuations where the stress values initially rise above the general level of stress and then sharply decrease. As for Type-B serrations, they are characterized by more frequent fluctuations about the general level of stress. Type-C serrations, on the other hand, can be described as sudden stress drops beneath the general level of stress [67]. Type-D serrations are characterized by plateaus in the stress-strain curve. Type-E serrations are similar to Type-A serrations, but with a more irregular structure, and occur with little or no working hardening [115]. The serrated flow can also exhibit mixtures of different serration types, e.g. A + B and B + C [74].

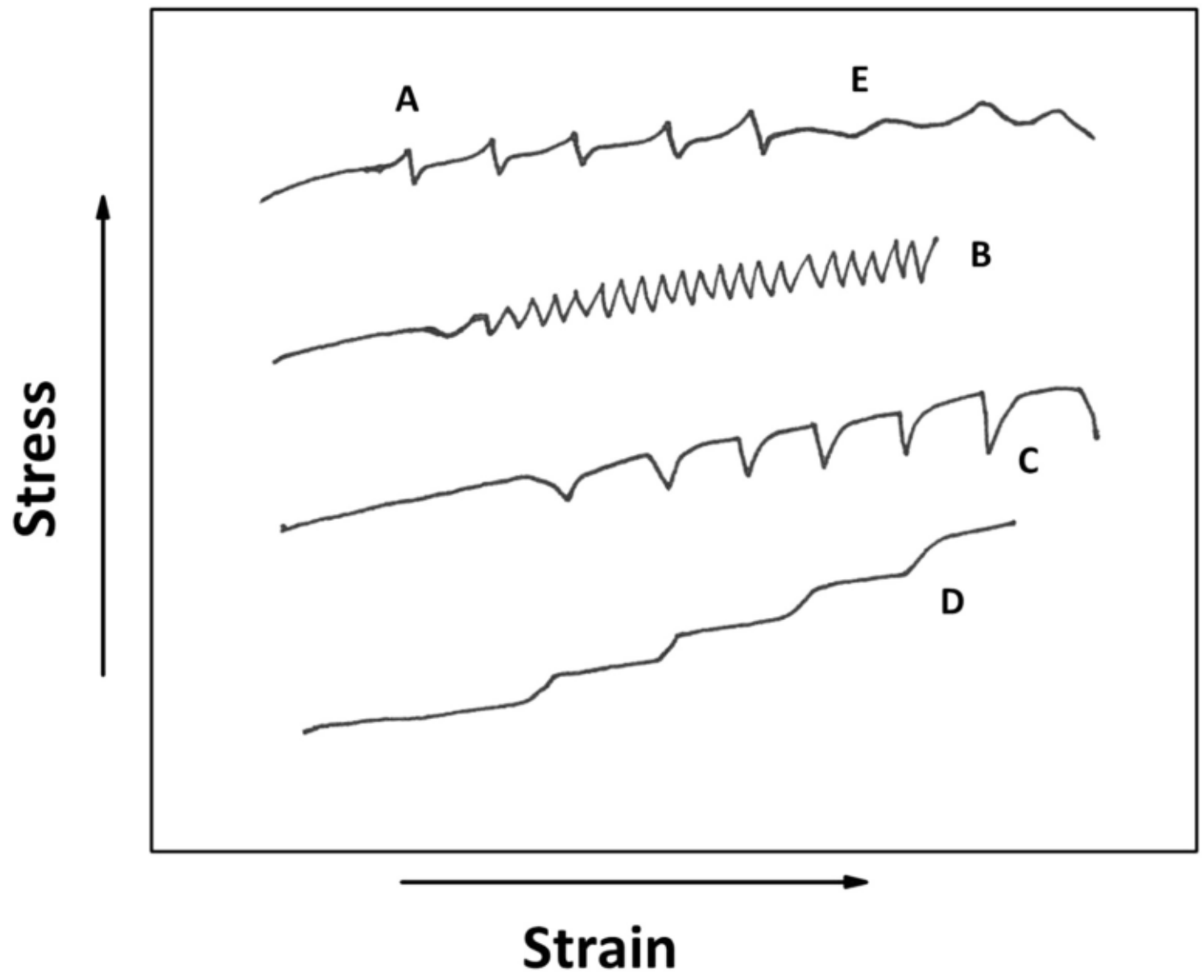

Fig. 1. The five types of serrations exhibited by HEAs, based on the literature. Figure from Chen et al. [67].

Although the serrated-flow phenomenon is a relatively-newer branch of study in the field of HEAs, the number of publications on the subject has been steadily increasing since 2011. It should be stated that there are some differences in the primary mechanisms of the serrated flow in conventional alloys and HEAs. For instance, it is believed that unlike conventional alloys where only certain atoms can catch and pin dislocations, all the atoms in an HEA can act as solutes, which pin them [6, 117]. The number of journal publications on this topic for years ranging from 2011 to 2020 (Web of Science Database) is presented in Fig. 2 [6]. As can be observed, there is a generally increasing trend in the number of articles published on this subject over time.

A number of different mechanisms can result in the serration behavior in HEAs during plastic deformation [6]. These mechanisms include mechanical twinning, solute pinning of dislocations (interstitial and substitutional), yielding across fracture surfaces in brittle materials, order-disorder phase transformations, and stress/strain-induced phase transformations [115, 118-121]. It has been suggested that the conventional interpretations of dislocation pinning that happens during DSA is applicable to HEAs as a similar weakening effect has also been observed in this alloy system [6, 91].

One of the first theoretical explanations for the interstitial solute-pinning phenomenon was provided by Cottrell and Bilby [122]. Here they stated that at sufficiently-high temperatures, interstitial solute-atoms diffuse and immobilize dislocation cores, thereby impeding their motion [121]. As for substitutional solutes, Herman suggested that they also can pin dislocations by vacancy-diffusion mechanisms [123]. During the pinning process, solute atoms initially migrate to above or below the dislocation lines, depending on their size [124]. If their size is smaller than the matrix atoms, they will move to sites above the dislocation line where compressive strains occur [6]. If the size of the solute atom is greater than the matrix atoms, they will migrate below the dislocation line where tensile strains occur. In this scenario, the tensile and compressive strain fields immobilize the moving dislocation. While the dislocation is trapped, the stress builds up until it reaches a threshold value where it can break loose and migrate. Once free, the dislocation migrates until it becomes pinned again. This cyclical pinning and unpinning process results in the serrated flow [125].

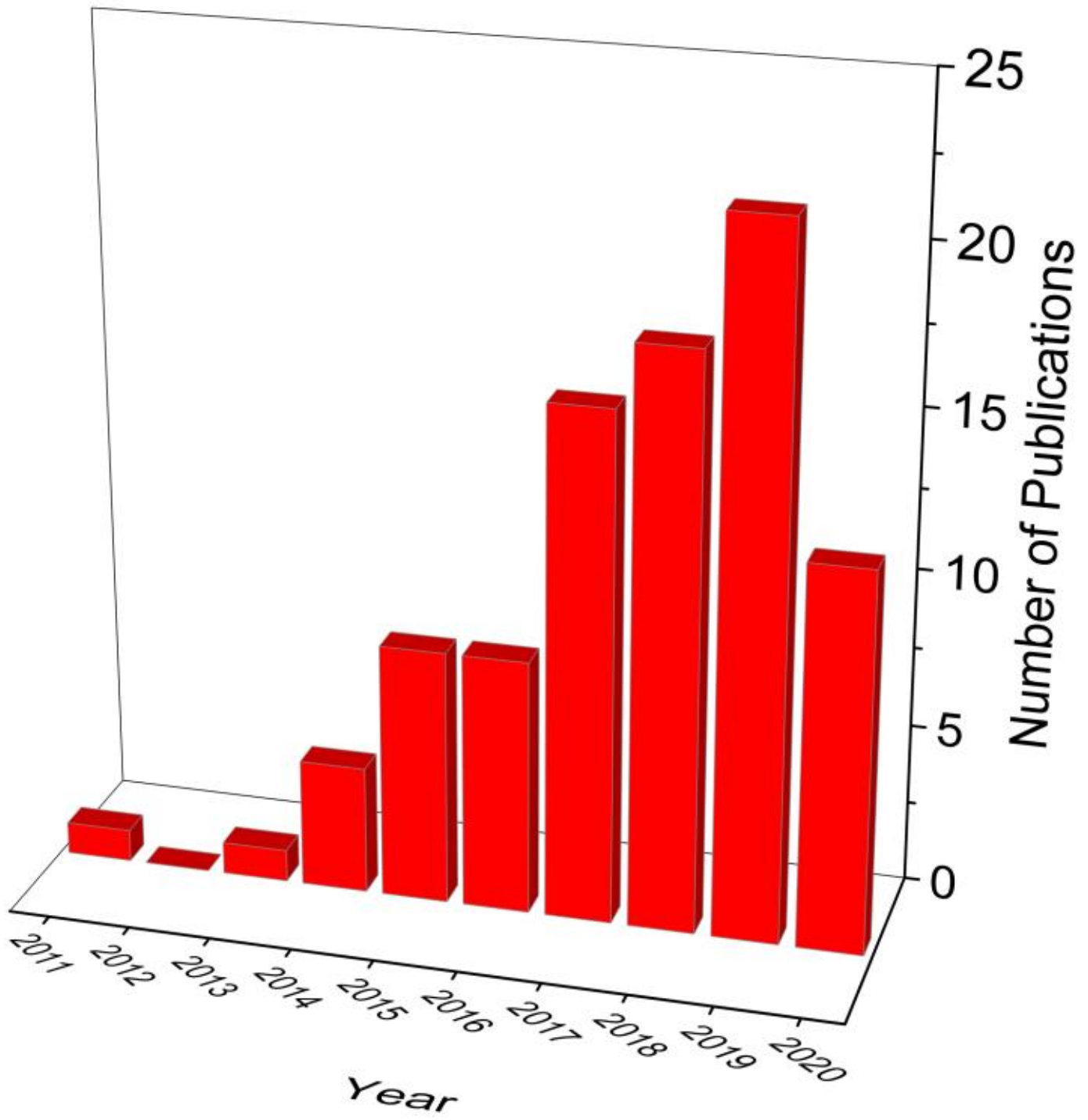

Fig. 2. The number of publications, which feature the serrated-flow phenomenon in HEAs, for a time period of 2011 - 2020 (via the Web of Science publication database).

Figure 3 illustrates the basic pinning-unpinning cycle [6]. During Section I, diffusing solute atoms catch and pin mobile edge dislocations. During Section II, the dislocation stays pinned until the stress reaches a critical value. Once a critical stress is attained, sufficient energy has built up and the dislocation can break free. Upon breaking loose, there is a sudden drop in the stress value, which corresponds to Section III in the figure. Subsequently, the dislocation migrates until it is re-pinned by migrating solute atoms, which initiates an increase in the stress (marked Section IV in Fig. 3). It is important to note that while the figure only considers the pinning of one dislocation, the serrated flow is still the outcome of collective pinning events, as will be discussed over the course of this chapter.

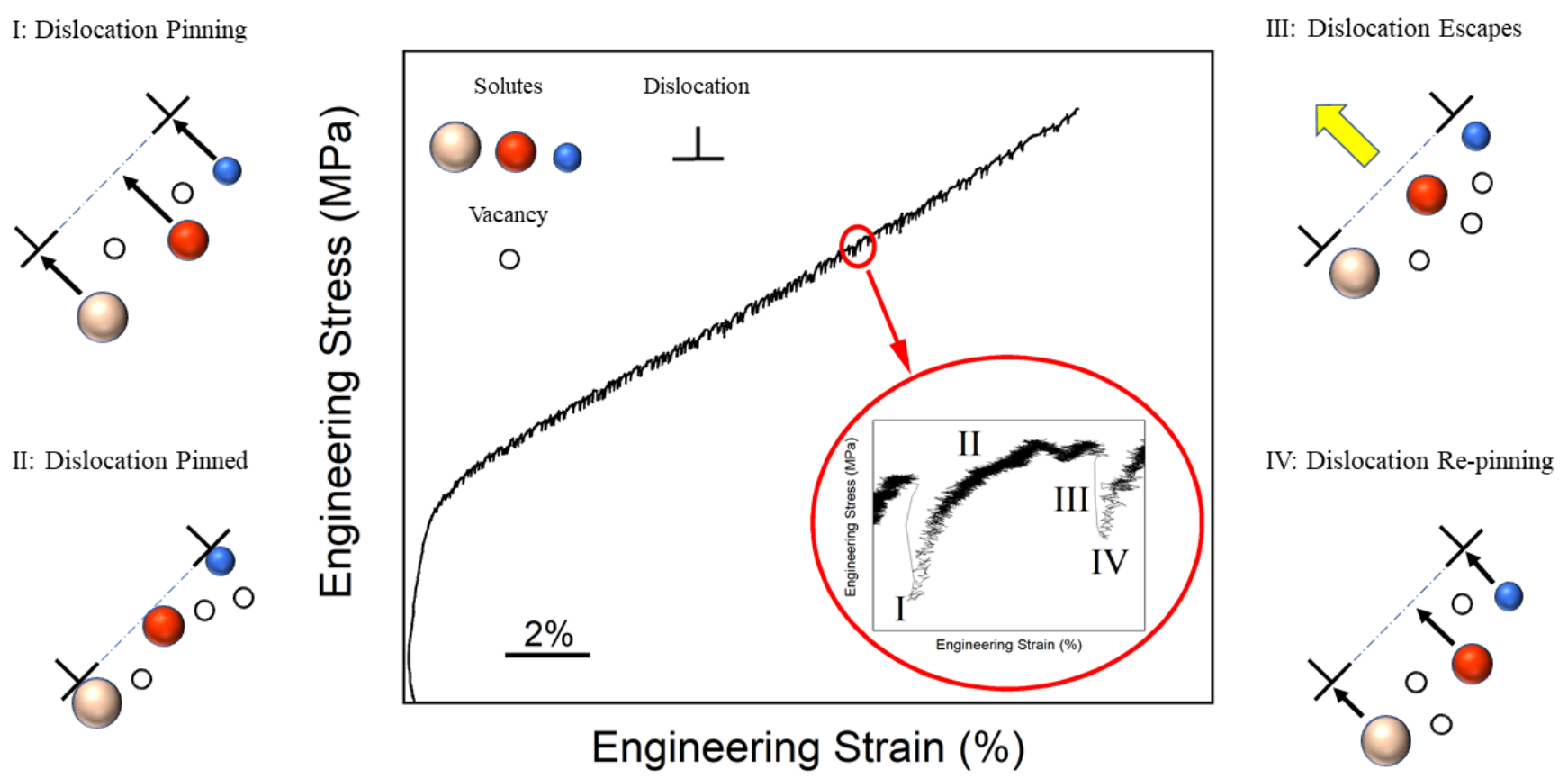

Fig. 3. The basic pinning and unpinning process of edge dislocations by solute atoms. The numbered Sections I - IV details this process that occurs during the serration event: I: solute atoms catch and pin a migrating dislocation, II: the stress increases as the dislocation is pinned, III: the stress build-up reaches a threshold, and the dislocation breaks loose, IV: solute atoms catch and re-pin the dislocation. Figure adapted from Refs. [6, 64, 126].

In HEAs, it has been suggested that any atom can act as a pinning solute [127]. Nevertheless, some studies have reported that only certain atoms in the matrix, such as Al, can lock dislocations [5, 91, 128]. In [128], the deformation behavior of CoCrFeNi and $Al_{0.3}$CoCrFeNi HEAs during compression was investigated. It was found that only the Al-containing HEA displayed serrations. The findings indicated that the serrations were a result of the frictional stress imposed upon the dislocations by the Al-containing solute atmospheres located in the vicinity of a moving dislocation core. In [129], serrations were observed in a carbon-doped CoCrFeMnNi HEA at room temperature (RT), which was atypical for that material system. It was suggested that the serrated flow resulted from the interactions between dislocations and carbon impurities. It has been suggested that an increase in the amount of solutes in an HEA can increase both the abundance and the atomic-size differences that enhance the pinning force in the lattice [6, 92]. Consequently, these microstructural modifications can increase the range of temperatures in which serrations can occur.

Besides solute atoms, there are other factors which contribute to the serrated-flow in HEAs. Twinning is one such factor that can cause serrations in HEAs since twins can hinder dislocation motion at twin boundaries [1, 3, 17, 18, 70]. Besides twinning, studies have indicated that phase structures and nanoparticles are important participants in the serrated-flow process in HEAs. For instance, a previous study reported that during compression of an $Al_{0.5}CoCrCuFeNi$ HEA, $L1_2$ particles can behave as obstacles for mobile dislocations [67]. Tabachnikova et al. [130] reported that sigma-phase particles and dendrite boundaries can impede dislocation motion, resulting in the serrated flow in $CoCrFeNiMnV_x$ HEAs

Previous studies have revealed that factors such as strain rate and temperature can significantly affect the serration behavior in HEAs [6]. An increase in the latter enhances the mobility of diffusing solute atoms, causing them to pin dislocations at a faster rate [64]. Such an increase in the pinning rate can sometimes lead to the transition of one serration type to another. For example, in an $Al_{0.5}CoCrCuFeNi$ HEA, the serrated flow transitioned from Type-A to C as the temperature increased [64]. Nevertheless, serrations can disappear once the temperature exceeds a certain limiting value. The disappearance of the serrations is due to a number of factors, including exceedingly-large atomic thermal vibrations which inhibit solutes from effectively locking dislocations or a prohibitively-high critical strain for the onset of serrations [1, 6, 65]. A decreasing temperature, on the other hand, reduces the solute mobility such that it is harder for them to catch and pin dislocations. Once below a certain temperature, the solutes will not be able to move fast enough to catch dislocations [131]. As for strain rates, the Types-A and C serrations are generally observed at relatively-higher and lower values, respectively [64, 65, 67, 91]. At the lower strain rate, the slower motion of dislocations allows solute atoms to pin dislocations more effectively. Such a scenario leads to the more efficient pinning and unpinning of dislocations that is characteristic of Type-C serrations [91].

## 2. Modeling and analytical techniques

Numerous modeling and analytical methods have been implemented to study the serration dynamics in HEAs. These techniques include complexity, multifractal, and statistical techniques [6, 65, 66, 70, 71, 76, 91, 132]. These techniques have been applied to model and analyze the serration behavior in HEAs since it is thought that serration curves consist of complex dynamics and scaling laws [6].

## 2.1 Mean-field theory and the mean-field interaction model

According to the mean-field theory (MFT) solid materials, such as HEAs, contains weak spots that are coupled elastically [9]. During the serrated flow, a slip avalanche occurs when a slipping weak spot triggers other weak spots to slip successively [44]. As soon as the stress at every point in the system drops beneath its current failure stress, these long-range interactions (and the serrated flow) will terminate [133]. In general, the slip avalanche statistics can be analyzed using relatively less complex probability distribution function (PDF) models [134]. Here, it is assumed that the long length-scale behavior of the slip statistics is not affected by the microstructural details of the material [6]. One such MFT model is comprised of an exponentially-decaying cutoff function that is multiplied by a power-law distribution of slip sizes [9, 44, 45, 51, 135, 136]:

$$D(S,q) = S^{-\kappa} D'(Sq^{\lambda}) \tag{1}$$

here, $D'(y)$ is a universal scaling function, $S$ is the avalanche-slip size, $\kappa$ and $\lambda$ are universal power-law exponents, and $q$ is an experimental parameter (strain rate, temperature) that is tunable. According to the model, the maximum avalanche-slip size, $S_{max}$, follows the relationship $S_{max} \sim q^{\lambda}$ [6]. The complementary cumulative distribution function (CCDF), $C(S,q)$, is useful in analyzing the serrated flow statistics. The CCDF, which evaluates the likelihood that avalanche sizes larger than $S$ will occur, is defined as [44, 91]:

$$C(S,q) = \int_{S}^{\infty} D(S',q)dS' \tag{2}$$

Substituting Eq. (1) into Eq. (2) and then integrating yields:

$$C(S,q) = q^{\lambda(\kappa-1)} C'(Sq^{\lambda}) \tag{3}$$

here, $C'(Sq^{\lambda})$ is another similar scaling function. From here, one can determine the universal scaling exponents ($\kappa$ and $\lambda$) by first plotting $C(S,q)q^{-\lambda(\kappa-1)}$ vs. $Sq^{\lambda}$ and subsequently tuning the exponents until the CCDF curves are superimposed. Once the universal scaling exponents determined, the distribution of stress drops can be predicted for specimens tested at other strain rates and temperatures.

Other forms of statistical analysis have also been applied to the serrated-flow behavior, such as the analysis of nanoindentation displacement burst sizes using the following cumulative PDF [137]:

$$P(>S) = AS^{-\beta} e^{-\frac{S}{S_c}} \tag{4}$$

where $A$ is a normalization constant, $S$ is the size of the displacement burst (or pop-in), $\beta$ is a scaling exponent, and $S_c$ is some cutoff value of $S$. A displacement burst is a plastic-deformation event that occurs during nanoindentation in which the nanoindenter tip undergoes a large displacement without a significant change in the load. According to the model, the magnitude of the pop-ins follows a power-law distribution if $S < S_c$. For $S > S_c$, the pop-in sizes decrease exponentially.

## 2.2 Chaos analysis

Chaos is a phenomenon that is characterized by its unpredictability and exponential sensitivity to small perturbations [138, 139], and has been observed in many different fields, such as weather and climate [140-142], electrical-circuit systems [143-145], financial markets [146-148], ecological systems [149-151], and physiological systems [152-154]. Chaos has also been found to occur during the serrated flows in alloys [29, 155, 156]. Ananthakrishna et al. [157] was one of the first to model the serrated-flow dynamics using a chaotic model. Such a system displayed period-doubling bifurcations that ultimately resulted in chaotic behavior [6]. Later investigations reported that for CuAl and AlMg alloys, the serrated flow that occurred during tension was characterized by chaotic dynamics [29, 155, 156].

An important parameter that can be used to determine whether dynamical behavior is chaotic is the largest Lyapunov exponent, which quantifies the system sensitivity to small perturbations [158]. Positive values indicate that the behavior is chaotic, while the behavior is not chaotic for negative values [70]. Below will be given a basic recipe on how to perform this type of analysis [6].

First define the reconstructed attractor (from the given time series), $X(t,m)$:

$$X(t,m) = [\sigma(t), \sigma(t + \tau), \ldots, \sigma(t + [m-1]\tau)] \quad (5)$$

here $\sigma$ is the stress magnitude at time $t = [1, 2, \ldots N - (m - 1)\tau]$, $N$ is the number of data points in the given set, $m$ is the embedding dimension, and $\tau$ is an arbitrarily-chosen delay time [66, 159]. Next, find the nearest neighbor (NN) to the initial point, $X_0(t_0, m)$, in which their separation distance (in a Euclidean sense) is written as:

$$L(t_0) = |X(t_0, m) - X_0(t_0, m)| \quad (6)$$

Now find the distance to the next point, $L'(t_1) = |X(t_1,m) - X_0(t_1,m)|$. If this new point, $X_0(t_1,m)$, is not a NN point of $X(t_1,m)$, then a new point, $X_1(t_1,m)$, is selected and evaluated. This step is then performed again. The angular separation between $L(t_i)$ and $L(t_{i+1})$ is minimized for each step as to decrease the influence on the orbit evolution when a NN point is selected [6, 70]. One repeats this process until the m-dimensional vector, $X(t_i)$, has traversed the entire data file [66, 159]. Next determine the largest Lyapunov exponent, $\lambda_1$:

$$\lambda_1 = \frac{1}{t_M - t_0} \sum_{k=1}^{M} Log_2 \frac{L'(t_k)}{L(t_{k-1})} \tag{7}$$

here $t_M$ is the final time point, and $M$ is the total number of the replacement steps. To reiterate, the dynamical behavior is chaotic when the exponent is positive, whereas a negative value signifies that the system will exhibit long-term stability.

## 2.3 Complexity analysis

Complexity algorithms have been used to analyze various dynamical phenomena, including biological signals, financial time series, and serrated-flow behavior during mechanical testing [14, 160-162]. The complexity algorithm measures the entropy or irregularity of a given time-series data, in which greater entropy values are indicative of a more irregular time series [160]. The approximate entropy (ApEn) technique is one such method that can measure the complexity (or entropy) time-series data [163]. The ApEn method has implemented to analyze physiological and serrated-flow data [66, 160, 164].

To perform the analysis, one does the following. For a given time series one creates the set $X(j) = [\xi(j), \xi(j+1), \ldots, \xi(j+m-1)]$ with $1 \leq j \leq N - m + 1$, where $m$ is the embedding dimension and $N$ is the number of data samples [165]. Next, evaluate the number of vectors, $n_j^m(r)$, within a distance, $d[X(j), X(k)] < r$ [66, 163]:

$$d[X(j), X(k)] = \max_{l=1,2,\ldots,m} \{|\xi(j+l-1) - \xi(k+l-1)|\} < r \tag{8}$$

It should be noted that $r$ is generally less than 0.2 times the standard deviation of $X$ [66]. Now define the following:

$$\Phi^m(r) = \frac{1}{N-m+1} \sum_{j=1}^{N-m+1} C_j^m(r) \tag{9}$$

here $\Phi^m(r)$ is the average degree of self-correlation [66], and $C_j^m(r) = n_j^m(r)/(N - m + 1)$ [66, 161]. From here, solve for the ApEn (for a given $m$ and $r$) as:

$$ApEn(m, r, N) = \Phi^{m}(r) - \Phi^{m+1}(r) \tag{10}$$

Another technique that has been used to analyze the complexity of a dynamical system is the refined composite multiscale entropy (RCMSE) method [166]. The RCMSE technique has been used to analyze different types of time-series data, including chaos, cognitive tasks, and serrated-flow behavior [26, 62, 89, 91, 166-168]. In terms of the serrated flow, previous investigations revealed that factors such as the microstructural composition can affect the complexity of the serration dynamics. As an example, in [26], the complexity of the serration behavior in steels increased with respect to the number of carbon impurities in the steel. It was suggested that an increase in the interstitial-carbon content led to a greater number of dislocation-locking interactions during serrations, which is indicative of more complex behavior. A similar link between the impurities and the complexity of the serrated flow was discussed in [117]. Also, the serration types exhibit differing amounts of dynamical complexity [32, 91] where Types-A and B serrations correspond to more complex dynamics, as compared to Type-C serrations [32]. Finally, it has been suggested that the degree of complexity exhibited by serrations in a material may be related to its ability to an applied load [62].

A basic description on how to implement the RCMSE method on the serrated-flow data is described below [166]. Given the serrated-flow time-series data, one omits the data from the elastic regime. From the remaining data, remove the underlying trend in the strain-hardening regime that is determined by a moving average or a polynomial fit (to an appropriate order) [169]. Next construct the coarse-grained time series, $y_{k,j}^{\tau}$ [26]:

$$y_{k,j}^{\tau} = \frac{1}{\tau} \sum_{i=(j-1)\tau+k}^{j\tau+k-1} x_i \quad ; \ 1 \le j \le \frac{N}{\tau} \qquad 1 \le k \le \tau \tag{11}$$

here $x_i$ is the ith data point of the detrended time-series data, $\tau$ is the scale factor, $N$ is the number of data points in the set, and $k$ designates at which data point to initiate the averaging procedure. Once $y_{k,j}^{\tau}$ is evaluated, create the set, $y_k^{\tau}$ [166]:

$$\boldsymbol{y_k^{\tau}} = \{ y_{k,1}^{\tau} \ y_{k,2}^{\tau} \ \ldots \ y_{k,M}^{\tau} \} \tag{12}$$

here $M$ is an integer below $N/\tau$, and each $y_{k,j}^{\tau}$ is determined from Eq. (11). Next, construct the template vectors of dimension, $m$:

$$\boldsymbol{y}_{\boldsymbol{k},\boldsymbol{i}}^{\boldsymbol{\tau},\boldsymbol{m}} = \left\{ y_{k,i}^{\tau} \;\; y_{k,i+1}^{\tau} \; .... \; y_{k,i+m-1}^{\tau} \right\} ; 1 \leq i \leq N - m ; 1 \leq k \leq \tau \tag{13}$$

For each $k$, use the following distance criteria to determine whether two distinct template vectors are a match:

$$d_{jl}^{\tau,m} = \left\| \boldsymbol{y}_{\boldsymbol{j}}^{\boldsymbol{\tau},\boldsymbol{m}} - \boldsymbol{y}_{\boldsymbol{l}}^{\boldsymbol{\tau},\boldsymbol{m}} \right\|_{\infty} = \max\left\{ \left| y_{1,j}^{\tau} - y_{1,l}^{\tau} \right| \ldots \left| y_{i+m-1,j}^{\tau} - y_{i+m-1,l}^{\tau} \right| \right\} < r \tag{14}$$

here $d_{jl}^{\tau,m}$ is the distance between two template vectors [170], and $r$ is typically chosen as 0.15$\sigma$, ($\sigma$ is the standard deviation of the data), which makes sure that the results are not dependent on the variance of the data [160, 161, 171]. Based on Eq. (14), two vectors will match when $d_{jl}^{\tau,m}$ is less than $r$. Now solve for the number of matching pairs for $m$ + 1 and total number of matching vectors, $n_{k,\tau}^{m}$ (for $m$ and $m$ + 1). To determine the RCMSE value (denoted also as the sample entropy) for the detrended serrated-flow time-series data, one uses the following formula [166]:

$$RCMSE(X, \tau, m, r) = Ln\left( \frac{\sum_{k=1}^{\tau} n_{k,\tau}^{m}}{\sum_{k=1}^{\tau} n_{k,\tau}^{m+1}} \right) \tag{15}$$

where $X$ is the time-series data and $\tau$, $m$, and $r$ are the parameters, as discussed above.

## 2.4 Multifractal modeling and analysis

The mathematics of the multifractal (MF) analysis, as applied to the serrated flow behavior, has been described in detail elsewhere [172]. Nevertheless, several words are worth adding to this description. The notion of fractal is now familiar to virtually everybody, due to multiple internet sites demonstrating beautiful and intricate geometrical patterns generated using different scaling rules [173]. This beauty is spawned by the property of self-similarity imposed by the scaling. However, it usually characterizes monofractal objects governed by a unique scaling law. When one looks at a serrated deformation curve, the observed serration pattern would hardly suggest such properties in many experimental conditions. This difficulty comes from the fact that real patterns and signals are often heterogeneous and require many fractal dimensions for their description, usually because the underlying geometrically fractal object carries some physical property, e.g., the amplitudes of serrations. The multifractal formalism, on the other hand, can analyze both the hidden fractal geometry as well as the distribution of the underlying physical properties on the fractal support [6, 91, 174]. Despite the incapacity of the intuition to assess the basic complexity, the application of the MF formalism has revealed that complex behavior of natural objects is not usually caused by random factors but by fractal properties. This ubiquity of scaling behaviors is not really surprising because such objects usually imply complex internal interactions.

In particular, the algorithms associated with the multifractal technique have been used to analyze the serrated flow in various alloy systems, such as bulk metallic glasses [175, 176], Al-Mg alloys [156, 172, 177-181], and HEAs [91]. This formalism provides a tool to investigate the connection between the characteristics of the serrated flow and its corresponding multifractal spectrum. For example, an investigation involving an Al-Mg alloy found that the multifractal characteristics of the serration dynamics were affected by the phase composition of the material [174]. Furthermore, the underlying multifractal and hierarchical structure was a consequence of the self-organization of dislocation motion [6]. Bharathi et al. [156] reported that the sharp increase in the multifractality (which is the range of the multifractal spectrum) of the stress-burst behavior was associated with the transition from Type-B to A serrations.

As a matter of fact, the interpretation of the quantitative results provided by the MF approach is not direct and the entire power of such analysis has not been explored so far. However, the examples given in both parts of this Chapter show that even at this stage, the MF spectra render indispensable information, e.g., through the evaluation of the effect of the material microstructure and experimental conditions on the occurrence of MF properties and also on the variability of fractal dimensions, which reflects the heterogeneity of the underlying deformation processes.

## 3. Serration studies in HEAs

### 3.1 Tension testing

#### 3.1.1 $Al_{0.5}$CoCrFeNi HEA

Niu et al. [5] investigated the plastic deformation of $Al_{0.5}$CoCrFeNi and CoCrFeNi alloys that were subjected to tension at strain rates and temperatures of $10^{-3}$ $s^{-1}$ - $10^{-4}$ $s^{-1}$ and 200 - 500 °C, respectively. The stress-strain data for the $Al_{0.5}$CoCrFeNi HEA is shown in Figs. 4(a)–(c). As indicated by the figures, the plastic-deformation behavior was comprised of Type-A, Type-A + B, and Type-B + C serrations (see Table 1). The findings reveal that for a given strain rate, the serrations transition from Type-A to Types-A + B, and then to Types-B + C with increasing temperature. The results of statistical analysis indicated that the stress-drop magnitudes are distributed according to a power law. Finally, it was determined that the DSA effect was responsible for the serrated flow instead of twinning or strain-induced transformations.

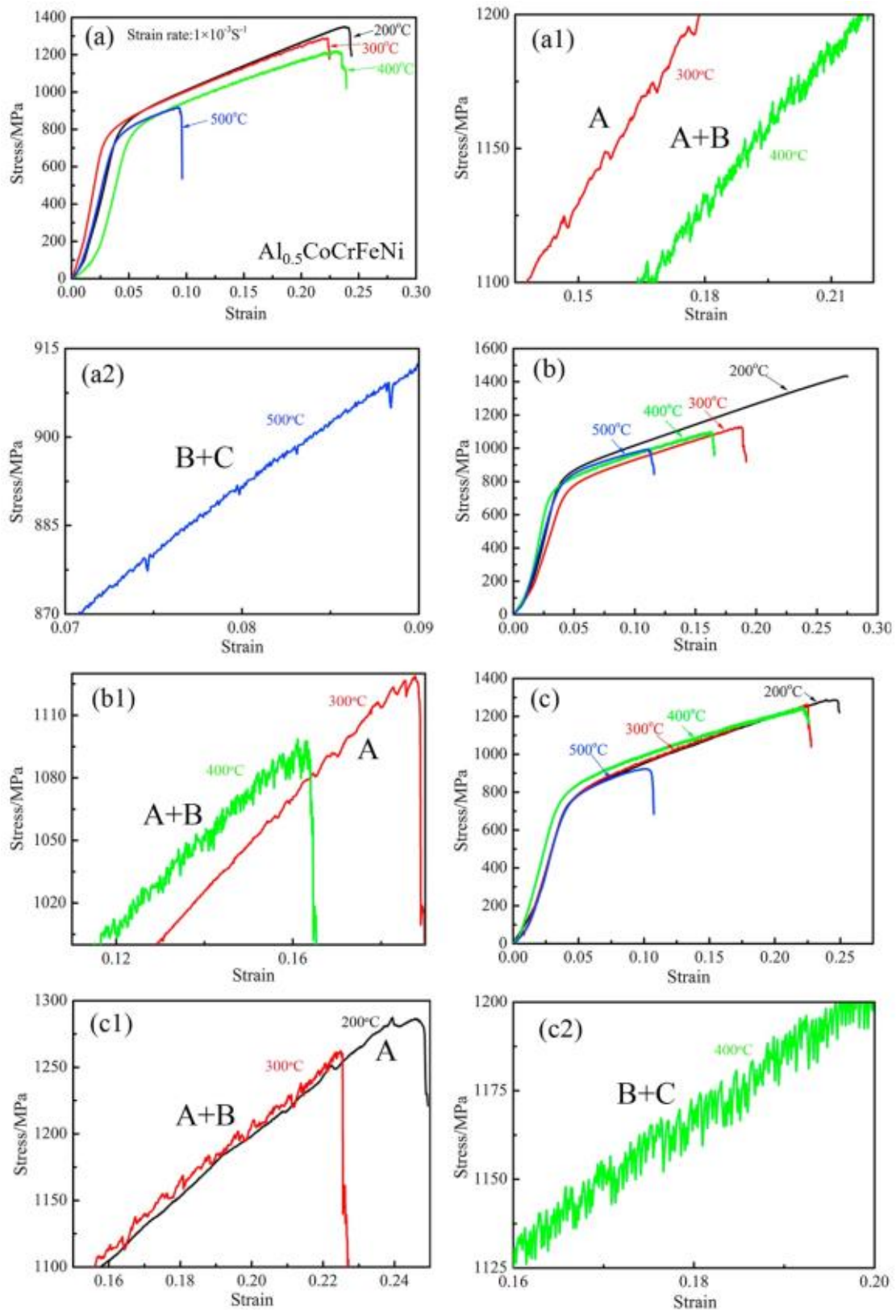

Fig. 4. The stress-strain data for the $Al_{0.5}$CoCrFeNi HEA subjected to tension at temperatures of 200 - 500 °C and strain rates of (a) $10^{-3}$ $s^{-1}$ [magnified sections of the associated serrated-flow curves are featured in (a1) and (a2)], (b) 5 × $10^{-4}$ $s^{-1}$ [magnified sections of the associated serrated-flow curves are featured in (b1)], and (c) $10^{-4}$ $s^{-1}$ [magnified sections of the associated serrated flow curves are featured in (c1) and (c2)]. Figures from Niu et al. [5].

Figures 5(a)-(b) present the results for the $Al_{0.5}$CoCrFeNi and CoCrFeNi alloys that underwent tension at a strain rate and temperature of $10^{-3}$ $s^{-1}$ and 400 °C, respectively. As compared to the CoCrFeNi alloy, the Al-containing HEA displayed pronounced serrations. From the results, the authors suggested that the Al played a substantial role in the serrated flow of the HEA due to the atomic-size differences between the Al and the other matrix atoms [91]. Additionally, embrittlement was observed in the Al-containing HEA during testing at higher temperatures and may be a consequence of the DSA of the specimen. Such behavior has also been observed in other Al-containing HEAs [290,291].

Table 1. List of the serration types for the $Al_{0.5}CoCrFeNi$ HEA specimens subjected to tension at strain rates of $10^{-4}$ - $10^{-3}$ $s^{-1}$and temperatures of 200 – 500 ºC. From Niu et al. [5].

| Strain Rate ($s^{-1}$) | Temperature (°C) | Serration Type |
|---|---|---|
| $1 \times 10^{-4}$ | 200 | A |
| | 300 | A + B |
| | 400 | B + C |
| $5 \times 10^{-4}$ | 300 | A |
| | 400 | A + B |
| $1 \times 10^{-3}$ | 300 | A |
| | 400 | A + B |
| | 500 | B + C |

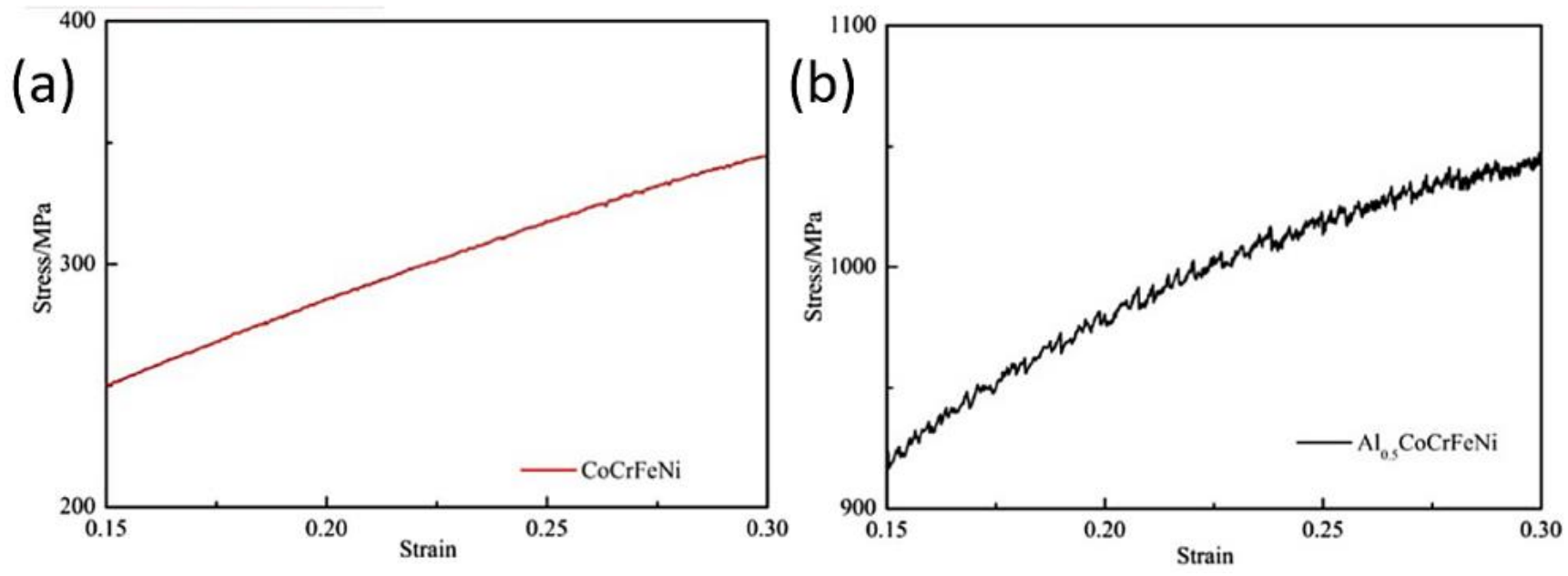

Fig. 5. The engineering stress vs. strain data for the (a) CoCrFeNi and (b) $Al_{0.5}CoCrFeNi$ HEAs that underwent tension testing at a strain rate of $10^{-3}$ $s^{-1}$ and temperature of 400 °C. Figures from Niu et al. [5].

### 3.1.2 $Al_5Cr_{12}Fe_{35}Mn_{28}Ni_{20}$ HEA

Antonaglia et al. [79] analyzed the serrated flow in an $Al_5Cr_{12}Fe_{35}Mn_{28}Ni_{20}$ HEA that underwent tension testing at a strain rate of $1 \times 10^{-4}$ $s^{-1}$ and temperatures of 573 and 673 K. The stress vs. displacement data is presented in Fig. 6(a). The specimens displayed serrations at both temperatures, and the serration magnitude increased as the specimen deformed. The corresponding CCDF curves are featured in Fig. 6(b). The curve for the 673 K condition is to the left of the one for the specimen tested at 573 K, which indicates that the magnitude of the largest stress-drop decreased with increasing temperature. The data also indicates that at 673 K, the specimen exhibited relatively-smaller slip sizes than at 573 K. Such behavior is characteristic of the simple mean-field model [133, 136, 182]. It was suggested that the smaller slip sizes is a consequence of the increased thermal-vibration energy of the pinning solute atoms at higher temperatures [79, 122]. In this scenario, the solutes have a greater predisposition to “shake away” from their low-energy sites for pinning [6].

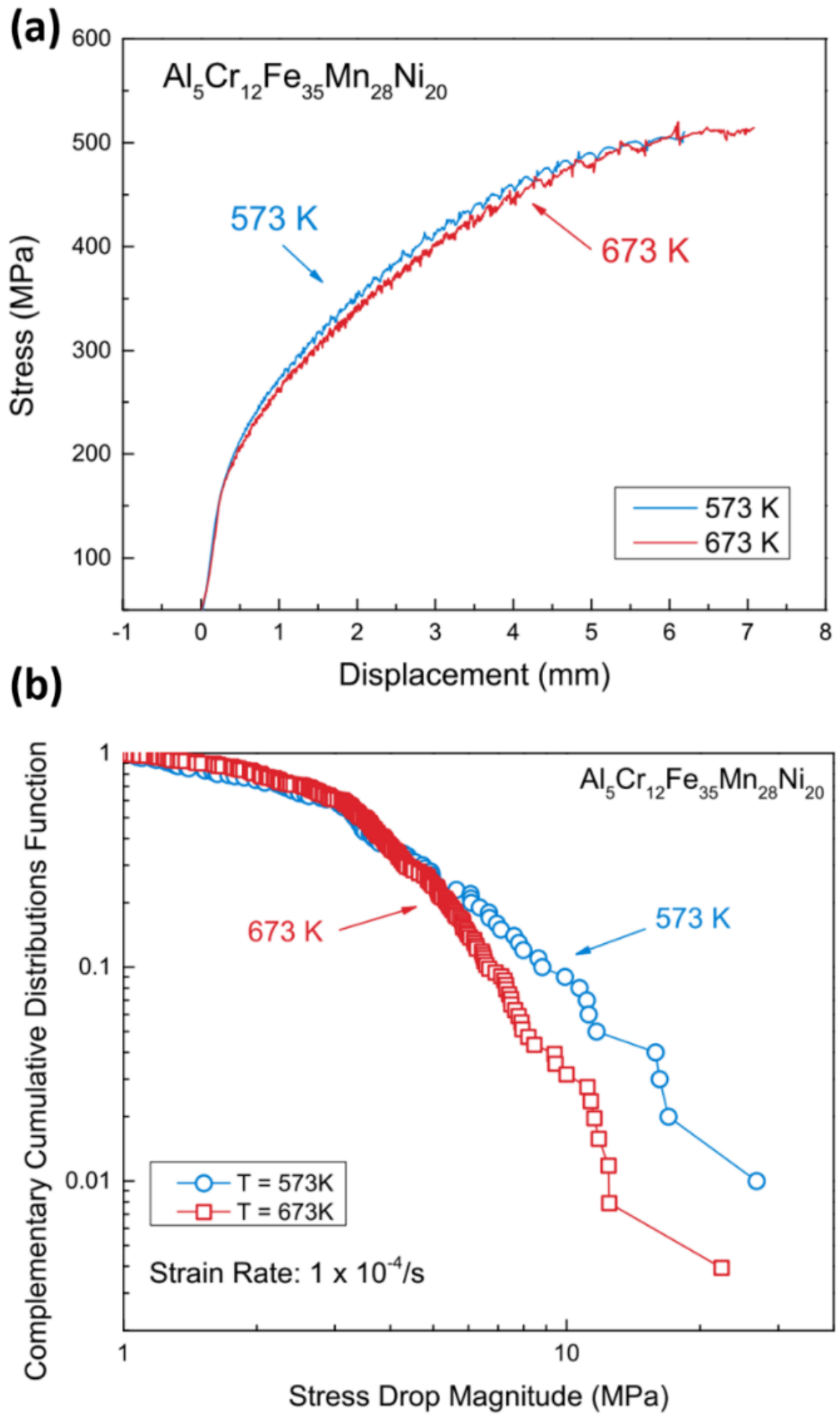

Fig. 6. (a) Stress vs. displacement data for the $Al_5Cr_{12}Fe_{35}Mn_{28}Ni_{20}$ HEA subjected to tension, and (b) the CCDF for the stress-drop magnitudes of the $Al_5Cr_{12}Fe_{35}Mn_{28}Ni_{20}$ HEA. The tests were performed at a strain rate of $1 \times 10^{-4}$ s$^{-1}$ and temperatures of 573 and 673 K. Figures from Antonaglia et al. [79].

### 3.1.3 $Al_xCrMnFeCoNi$ (x = 0, 0.4, 0.5, and 0.6) HEA

Xu et al. examined the deformation behavior of $Al_xCrMnFeCoNi$ (x = 0, 0.4, 0.5, and 0.6) HEAs [183]. Here, specimens underwent tension tests at a strain rate of $3 \times 10^{-4}$ $s^{-1}$ and temperatures of RT - 700 °C. No serrated flow was observed for the $Al_xCrMnFeCoNi$ at RT. Figures 7(a)-(c) display the serrated flow for the specimens tested at temperatures of 300 – 600 °C. The results indicated that there was a temperature dependence of the serration type. In general, the serration type evolved from Types-A + B at 300 °C to Type-C at 500 and 600°C. Table 2 summarizes the serrated-flow behavior. A similar trend has been observed in other HEAs [91]. Furthermore, it was found that the critical strain first decreased and then increased with temperature. Such behavior is indicative of a change from the normal to the inverse behavior of the critical strain [91]. Importantly, this trend in the critical strain increased in a linear fashion with respect to temperature [184]. Finally, serrations that were observed at 600 °C for the $Al_{0.4}CrMnFeCoNi$ HEA were not seen in the $Al_{0.5}CrMnFeCoNi$ and $Al_{0.6}CrMnFeCoNi$ HEAs. The authors surmised that the lack of serrations in the HEAs with a higher Al content may be a result of stronger pinning effects caused by the increased Al atoms and the BCC phase particles.

Table 2. The serration type for temperatures of 300 - 600°C (strain rate of $3 \times 10^{-4}$ $s^{-1}$) for the $Al_xCrMnFeCoNi$ HEA (x = 0.4, 0.5, and 0.6). From Xu et al. [183]).

| Alloy | Temperature (°C) | Serration Type |
|---|---|---|
| $Al_{0.4}CrMnFeCoNi$ | 300 | A + B |
| | 400 | B |
| | 500 | B + C |
| | 600 | C |
| $Al_{0.5}CrMnFeCoNi$ | 300 | A + B |
| | 400 | B |
| | 500 | B + C |
| $Al_{0.6}CrMnFeCoNi$ | 300 | A + B |
| | 400 | B |
| | 500 | C |

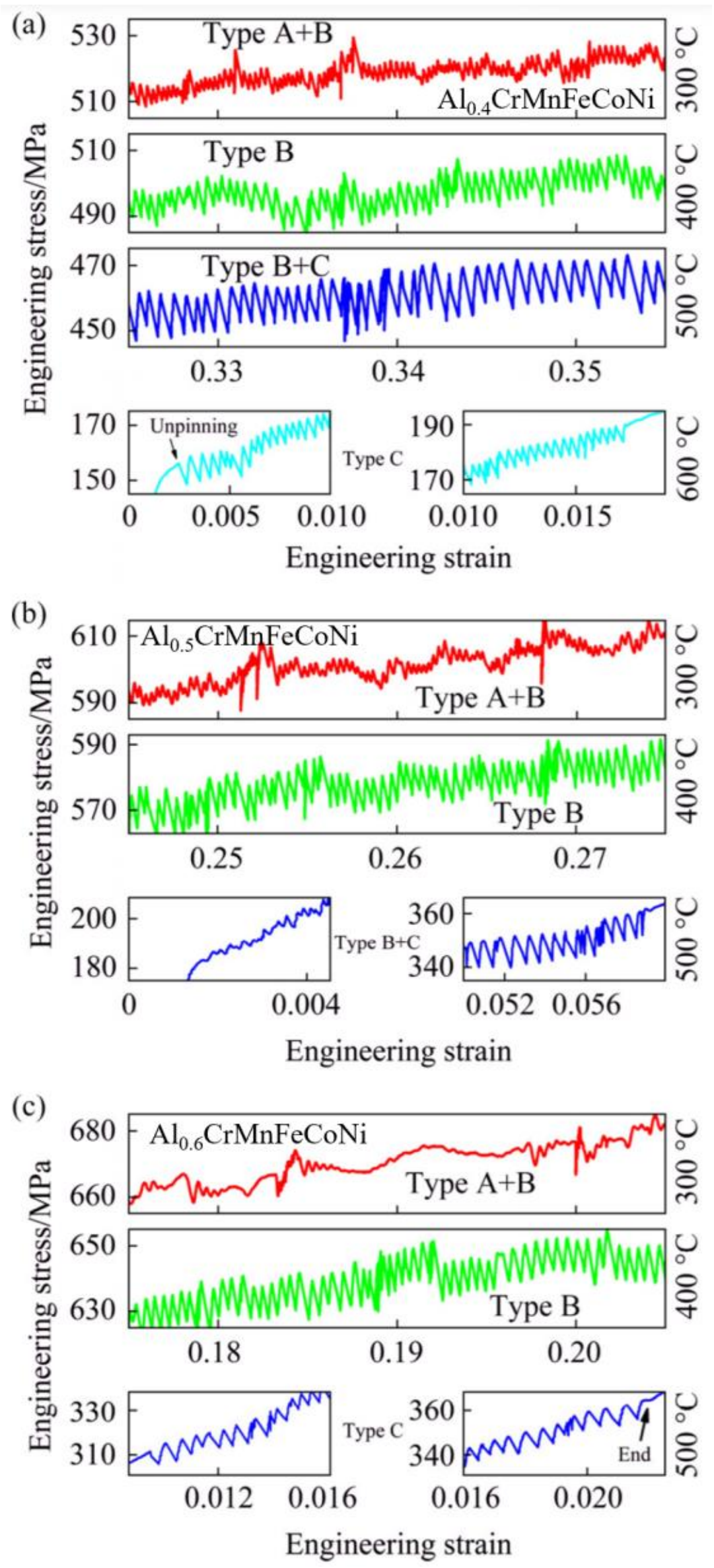

Fig. 7. The engineering stress-strain data for the (a) $Al_{0.4}$CrMnFeCoNi HEA, (b) $Al_{0.5}$CrMnFeCoNi HEA, and (c) $Al_{0.6}$CrMnFeCoNi HEA specimens subjected to tension testing at a strain rate of $3 \times 10^{-4}\ s^{-1}$ and temperatures of 300º - 600ºC. Figures from Xu et al. [183].

### 3.1.4 CoCrFeNi HEA

Liu et al. [70] analyzed the serrated flow in a CoCrFeNi HEA that was subjected to tension tests at a strain rate of $10^{-3}$ $s^{-1}$ and temperatures of 4.2 - 293 K. Figures 8(a)–(d) display the results of the microstructural characterization for the as-cast specimen. The optical microscopy results show that the specimen consisted of nearly-equiaxed grains that had a mean size of about 13 µm [see Fig. 8(a)]. The results of the X-ray diffraction (XRD) characterization reveal that the as-cast specimen contained an FCC structure [see Fig. 8(b)]. Finally, the transmission electron microscopy (TEM) characterization, as displayed in Figs. 8(c)-(d), reveals the presence of dislocations and {111} twins in the as-cast specimen.

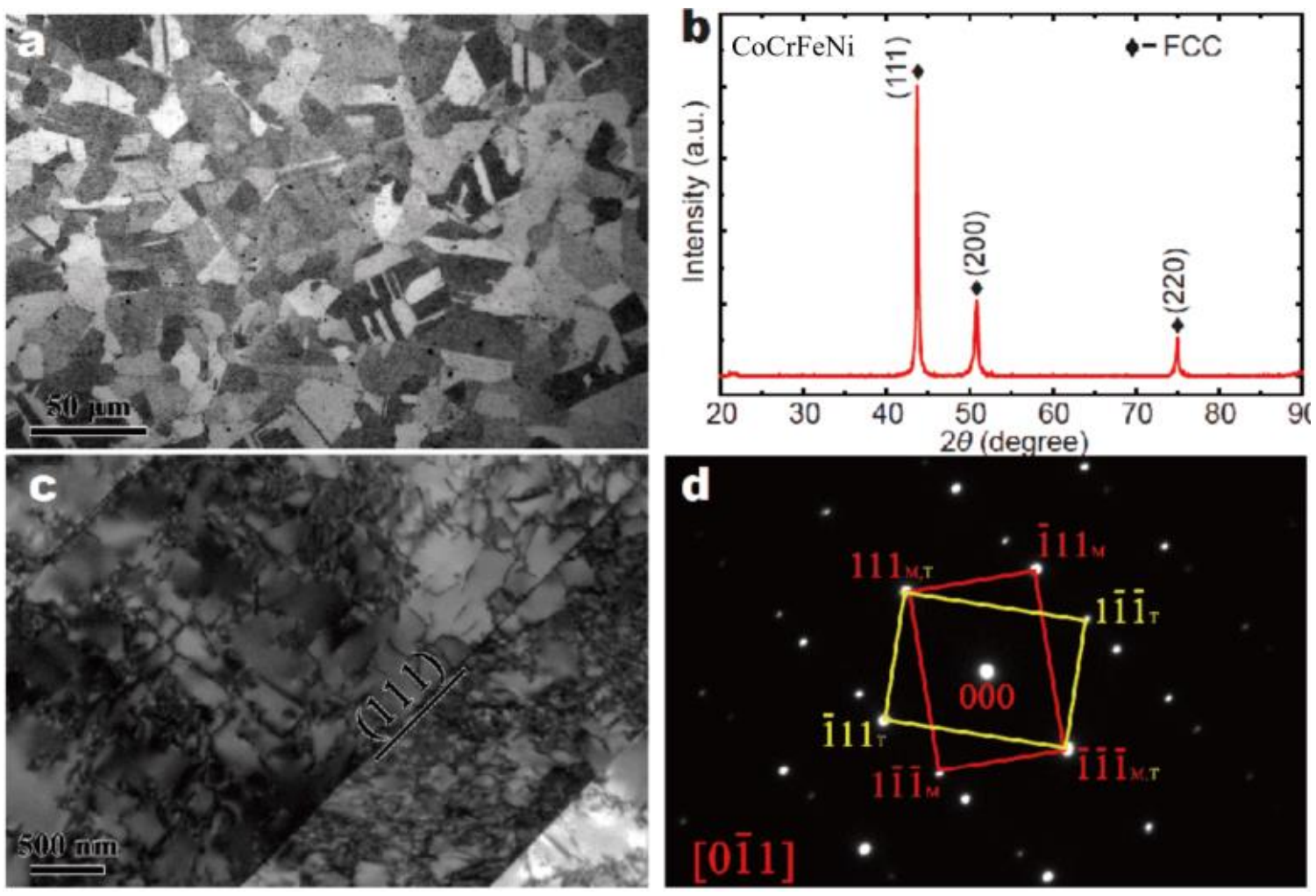

Fig. 8. The (a) optical microscopy image, (b) XRD pattern, (c) TEM image, and (d) the selected area electron diffraction (SAED) results for the as-cast CoCrFeNi HEA. Figures from Liu et al. [70].

The engineering stress-strain data for the specimens that underwent tension at a strain rate of $10^{-3}$ $s^{-1}$ and temperatures of 4.2 - 293 K is displayed in Fig. 9. In the specimens tested at temperatures of 4.2 and 20 K, serrations were clearly present in the stress-strain curves. The largest Lyapunov exponent was determined to be 0.05 and 0.001 for 4.2 and 20 K, respectively, which indicates that the serrated flow exhibited chaotic dynamics. TEM characterization revealed that an FCC to hexagonal-closed-packed (HCP) phase transformation occurred in the specimen during testing at 4.2 K, and was believed to play a role in the unstable behavior of the specimen at the above condition.

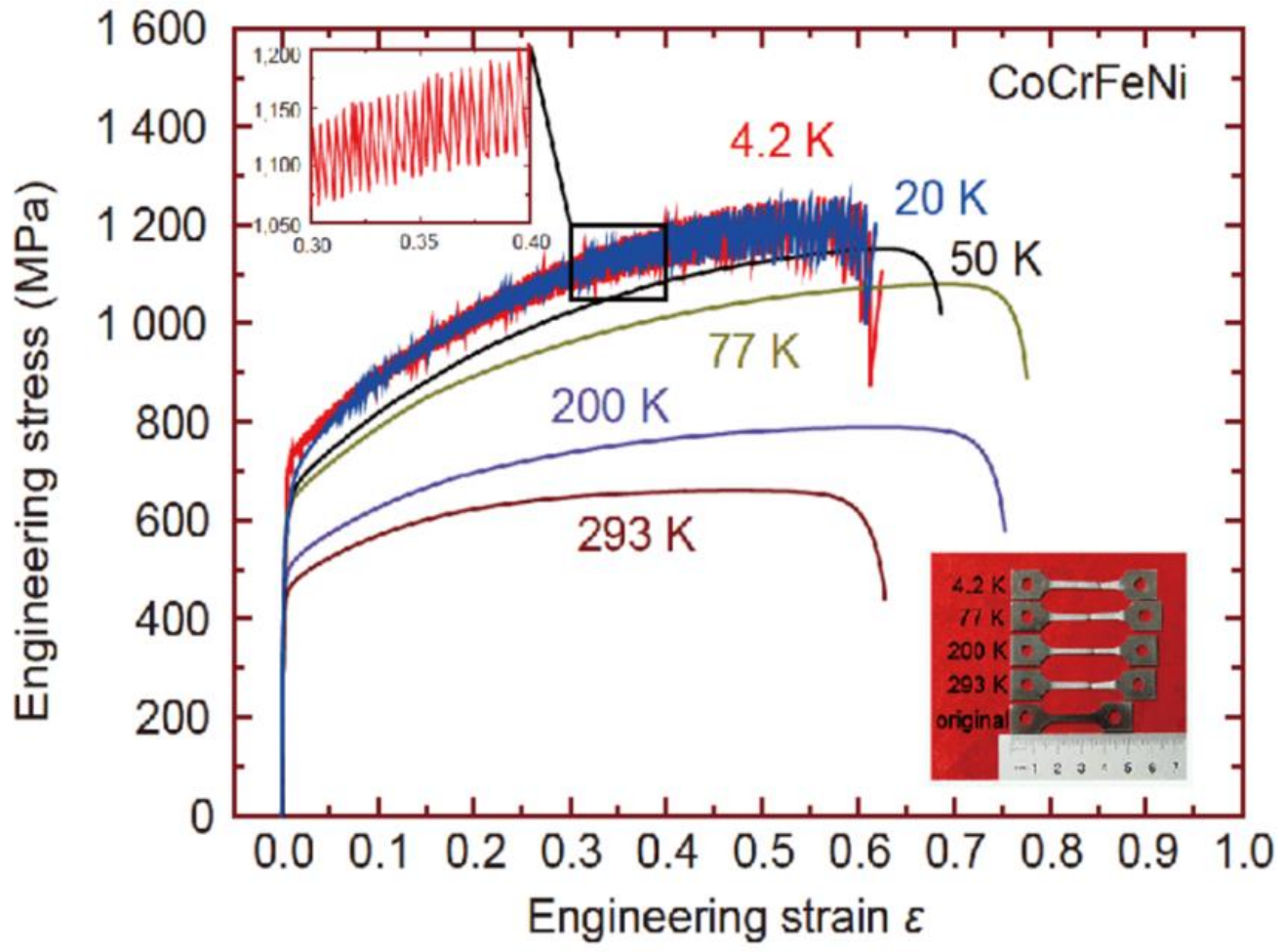

Fig. 9. The engineering stress vs. engineering strain data for the CoCrFeNi HEA subjected to tension at a strain rate of $10^{-3}$ $s^{-1}$ and temperatures of 4.2 - 293 K. A magnification of the serration behavior (strains of 0.3 - 0.4) for the specimen tested at 4.2 K is shown in the inset of the upper left-hand corner of the graph. An image of the test specimens, from before and after the tensile tests (at the corresponding test temperatures), are displayed in the inset that is located in the bottom right-hand corner of the graph. Figure from Liu et al. [70].

### 3.1.5 CoCrFeMnNi HEA (Cantor alloy)

In [65], the serrated flow of the CoCrFeMnNi HEA, CoCrFeNi HEA, CoFeNi medium-entropy alloy (MEA), CoNi low-entropy alloy (LEA), and pure Ni were compared [6]. Tension tests were performed on the specimens using strain rates of $1 \times 10^{-5}$ to $1 \times 10^{-2}$ $s^{-1}$ and temperatures of 300 - 700 °C. The results revealed that when the strain rate was $1 \times 10^{-4}$ $s^{-1}$, serrations were not observed in the pure Ni and CoNi alloy, although they did occur in the more chemically-complex alloys. It was surmised that the chemically-complex structure of the HEA allows atoms to pin dislocations at lower temperatures and prevents larger thermal vibrations from destroying the pinning effect [65].

Figure 10 presents the stress vs. strain data for the CoCrFeMnNi HEA specimens subjected to tension at a strain rate of $1 \times 10^{-4}$ $s^{-1}$ and temperatures of 300 - 600 °C [92]. Table 3 displays the serration type for the experiments conducted at strain rates of 1 × $10^{-4}$ - 1 × $10^{-2}$ $s^{-1}$ and temperatures of 300 - 600 °C. It was found that plastic-deformation behavior was comprised of Type-A, Type-B, and Type-C serrations. Furthermore, the lowest strain rate was the only condition where all three types of serrated flow were observed, whereas only Type-A serrations were found at the highest strain rate. Also, for the lowest strain rate, Type-A serrations were observed at the lowest temperature, while Type-B serrations were found to occur at the temperatures of 400 and 500 °C. Type-C serrations were seen at the highest temperature. Similar behavior was also observed during compression of a $Al_{0.5}$CoCrCuFeNi HEA, where it was theorized that Type-C serrated flow was the outcome of the efficient pinning and unpinning of dislocations by mobile solute atoms [91].

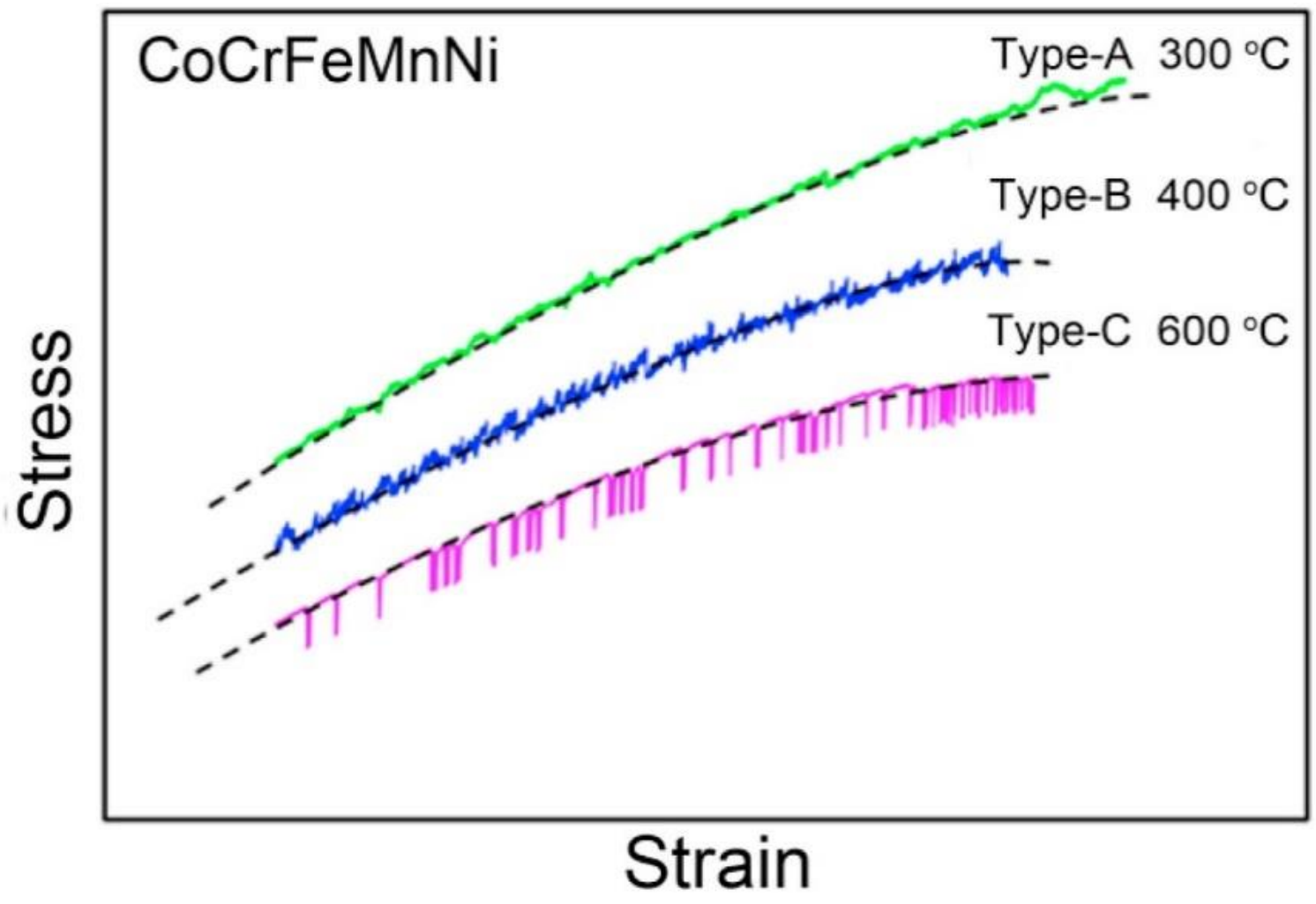

Fig. 10. The stress vs. strain data for the CoCrFeMnNi HEA that underwent tension at temperatures of 300 - 600°C and a strain rate of 1 × $10^{-4}$ $s^{-1}$. Figure from Tsai et al. [92].

In [74], the plastic-deformation behavior of a homogenized, cold-rolled, and recrystallized CoCrFeMnNi HEA was investigated. The specimens were homogenized at 1,100 °C for 24 h (in vacuum), and the thickness was then reduced by 40 % via cold-rolling. The specimens were then recrystallized at 900 °C for 1 h. Flat-dog-boned tensile specimens (25-mm gage length and 5.6 mm × 1.5 mm) were fabricated from the sheets. Specimens underwent tension testing at strain rates and temperatures of 1.0 × $10^{-5}$ - 5.0 × $10^{-3}$ $s^{-1}$ and RT - 800 °C, respectively. Figures 11(a)-(b) present the engineering stress vs. strain data for a strain rate of 3.0 × $10^{-4}$ $s^{-1}$ and temperatures of RT - 800 °C. For temperatures of 300 - 600°C, distinct serrations can be observed in the figures. Here, Type-A, Types-A + B, Type-B, and Type-C serrations were present at temperatures of 300, 400, 550, and 600°C, respectively. The serrated flow was not observed outside this temperature range. At higher temperatures, the thermal vibration of atoms prevents the effective locking of dislocations [67, 185]. As for temperatures below 300°C, the relatively low mobility of the solute atoms at these temperatures prevent them from catching and pinning moving dislocations [65].

Table 3. The serration type for the CoCrFeMnNi HEA that underwent tension at temperatures of 300 to 600 °C and strain rates $1 \times 10^{-4}$ - $1 \times 10^{-2}$ $s^{-1}$. From Carroll et al. [65]).

| **Strain Rate ($s^{-1}$)** | **Temperature (°C)** | **Serration Type** |
|---|---|---|
| $1 \times 10^{-4}$ | 300 | A |
| | 400 | B |
| | 500 | B |
| | 600 | C |
| $1 \times 10^{-3}$ | 300 | A |
| | 400 | A |
| | 500 | B |
| | 600 | B |
| $1 \times 10^{-2}$ | 400 | A |
| | 500 | A |
| | 600 | A |

Figures 12(a)-(b) present the engineering stress-strain data for the specimens subjected to tension at strain rates of $1.0 \times 10^{-5}$ - $5.0 \times 10^{-3}$ $s^{-1}$ and a temperature of 500 °C. At the lowest strain rate, Type-B + Type-C serrations were observed. In contrast, Type-A serrations were observed at $5.0 \times 10^{-3}$ $s^{-1}$. Types-A + B and Type-B serrations were observed at strain rates of $1.0 \times 10^{-5}$ and $1.0 \times 10^{-4}$ $s^{-1}$, respectively. Table 4 summarizes the serration type for the different experimental conditions.

The results also indicated that both the strain rate and temperature significantly affected the critical plastic strain for the onset of serrations. Findings indicate that the critical strain increased with respect to the strain rate. For a strain rate of $3.0 \times 10^{-4}$ $s^{-1}$, an increase in the temperature led to a decrease in the critical strain. It was also determined that for this experimental condition, there were two distinct temperature regimes with different activation energies for the serrated flow. Using the formalism proposed by McCormick [186], the authors calculated the activation energies for the two regions. One region corresponded to temperatures of 300 - 500 °C in which the activation energy was 116 kJ/mol. In this temperature range, dislocation pinning by solutes (via pipe diffusion) was responsible for the serrated flow. In the second region, which consisted of temperatures in the range of 500 - 600 °C, the serration activation energy was 296 kJ/mol. It was suggested that a cooperative lattice-diffusion mechanism (Ni was the rate-controlling constituent) was primarily responsible for the repeated pinning of dislocations.

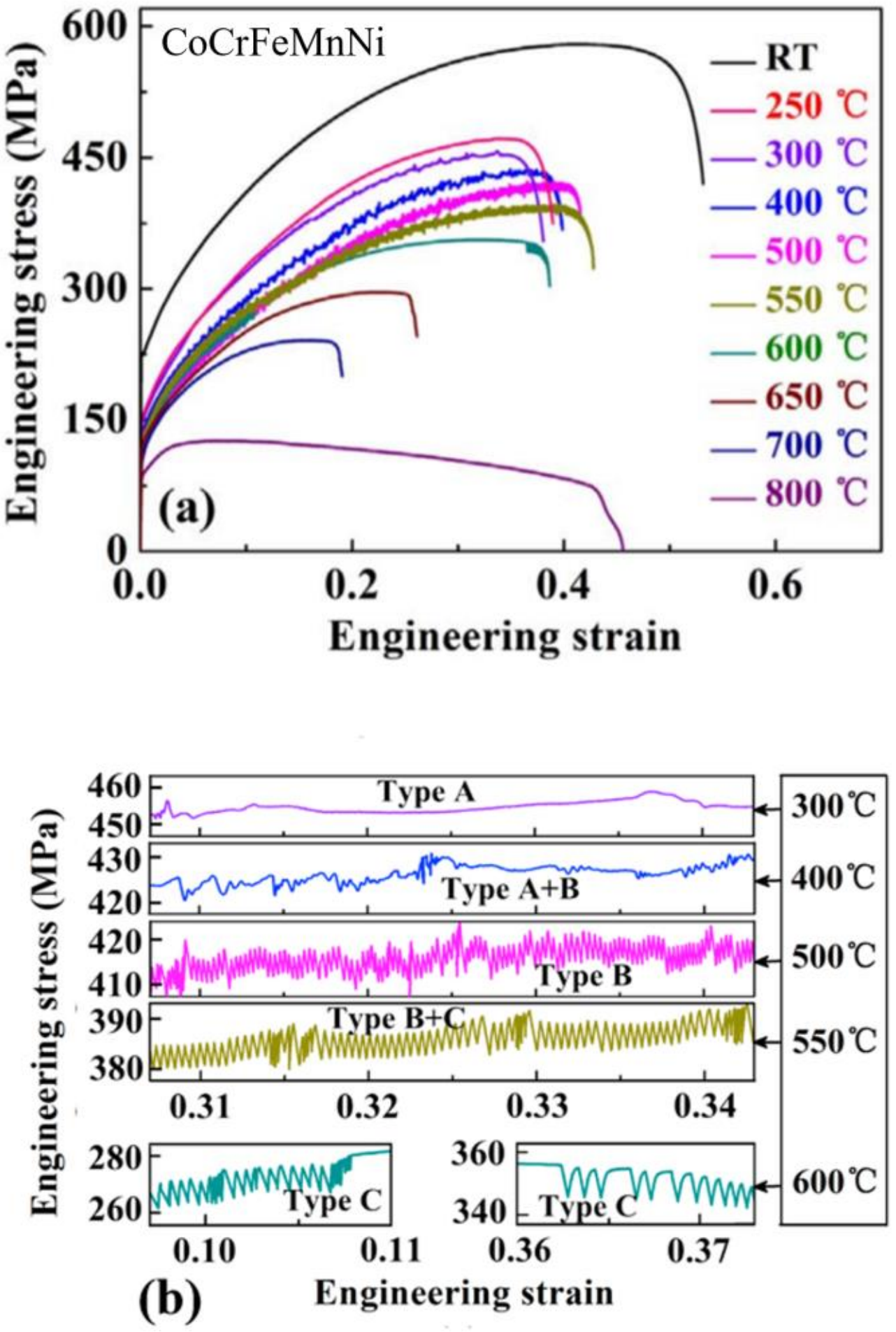

Fig. 11. (a) Engineering stress vs. strain data for the CoCrFeMnNi HEA subjected to tension under a strain rate of $3 \times 10^{-4}$ $s^{-1}$ and temperatures of RT - 800 °C and (b) the magnified regions of the curves that feature the serrated flow. Figures from Fu et al. [74].

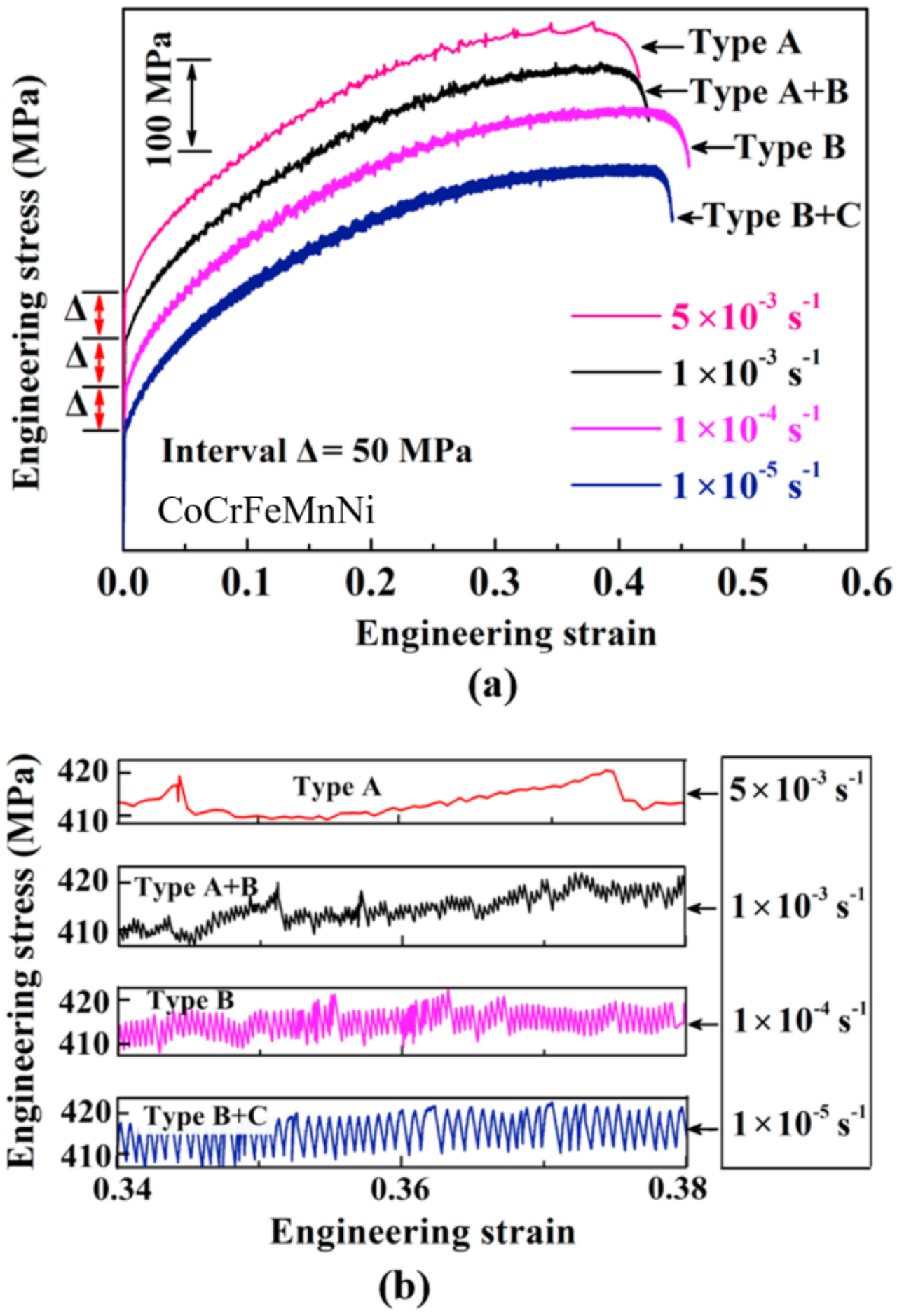

Fig. 12. (a) The engineering stress-strain data for the CoCrFeMnNi HEA specimens subjected to tension at a temperature of 500 °C and strain rates of $1.0 \times 10^{-5}$ - $5.0 \times 10^{-3}$ $s^{-1}$and (b) the magnified regions from (a) for engineering strain values of 0.34 - 0.38. Figures from Fu et al. [74].

Table 4. Summary of the serration type for the CoCrFeMnNi HEA that was subjected to tension at temperatures and strain rates of 300 °C - 600 °C and $1 \times 10^{-5}$ $s^{-1}$ - $5 \times 10^{-3}$ $s^{-1}$, respectively. From Fu et al. [74].

| Strain Rate ($s^{-1}$) | Temperature (°C) | Serration Type |
|---|---|---|
| $1 \times 10^{-5}$ | 500 | B + C |
| $1 \times 10^{-4}$ | | B |
| $3 \times 10^{-4}$ | 300 | A |
| | 400 | A + B |
| | 500 | B |
| | 550 | B + C |
| | 600 | C |
| $1 \times 10^{-3}$ | 500 | A + B |
| $5 \times 10^{-3}$ | | A |

Tirunilai et al. [187] studied the mechanical deformation of pure Cu and CoCrFeMnNi HEA that was tension tested at strain rates of $6 \times 10^{-5}$ - $1 \times 10^{-3}$ $s^{-1}$ and temperatures of 4.2 K - RT (295 K). Micrographs of the post-fracture surface indicated that the significant dislocation activity and deformation twinning had occurred during testing. Figure 13(a) presents the engineering stress-strain data for a strain rate of $3 \times 10^{-4}$ $s^{-1}$ and temperatures of 4.2 K - RT. For temperatures of 4.2 K and 8 K, the serrated flow could be observed, while there were no apparent serrations at 77 K and RT. For the specimen tested at 4.2 K, the ductility remained relatively high. Deformation twinning occurred at temperatures of 4.2 and 77 K, although it was concluded that this behavior had no effect on the serrated flow. Figure 13(b) shows the engineering-strain data for the pure Cu and HEA specimens subjected to tension at a strain rate and temperature of $3 \times 10^{-4}$ $s^{-1}$ and 4.2 K, respectively. For the HEA, the serration amplitude was ~ 150 MPa, while for the pure Cu it was 5 MPa. The comparatively larger serrations that occurred in the HEA could have been due to several different factors such as the heat capacity, work-hardening, and thermal conductivity of the material.

The stress-strain data for the HEA specimens that underwent tension at 8 K and strain rates of $6 \times 10^{-5}$ - $1 \times 10^{-3}$ $s^{-1}$ is presented in Fig. 14. From the results, it can be ascertained that the serration behavior was affected by the strain rate. For instance, the specimens compressed at a strain rate of $1 \times 10^{-3}$ $s^{-1}$ underwent a greater amount of strains between stress drops, as compared to the other strain rates. Furthermore, Type-B serrations for strain rates of $6 \times 10^{-5}$ and $3 \times 10^{-4}$ $s^{-1}$, and possibly Type-D serrations at a strain rate of $1 \times 10^{-3}$ $s^{-1}$, could be observed.

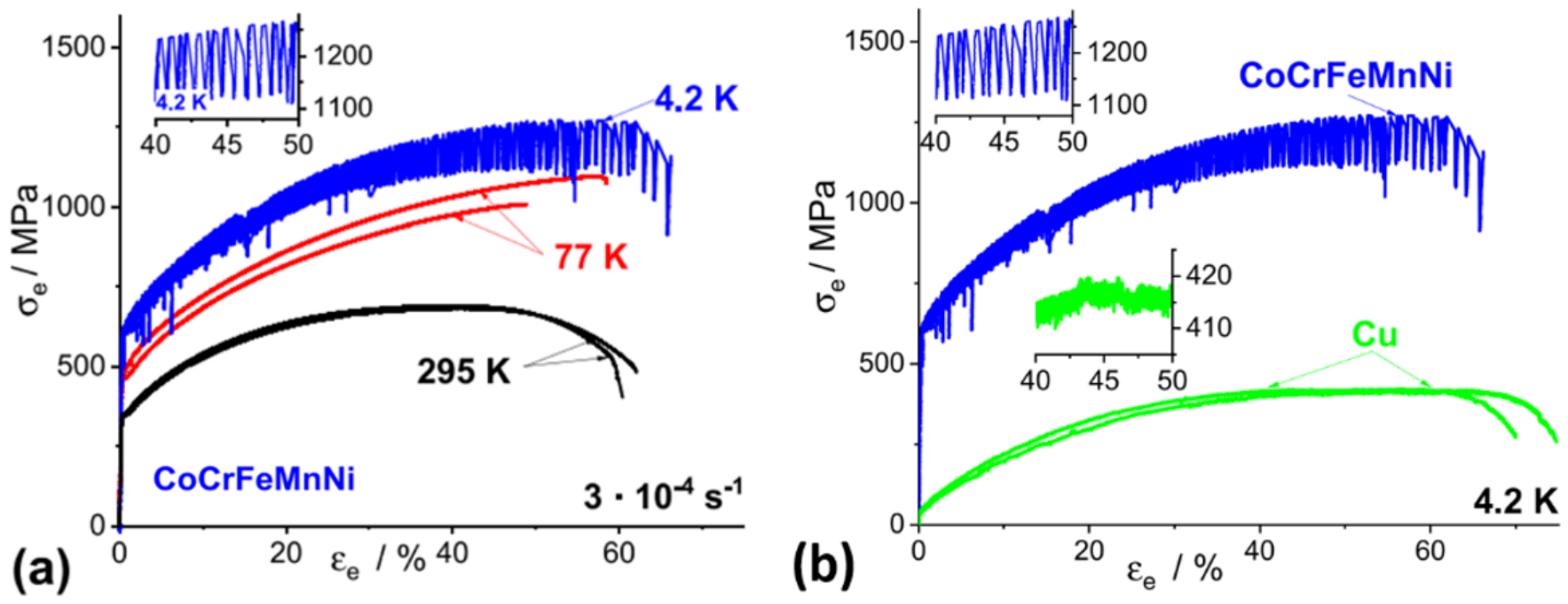

Fig. 13. The stress vs. strain data (during tension) for the (a) CoCrFeMnNi HEA at temperatures of RT - 4.2 K and (b) CoCrFeMnNi HEA and pure Cu at a temperature of 4.2 K. Figures from Tirunilai et al. [187].

Their findings led to some important observations. Firstly, the stress drops were not associated with a significant plastic deformation. Secondly, only a few of the stress drops were associated with a macroscopic elongation of the specimen. Thirdly, although the serrations may have been the result of dislocations overcoming an obstacle, it was not clear why the associated strain change was so small. It was hypothesized that the relatively-small strain changes may result from the strong localization of the individual events that macroscopically offset the elastic unloading of the specimen during the stress drop. Lastly, the influence of the thermal softening via adiabatic heating on the deformation behavior appears to be insignificant, considering the time scales involved.

In a subsequent investigation, Tirulini et al. investigated the deformation behavior of CoNi, CoCrNi, and CoCrFeMnNi alloys [188]. For the experiments, specimens were deformed at a strain rate of $3 \times 10^{-4}$ $s^{-1}$ and cryogenic temperatures of 4.2 K - 35 K. Figure 15 shows the engineering stress vs. engineering strain data for the CoCrFeMnNi HEA in which serrated flow can be observed at all of the temperatures. The experiments yielded several important results. It was found that temperature does have a significant effect on the serrated flow behavior, as the onset strain for serrations increased with temperature. Furthermore, the authors reported that neither deformation twinning nor ε-martensite formation are critical for, nor have an influence, on the serrated flow. Finally, the results suggested that the thermomechanical model of serrations was not valid as a cause for the stress-drop behavior that was exhibited by the specimens.

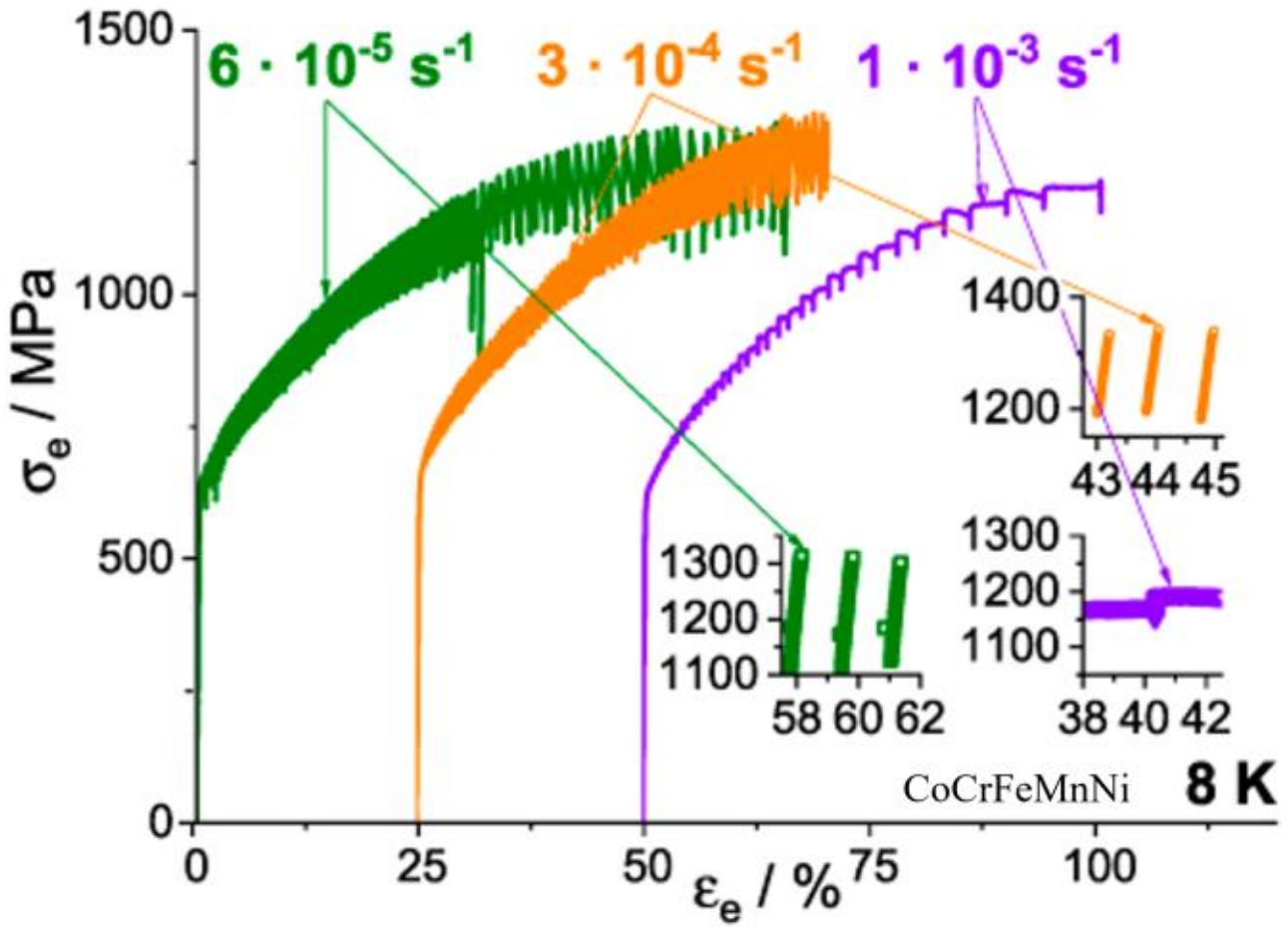

Fig. 14. The engineering stress vs. strain data for the CoCrFeMnNi HEA subjected to tension at strain rates of $6 \times 10^{-5}$ - $1 \times 10^{-3}$ $s^{-1}$ and a temperature of 8 K. Figure from Tirunilai et al. [187].

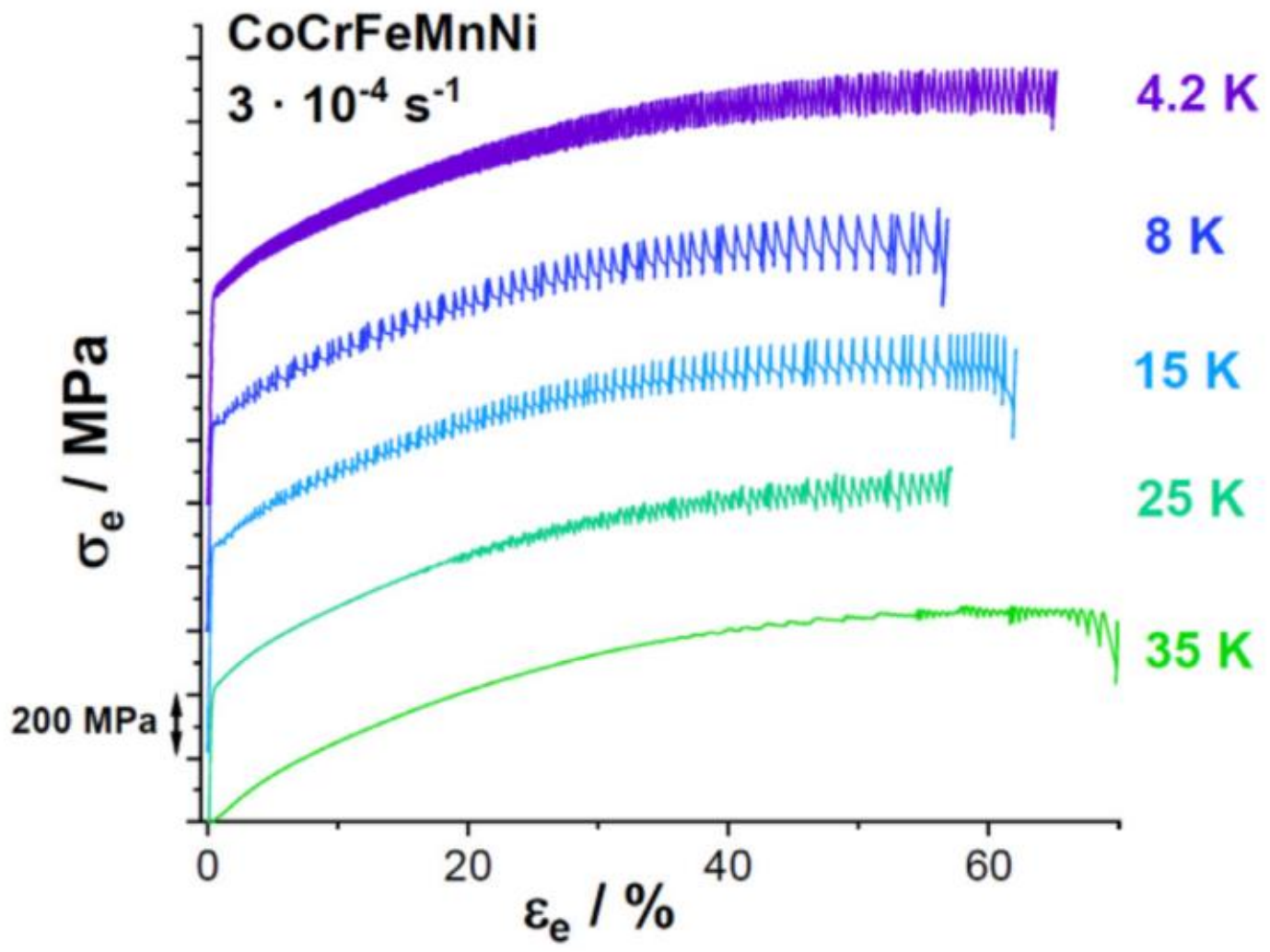

Fig. 15. Engineering stress ($\sigma_e$) – engineering strain ($\varepsilon_e$) data for the CoCrFeMnNi HEA deformed at a strain rate of $3 \cdot 10^{-4}$ $s^{-1}$ and temperatures of 4.2 K - 35 K. Figure from Tirunilai et al. [188].

In terms of the serration dynamics, the rate of stress drops was relatively higher at lower temperatures and was associated with a higher dislocation velocity.

It was surmised that this higher velocity was due to a drop in the viscous dampening of dislocations, which results from the vanishing phonon scattering [189, 190]. Figure 16 exhibits the deformation behavior of the CoNi, CoCrNi, and CoCrFeMnNi alloys during tension at a strain rate of $3 \times 10^{-4}$ $s^{-1}$ and temperature of 8 K. The binary alloy deformed significantly less and exhibited markedly smaller stress drops, as compared to the other two alloys. This result indicates that the serration behavior has a compositional dependence. Furthermore, the authors hypothesized that the difference in the serrated flow was possibly related to the stacking fault energy and/or the solute-dislocation interaction. In terms of the latter, the more compositionally complex alloys exhibited a severe solid-solution strengthening effect, which resulted in more pronounced serrations in the CoCrNi and CoCrFeMnNi alloys.

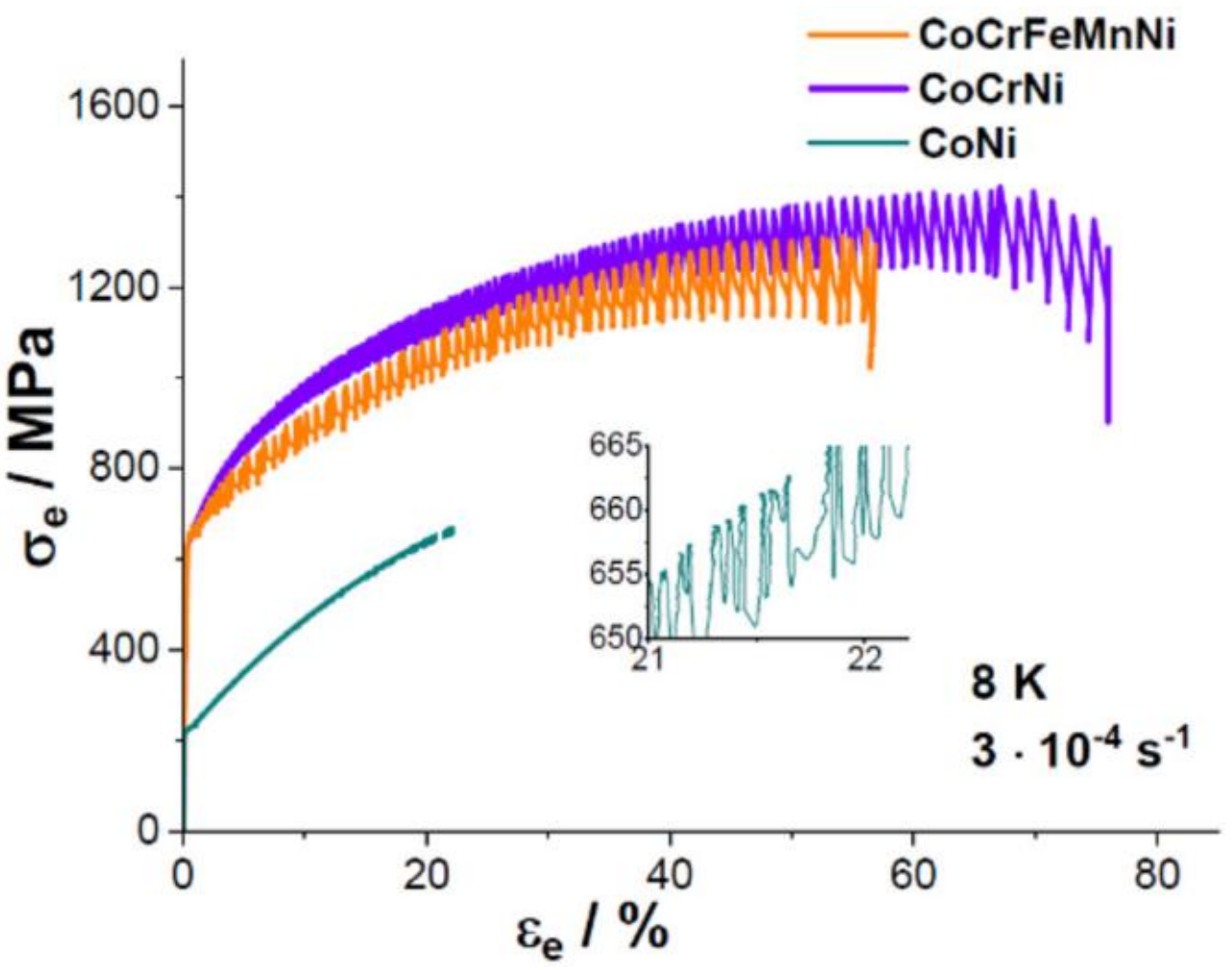

Fig. 16. $\sigma_e - \varepsilon_e$ curves of CoCrFeMnNi, CoCrNi, and CoNi alloys deformed at 8 K and a strain rate of $3 \cdot 10^{-4}$ $s^{-1}$. Figure from Tirunilai et al. [188].

In another study, Pu et al. [191] studied the mechanical behavior of the Cantor alloy that was tension tested at a strain rate of $10^{-3}$ $s^{-1}$ and temperatures of 4.2 – 293 K. Figure 17 exhibits the engineering stress vs. engineering strain data for the specimens tested under the prescribed conditions. It is apparent that only the specimen tested at 4.2 K displayed serrations. Using nonlinear analysis, the authors found that the underlying dynamics of the serrated flow at 4.2 K was low dimensional chaotic behavior that is characterized by a positive Lyapunov exponent and a finite correlation dimension. It was also determined that the mode of heat transfer at 4.2 K was due to electronic effects.

The TEM characterization revealed that there were significant microstructural differences between the specimens deformed at 4.2 K and 77 K. Figures 18 (a)-(d) display the TEM micrographs of the specimens tested at the above temperatures. The findings revealed that at 4.2 K, the specimen contained a significantly greater number of Lomer–Cottrell (L-C) locks, as compared to the higher temperature. These L-C locks form by the interaction of two leading Shockley partials that have dissociated from two perfect dislocations on two intersecting slip planes [191, 192]. From the results, it was thought that the serrated flow is caused by an interaction between dislocation inertial motion with L-C locks. Here, the L-C locks act as obstacles for moving dislocations, which causes them to pile-up until a critical stress is attained. Subsequently, the pile-up destabilizes, resulting in a stress drop.

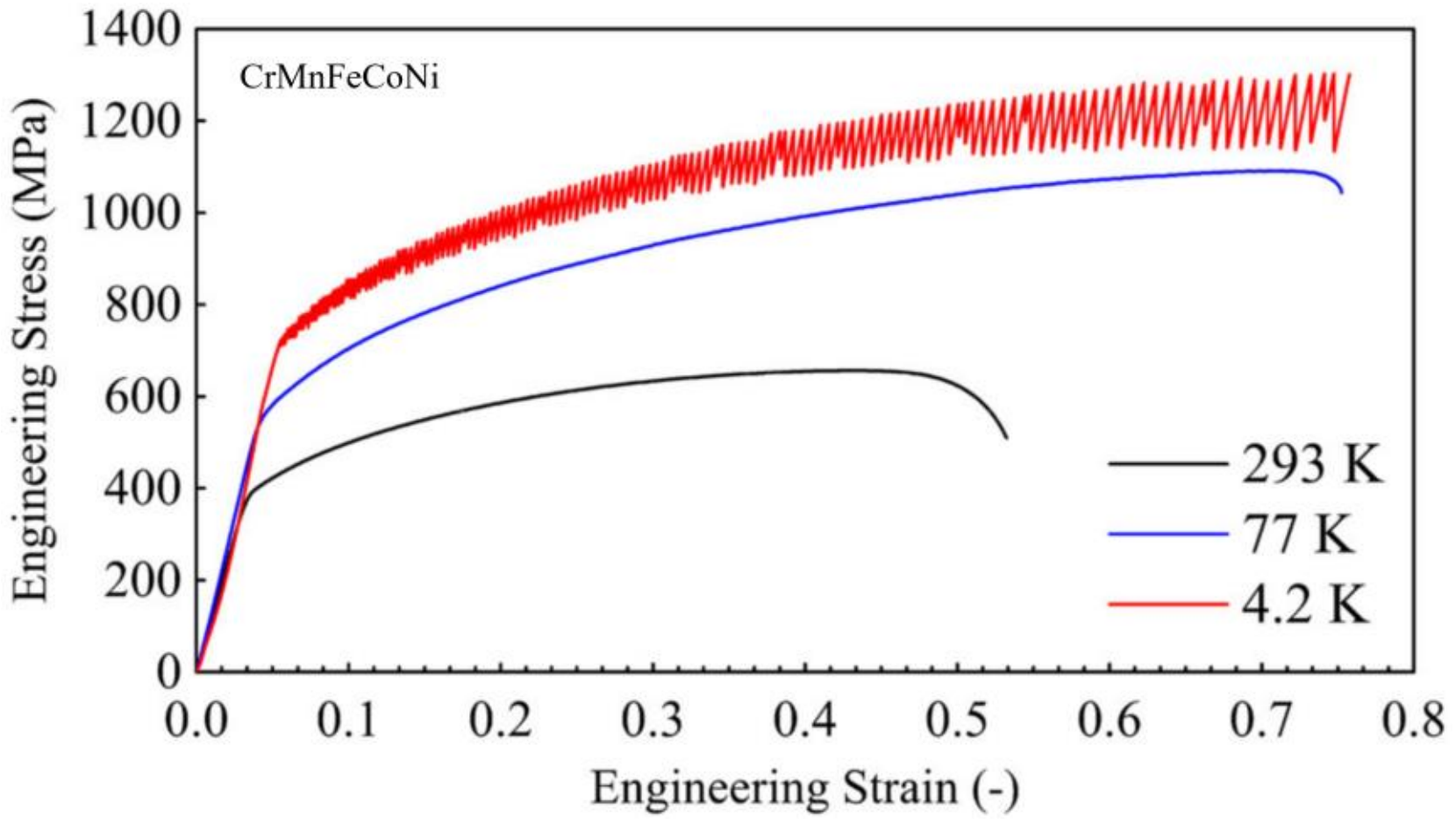

Fig. 17. Engineering stress–strain curves of the CrMnFeCoNi HEA tested at 4.2, 77, and 93 K and a strain rate of $10^{-3}$ $s^{-1}$. Figure from Pu et al. [191].

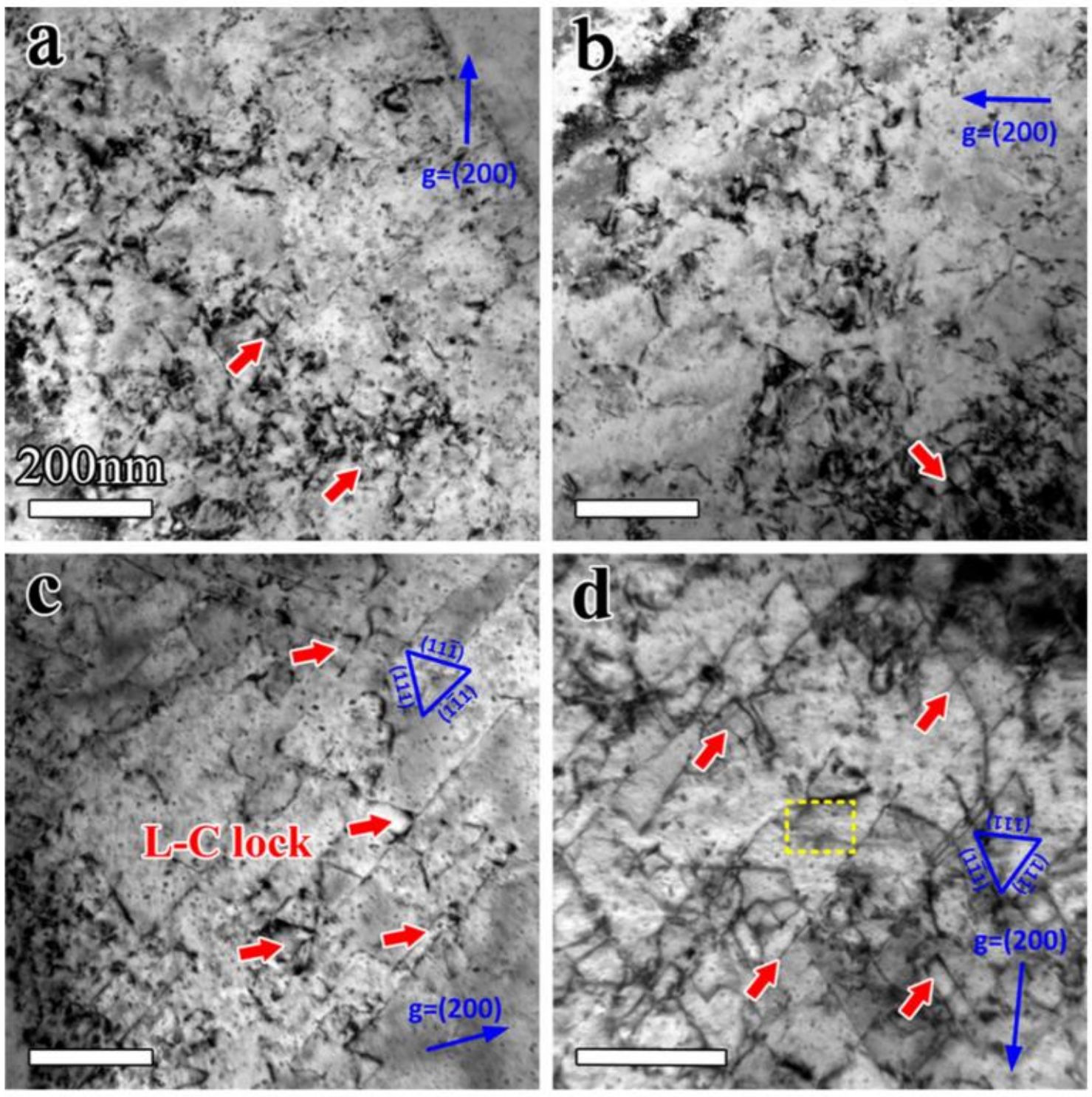

Fig. 18. TEM micrographs of the CrMnFeCoNi HEA at a strain of 8 % (strain rate of $10^{-3}$ $s^{-1}$) taken under the same diffraction condition of $g = (200)$, which features (a, b) a small amount of L-C locks that were present at 77 K, and (c, d) a large number of L-C locks that were observed at 4.2 K, as labelled by red arrows. All scale bars are 200 nm. Figures from Pu et al. [191].

Naeem et al. [193] investigated the deformation behavior of the Cantor alloy that underwent tension at a strain rate of $2.67 \times 10^{-5}$ $s^{-1}$ and temperatures of 15 K - 295 K. Serrations were only observed during the test at 15 K. The results also indicated that there were a rich variety of deformation mechanisms in the HEA at 15 K. Initially, the deformation mechanism was dislocation slip, which was followed by stacking faults and twinning, and then finally inhomogeneous deformation in the form of a serrated flow. It was determined that the combination of the stacking faults, twinning, and serrations resulted in minimum necking and more significant uniform elongation.

In [129], the impact of carbon impurities on the plastic-deformation behavior of a CoCrFeMnNi HEA was investigated. For the experiments, as-received and carbon-doped (0.93 atomic %) specimens underwent tension at a strain rate of $1.6 \times 10^{-3}$ $s^{-1}$ at RT. Figures 19(a)-(b) compare the true stress vs. strain of the two specimens, which shows that serrations (Type-A) were only exhibited by the doped specimen. Figure 19(b) features a more in-depth view of the serration behavior displayed by the doped specimen, which appears to consist of a stair-step type pattern indicative of Type-D serrations (see Fig. 1) [6]. The observable serrations in the carbon-doped HEA was thought to be a consequence of the interactions among the stacking faults (SFs), moving dislocations, and solutes [194, 195]. During a serration event, the passage of a dislocation leads to the reorientation of a carbon atom from octahedral sites and SFs into a tetrahedral interstice. This interaction results in the destruction of the ordered structure and accompanying stress increase. The SF energy will then be reduced, resulting in the pronounced increase in the plastic deformation that corresponds to the plateau region of the true stress curve in Fig. 19(b).

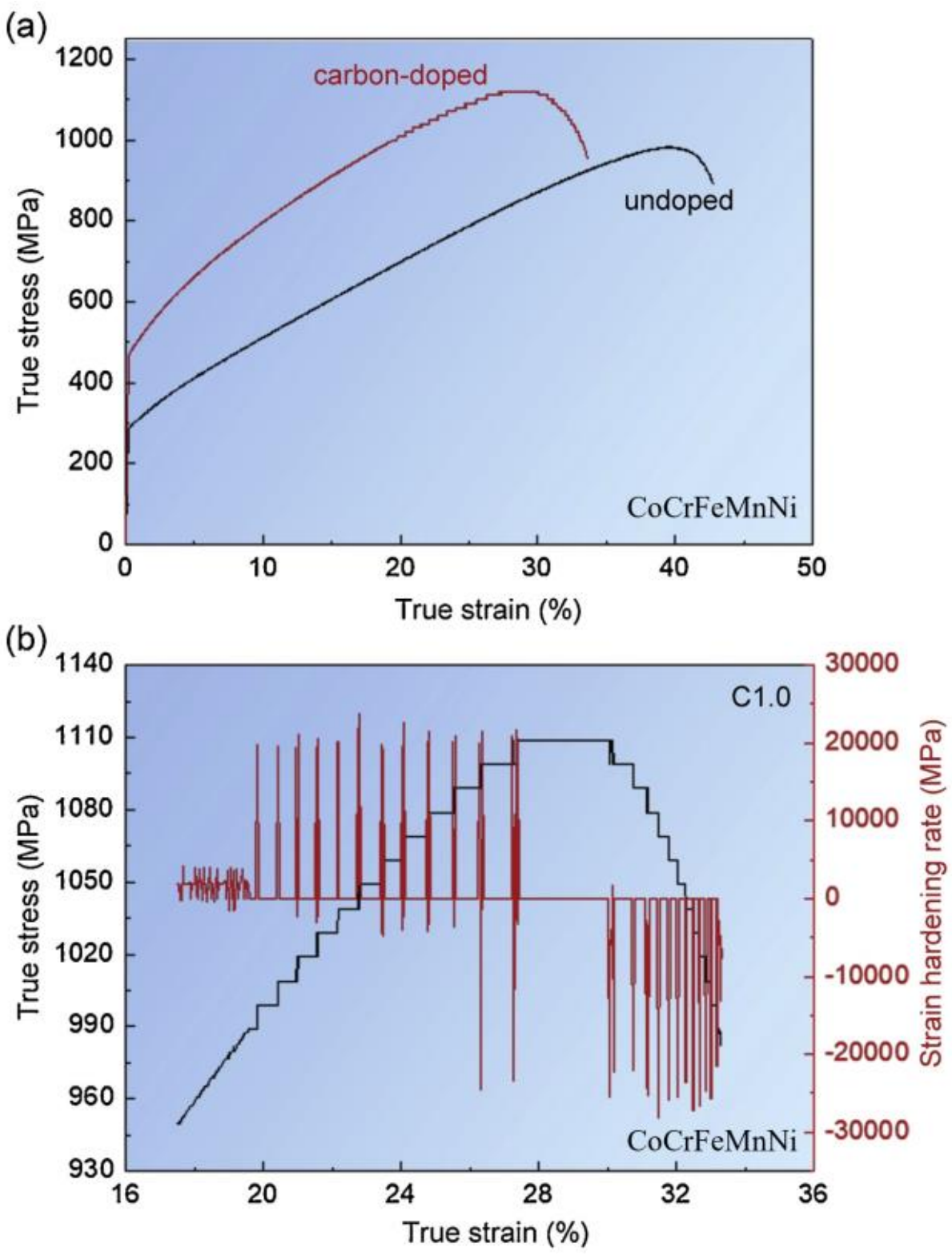

Fig. 19. (a) True stress vs. strain data for the as-received and carbon doped (0.93 at.% carbon) CoCrFeMnNi HEA subjected to tension at RT and a strain rate of 1.6 × $10^{-3}$ $s^{-1}$ and (b) an expanded image for the strain range of ~ 17.5 – 33.5%, which features both the true stress and strain hardening rate vs. true strain data for the carbon-doped HEA. Figures from Guo et al. [129].

### 3.1.6 HfNbTaTiZr HEA

In [196], the plastic-deformation behavior of a BCC HfNbTaTiZr HEA was examined. For the experiment, specimens underwent tension testing at a strain rate of $1 \times 10^{-4}$ $s^{-1}$ and temperatures of 77 - 673 K. Figure 20 displays the resulting engineering stress vs. strain data for the HfNbTaTiZr HEA. It was found that the yield and flow stresses decrease with an increase in the temperature. On the other hand, the ductility was found to increase with temperature. Furthermore, the serrated flow only occurred in the specimen tested at 673 K. For temperatures of 573 and 673 K, the strain hardening was the result of both forest and DSA hardening. It was suggested that although there were no serrations observed in the specimen tested at 573 K, DSA was still a likely-strengthening mechanism, since serrations may not always be visible during the plastic deformation [197].

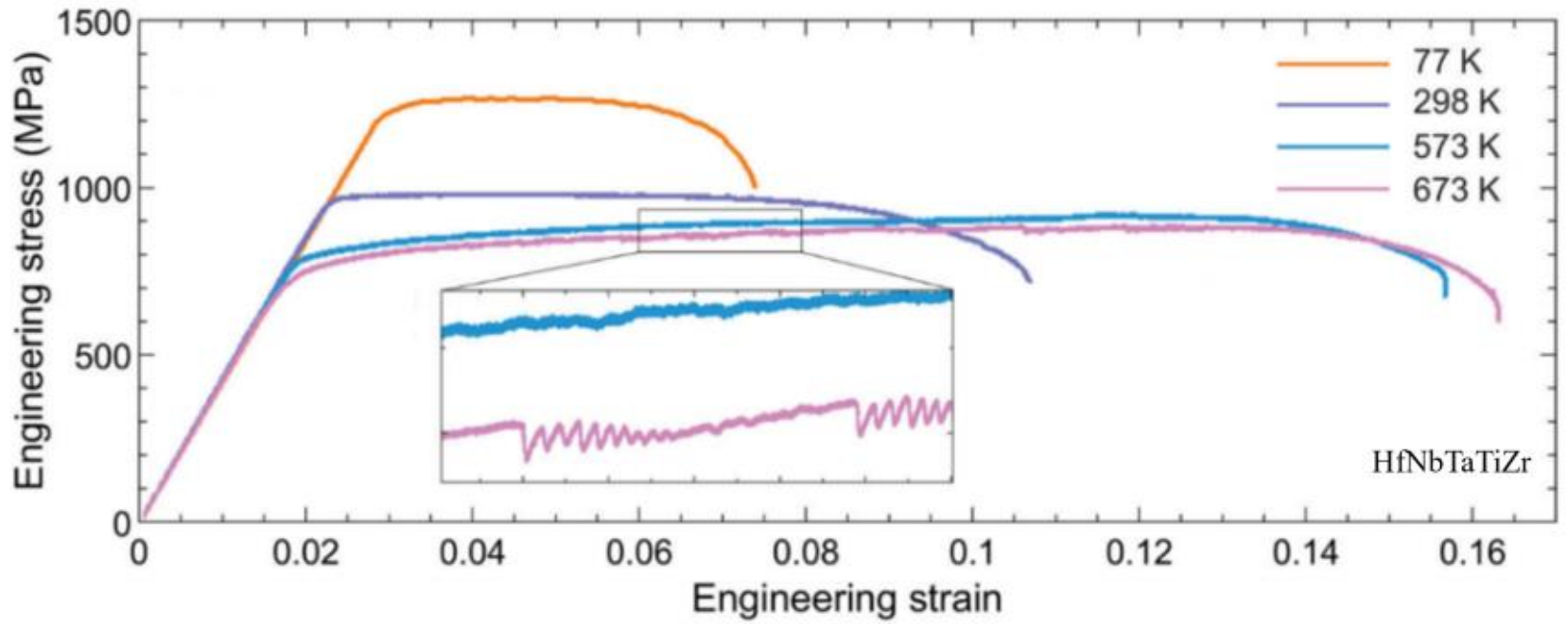

Fig. 20. The engineering stress-strain data for the HfNbTaTiZr BCC HEA subjected to tension at a strain rate of $1 \times 10^{-4}$ $s^{-1}$ and temperatures of 77 - 673 K. Figure from Chen et al. [196].

## 3.2 Compression testing

### 3.2.1 $Ag_{0.5}$CoCrCuFeNi HEA

Laktionova et al. [198] examined the plastic-deformation of an $Ag_{0.5}$CoCrCuFeNi HEA subjected to compression testing at a strain rate of $4 \times 10^{-4}$ $s^{-1}$ and temperatures of 4.2 - 300 K (- 269 – 27 °C). Specimens consisted of cylindrical geometry with a length and diameter of 4 mm and 2 mm (aspect ratio of 2), respectively. Helium vapor was used to lower the temperature to cryogenic conditions (below 77 K). The resulting load-time data were converted to stress-strain curves, which can be observed in Fig. 21. From the results, the following can be observed. Firstly, it is evident that serrations (Type-B) were only observed at cryogenic temperatures of 4.2 and 7.5 K. Secondly, the serration magnitude increased with respect to the amount of strain deformation, where it went from 20 MPa at a strain of ~ 2% to 67 MPa at a strain of ~ 23%. Finally, the critical strain rose with temperature, where it increased from a strain of 1.2% at 4.2 K to 4.5% at 7.5 K.

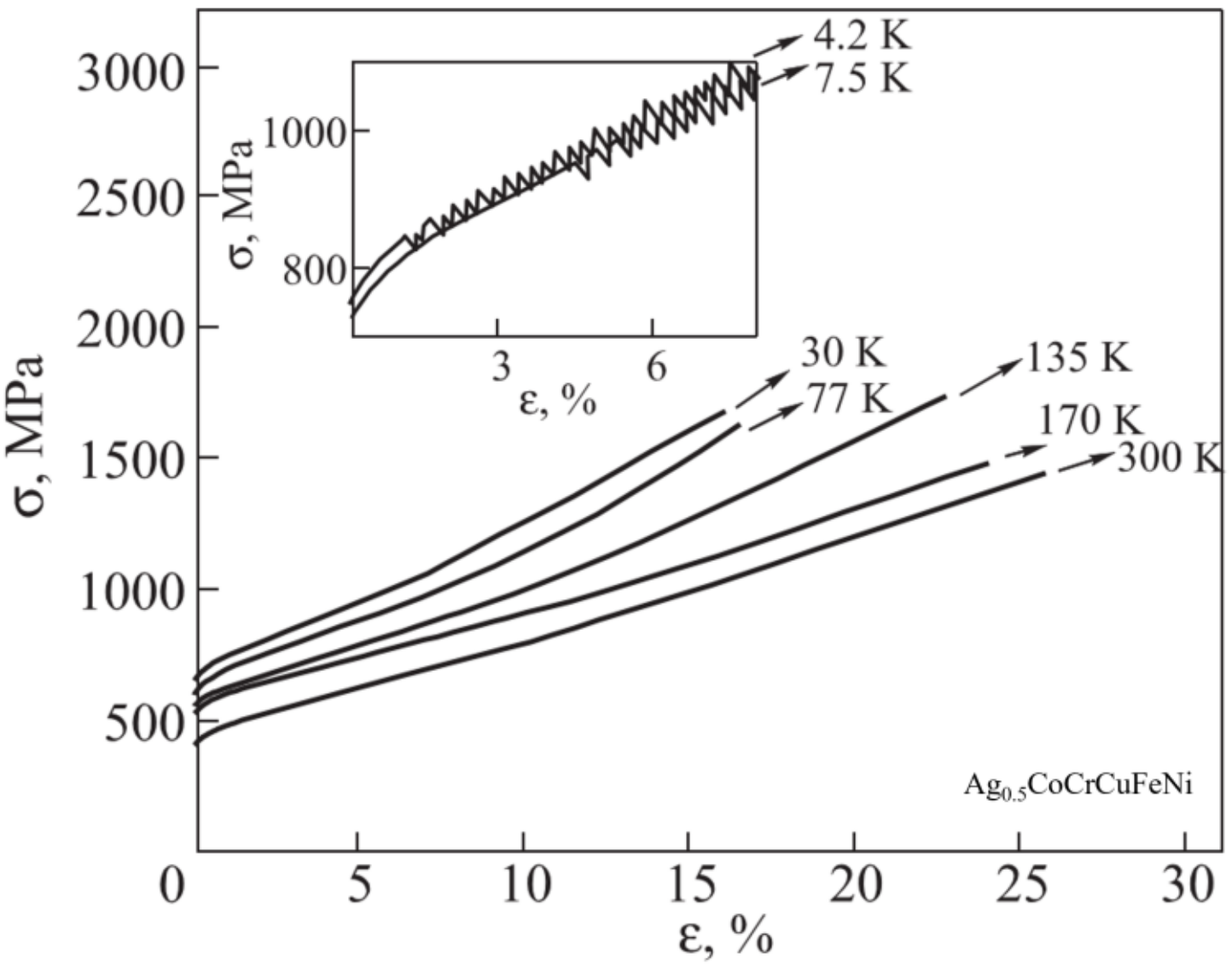

Fig. 21. The stress vs. strain data for the $Ag_{0.5}CoCrCuFeNi$ HEA specimens compressed at a strain rate of 4 × $10^{-4}$ $s^{-1}$ and temperatures of 4.2 - 300 K. The inset shows a magnified region of the graph which consists of the serrations that occurred at temperatures of 4.2 and 7.5 K. Figure from Laktionova et al. [198].

### 3.2.2 $Al_{0.1}CoCrFeNi$ HEA

In [199], the deformation mechanisms of an $Al_{0.1}CoCrFeNi$ HEA that underwent compression was investigated. Here, specimens were subjected to a strain rate of 2 × $10^{-4}$ $s^{-1}$ and temperatures of 77 K - 298 K was investigated. Although serrations were not reported in the specimens tested at 200 K and 298 K, they were observed at 77 K. Figures 22(a)-(d) present the results of the TEM characterization of the microstructure after testing. Figures 22(a)-(b) display the microstructures of the compressed specimens for temperatures of 200 and 298 K, respectively. Further analysis indicated that the plastic deformation occurred primarily via the planar slip of ½ < 110> type dislocations on {111}-type planes [6]. Such a finding suggests that the plastic deformation was a consequence of the dislocation glide at these temperatures. Such behavior has been previously observed in other FCC solid solutions [200]. Figures 22(c)-(d) show the results of the TEM characterization for the specimen compressed at 77 K. The results of the characterization revealed that nanoscale deformation twins (widths on the order tens of nanometers) led to {111} <110> primary slip systems that were primarily responsible for the serrated flow.

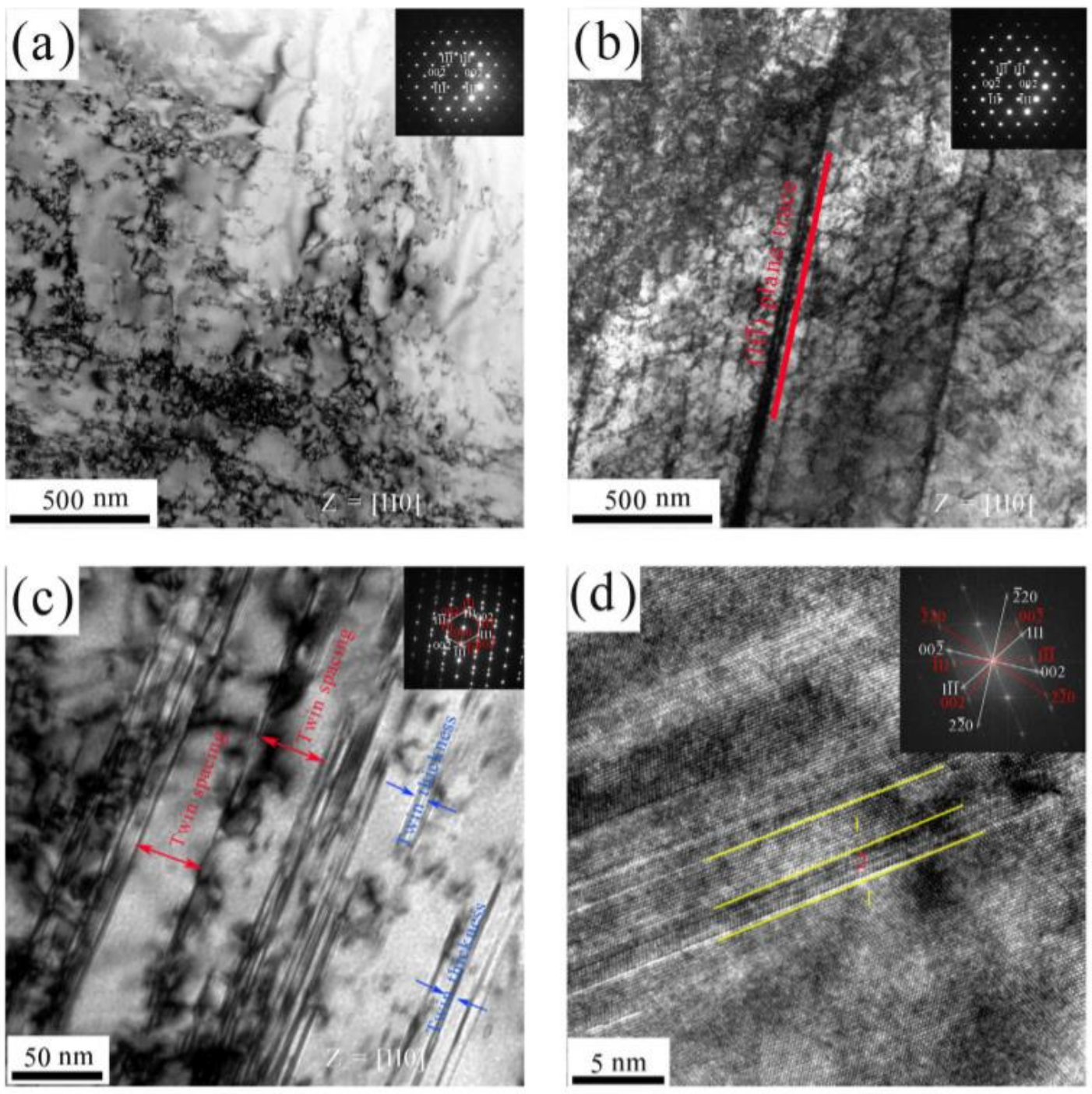

Fig. 22. TEM BF images of the $Al_{0.1}CoCrFeNi$ alloy with the corresponding SAED pattern and fast Fourier transform [zone axis (110)] for the conditions of (a) 25 °C , where the dislocation-cell formation occurred for a strain of 50%, (b) − 73 °C with a (111)-type slip plane present, (c) − 196 °C, where nanotwins can be observed, and (d) the high-resolution TEM (HRTEM) image for (c) which shows the region containing two deformation twins (~ 2 nm thickness). Figures from Xia et al. [199].

Hu et al. [71] utilized the MFT to analyze the serration dynamics of compressed $Al_{0.1}CoCrFeNi$ HEA nanopillars. The compression tests were performed, using a Hysitron PI95 picoindenter equipped with a 2-μm flat punch diamond indenter. For the experiments, an applied strain rate of ~ $1 \times 10^{-3}$ $s^{-1}$ was used. During the test, the time-dependent load and displacement data were obtained for a data acquisition rate of 500 Hz. Direct TEM imaging was employed to characterize the pillars (diameters of 500 – 700 nm) during the tests.

The stress vs. time data for the compression test is shown in Fig. 23(a), which reveals that multiple slip events occurred during the tests. The deformation process was separated into three distinct stages, which are labelled I, II, and III in the figure. The first stage consisted of few or no stress drops, the second stage contained medium-sized stress-drops, and the third stage was characterized by relatively-larger stress drops. Figures 23(b)-(d) display the microstructural evolution (as observed via TEM) of the specimen during the three stages of the deformation, respectively. The SEM image of the deformed nanopillar is displayed in Fig. 23(e). The deformation data and the microscopy characterization revealed the following for the distinct deformation stages. It was found that during the first stage, the serrations were attributed to the wave-like propagation of dislocations from the top to the bottom of the pillar [6]. As for the second stage, the serration events corresponded to small dislocation avalanches. The third stage, on the other hand, was associated with dislocation avalanches which result in large crystal slips.

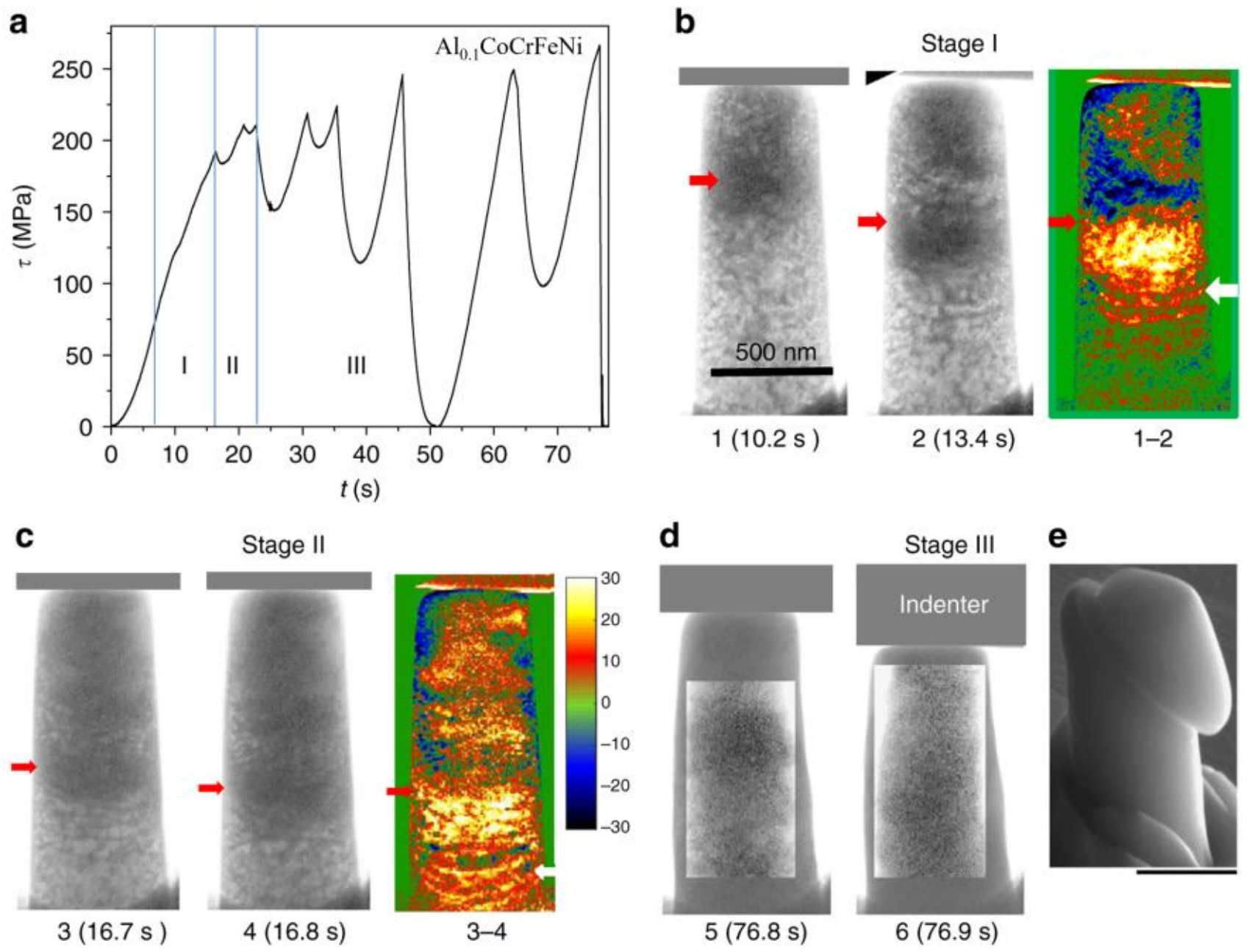

Fig. 23. The *in situ* micropillar compression and corresponding TEM-imaging results for the $Al_{0.1}CoCrFeNi$ HEA for the (a) stress-time data where Stages I, II, and III are demarcated in the graph with the corresponding TEM images of the microstructure during (b) Stage I, (c) Stage II, and (d) Stage III (with color-coded deformation stages) of the serrated flow, and (e) the SEM image of the deformed nanopillar, which shows the large crystal-slip that occurred during deformation. Figures from Hu et al. [71].

Figures 24(a)-(b) display the stress-binned CCDF for slip sizes, *S*, as a function of the stress level and its corresponding collapsed graph. The CCDF for this model was $D(S, \sigma) = S^{-(\kappa-1)}g[S(\sigma-\sigma_c)^{1/\beta}]$, in which $\sigma$ is the applied stress, $\sigma_c$ is the failure stress, and $\kappa$ and $\beta$ are critical exponents. Also, $g$ is defined as a universal scaling function, $x^{\kappa-1}\int_x^{\infty} e^{-At}t^{-\kappa}dt$ (here $A = 1.2$) [182]. Figure 24(a) presents the CCDF of slip sizes for different stress levels over the maximum stress. It is apparent that the curves shifted to the right with the increasing stress. The scale collapse of the four curves is displayed in Fig. 24(b), where the critical exponents ($\kappa$ and $\beta$) were determined to be 1.5 and 0.5, which is consistent with the predicted MFT [182].

The results revealed a few important findings regarding the serrated-flow phenomenon in HEAs. The avalanche mechanism responsible for the serrated flow in HEAs is a consequence of the interactions between the dislocation pileup and dislocation band, as well as between the pileup and dislocation centers [71]. Moreover, a coarse-grained model can be used to predict the distribution of the stress-drop magnitudes [133, 182]. Finally, the serrated flow exhibits the tuned critical behavior instead of self-organized criticality.

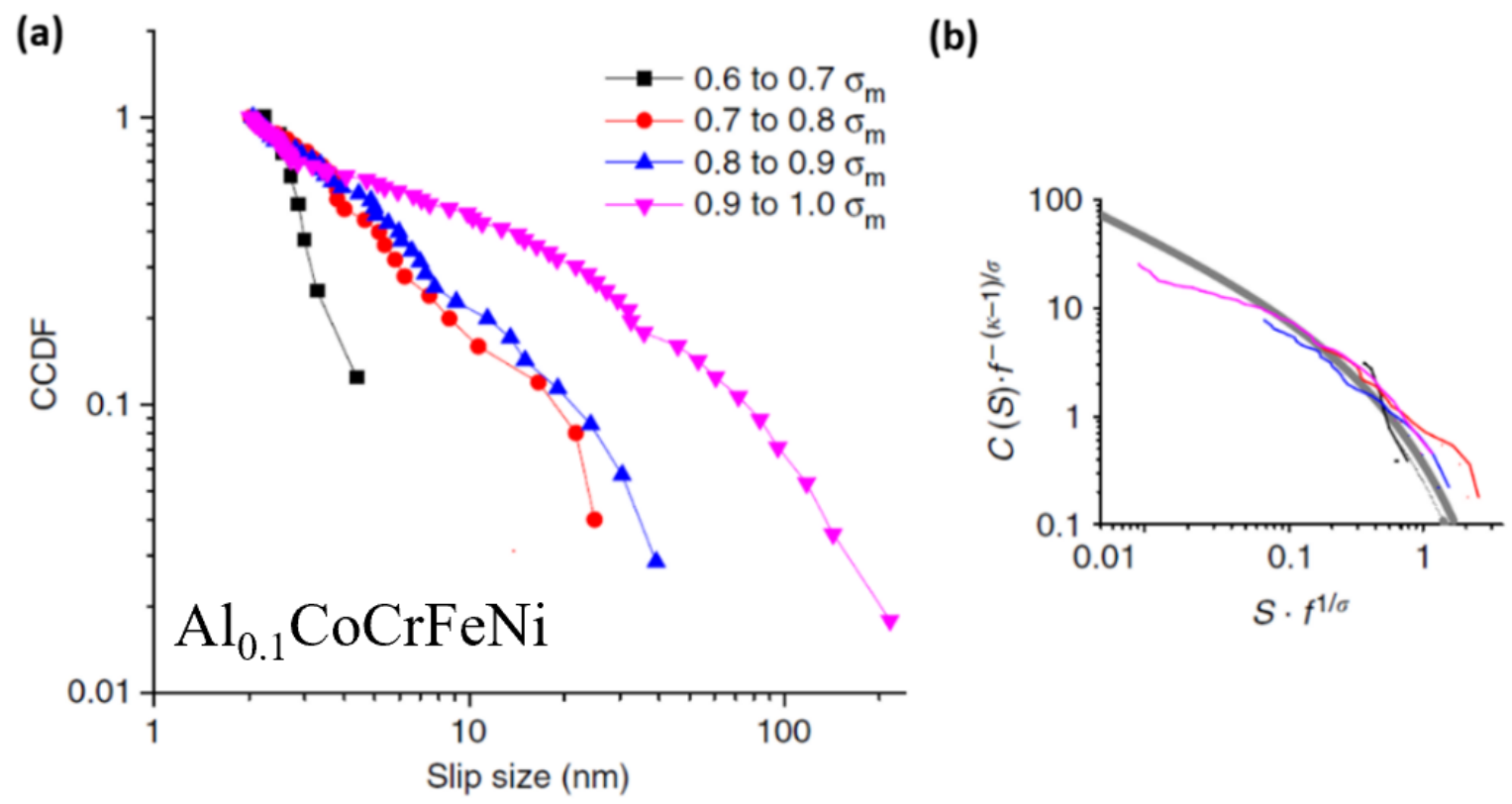

Fig. 24. (a) The stress-binned CCDF, as a function of slip sizes, for different fractions of the maximum stress that underwent compression at a strain rate of $1 \times 10^{-3}$ $s^{-1}$ and (b) The corresponding scale collapse with the predicted scaling function. Figures from Hu et al. [71].

### 3.2.3 $Al_{0.3}CoCrFeNi$ HEA

In [128], the mechanical response of a single-crystal CoCrFeNi and $Al_{0.3}CoCrFeNi$ HEAs was examined. Here, specimens underwent compression at a strain rate of $1.7 \times 10^{-4}$ $s^{-1}$ and temperatures of 93 - 1,273 K (− 180 - 1,000 °C). Figure 25 exhibits the stress-strain data for the experiments. At 873 K (600 °C), the $Al_{0.3}CoCrFeNi$ HEA displayed serrations while the CoCrFeNi HEA did not. Furthermore, serrations were detected in the Al-containing HEA at intermediate temperatures of 600 and 800 °C. It was suggested that the serrated flow resulted from the Al-containing solute atmospheres. In this scenario, the atmospheres are created near a moving dislocation core, thereby increasing the frictional stress on the dislocations. Importantly, the finding of this study suggests that in HEAs, perhaps not every solute atom can participate in dislocation locking.

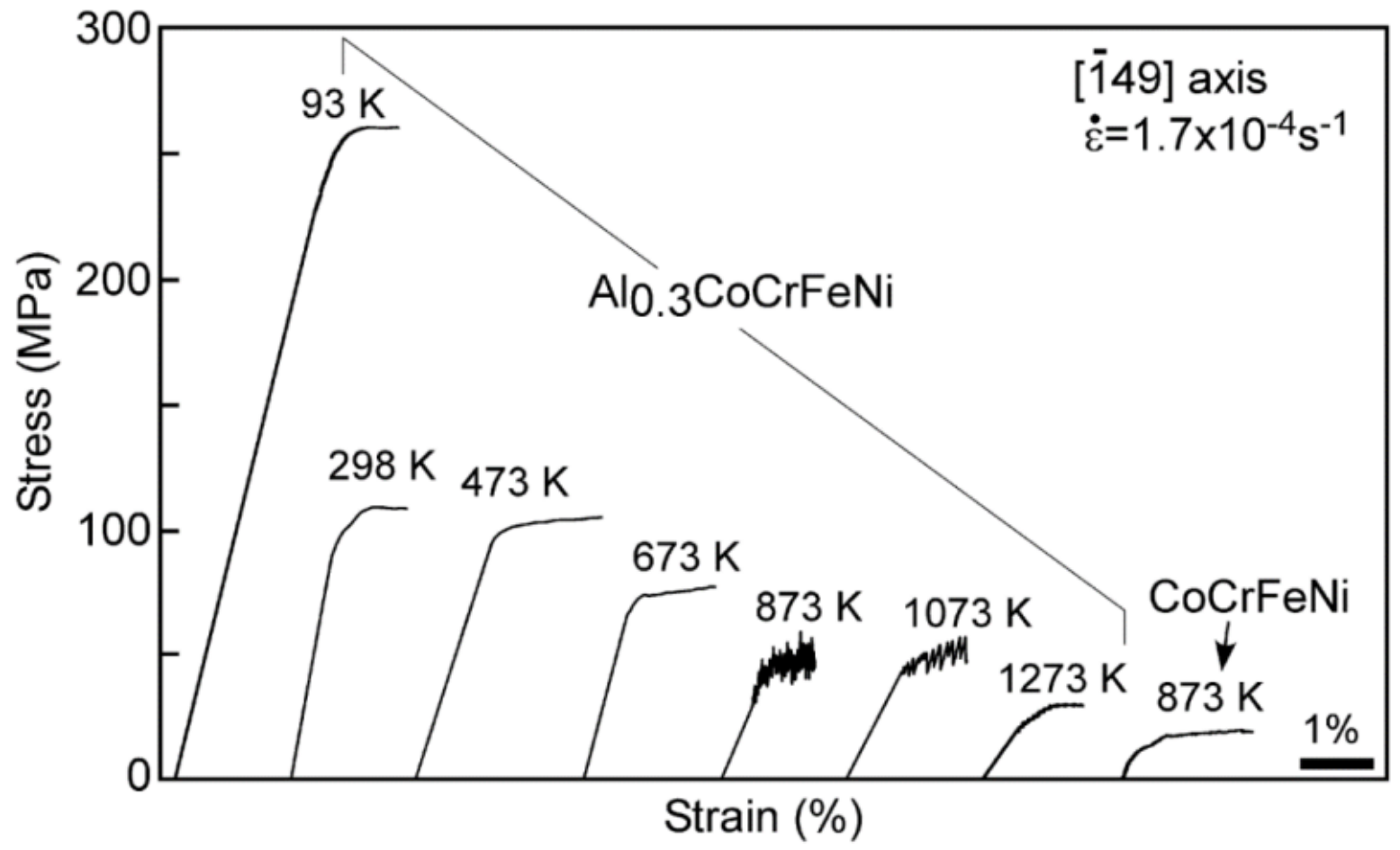

Fig. 25. The stress vs. strain data for the $Al_{0.3}CoCrFeNi$ and CoCrFeNi single crystal HEAs subjected to compression testing at a strain rate of $1.7 \times 10^{-4}$ $s^{-1}$ at temperatures of 93 K - 1,273 K (- 180 to 1,000°C). Figure from Yasuda et al. [128].

A similar study was performed in [81], where the plastic-deformation behavior of the $Al_{0.3}CoCrFeNi$ HEA (under compression) was investigated at a strain rate of $10^{-3}$ $s^{-1}$ and temperatures of 400 - 900 °C. The true stress-true strain behavior is featured in Figs. 26(a)–(d). There were apparent serrations in the deformation curves at temperatures of 400 – 800 °C. Table 5 displays the serration type for the different temperature conditions. Type-C serrations were observed at temperatures of 400, 500, and 600 °C, whereas Types-B and C serrations were both observed at 700 °C. At 800 °C, Type-B serrations were detected. Importantly, an increase in the temperature led to a decrease in the number of relatively-larger stress drops. Type-A serrations were not observed at any temperature, which contrasts with previous investigations [1]. It was also suggested that Type-B serrations were a result of the repeated pinning and unpinning of migrating dislocations [6]. Finally, it was thought that Type-C serrations result from microstructural changes induced by the nucleation and growth of twins [5, 128].

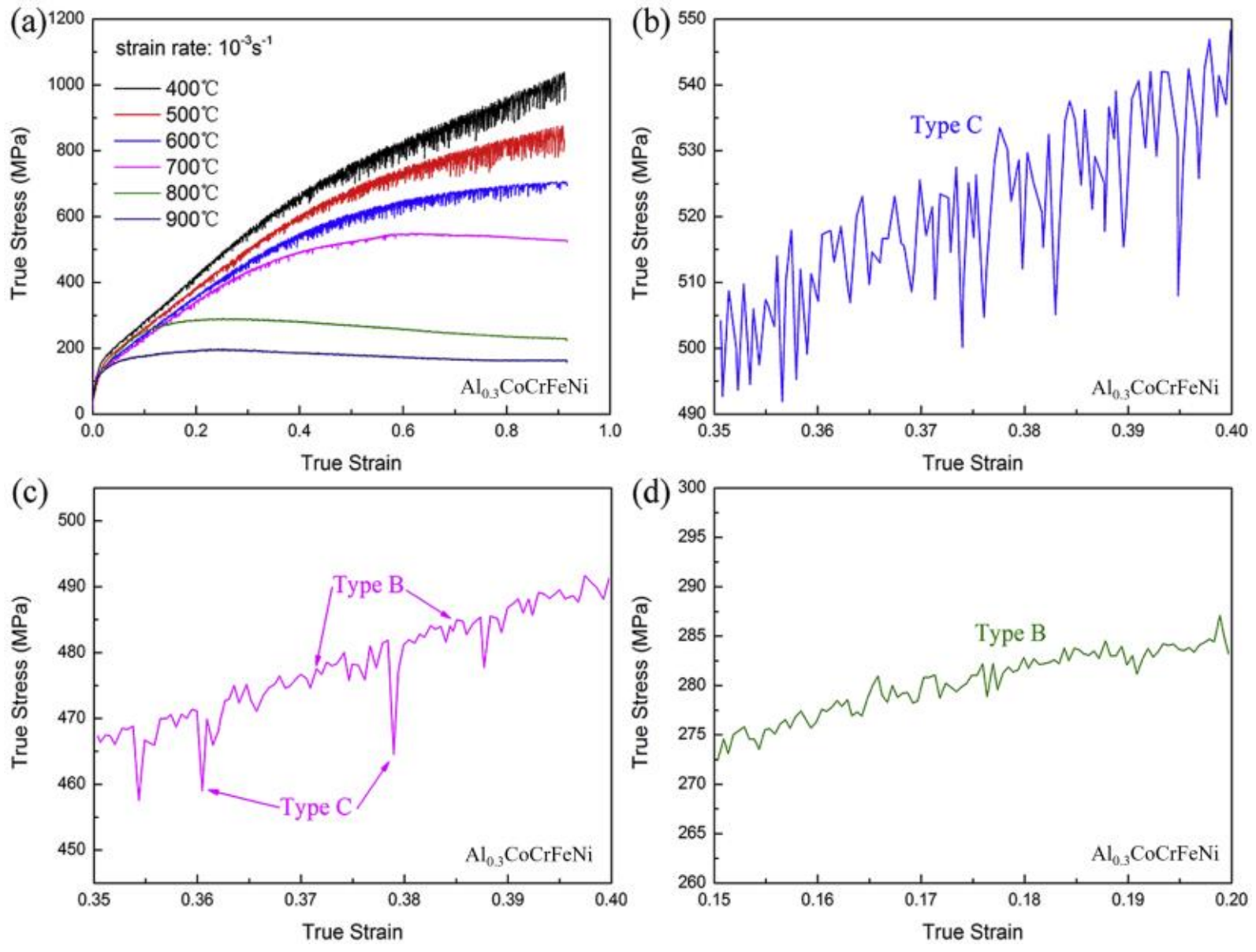

Fig. 26. (a) True stress vs. strain data for the $Al_{0.3}CoCrFeNi$ HEA compressed at a strain rate of $1 \times 10^{-3}$ $s^{-1}$ and temperatures of 400 - 900 °C, and the corresponding magnified regions for the temperature conditions of (b) 600 °C, (c) 700 °C, and (d) 800 °C. Figures from Zhang et al. [81].

Table 5. The serration type displayed by the $Al_{0.3}CoCrFeNi$ HEA during compression at temperatures of 400 - 800 °C and a strain rate of $1 \times 10^{-3}$ $s^{-1}$. From Zhang et al. [81].

| **Strain Rate ($s^{-1}$)** | **Temperature (°C)** | **Serration Type** |
|---|---|---|
| $1 \times 10^{-3}$ | 400 | C |
| | 500 | C |
| | 600 | C |
| | 700 | B + C |
| | 800 | B |

### 3.2.4 $Al_{0.5}CoCrCuFeNi$ HEA

In [91], the serrated flow of the $Al_{0.5}CoCrCuFeNi$ HEA was analyzed, using the multifractal and RCMSE techniques. For the experiments, specimens underwent compression at strain rates of $5 \times 10^{-5}$ $s^{-1}$ - $2 \times 10^{-3}$ $s^{-1}$ and temperatures of 400 - 600 °C. Figures 27(a)-(c) display the stress vs. strain curves of the experiments, and it is apparent that the serrated flow was observed at all of the experimental conditions. Table 6 displays the serration type for the prescribed strain rates and temperatures. From the figure and table, it is evident that the serration type was affected by the strain rate and temperature. For example, Type-C serrations were detected at 600 °C, whereas Types-A and B serrations were observed at 400 and 500 °C.

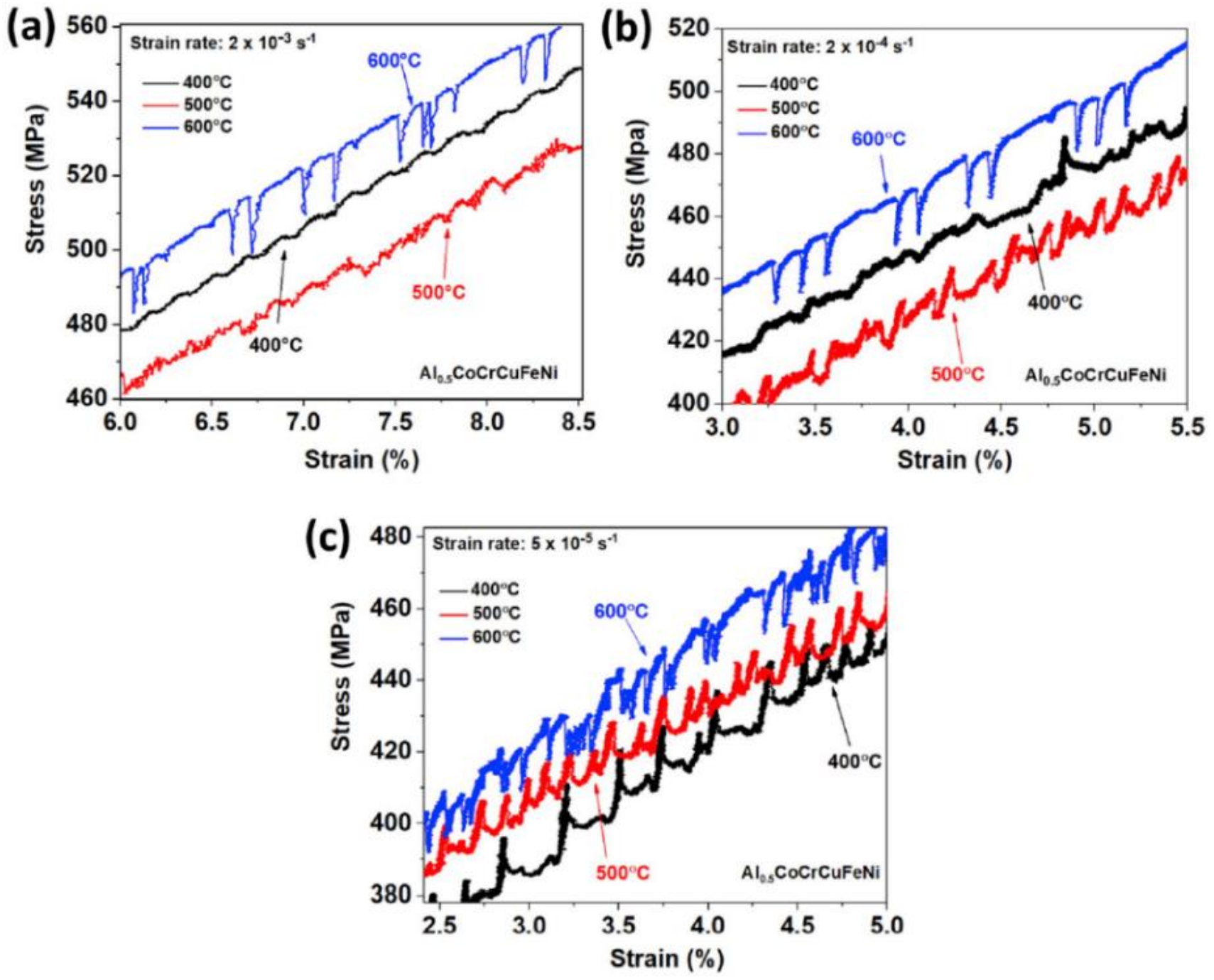

Fig. 27. Graph of the stress vs. strain data for the $Al_{0.5}CoCrCuFeNi$ HEA specimens compressed at temperatures of 400 - 600°C and strain rates of (a) $2 \times 10^{-3}$ $s^{-1}$, (b) $2 \times 10^{-4}$ $s^{-1}$, and (c) $5 \times 10^{-5}$ $s^{-1}$. Figures from Brechtl et al. [91].

Table 6. Serration types for the $Al_{0.5}CoCrCuFeNi$ HEA that was compressed at temperatures of 400 – 600 °C and strain rates of $5 \times 10^{-5}$ s$^{-1}$ - $2 \times 10^{-3}$ s$^{-1}$. From Brechtl et al. [91].

| Strain Rate (s$^{-1}$) | Temperature (°C) | Serration Type |
|---|---|---|
| $5 \times 10^{-5}$ | 400 | A |
| | 500 | B |
| | 600 | C |
| $2 \times 10^{-4}$ | 400 | A |
| | 500 | B |
| | 600 | C |
| $2 \times 10^{-3}$ | 400 | A |
| | 500 | A |
| | 600 | C |

The serrated flow was analyzed, using the RCMSE and multifractal methods. Figures 28(a)-(c) display the results of the RCMSE analysis. It was found that the sample-entropy values were the lowest for the specimens tested at 600 °C, which indicates that the serrations exhibited the least complex behavior in this condition. In contrast, the sample-entropy curves displayed the highest values at 500 °C (scale factors greater than 2), suggesting that the serration dynamics exhibited the most complex behavior. Based on the results from Fig. 28 and Table 6, it can be said that the Type-C serrations exhibit comparatively less complex behavior, as compared to Types-A and B serrations. It was suggested that the serrations that occurred at the highest temperature consisted of the repeated pinning and unpinning of dislocations by solutes [64, 91]. Such behavior can be compared to a falling row of dominoes, which is simple in nature [91]. As for the Types-A and B serrations, the more complex serration behavior may correspond to more complex interactions and deterministic chaos (Type-B serrations) [14]. The interactions could encompass intricate exchanges between solute atoms, dislocation lines, and precipitates [6, 91]. A similar conclusion (less complexity) was drawn for Type-C behavior of the PLC effect in conventional alloys, using the analyses of statistical distributions and multifractality. It was interpreted in terms of approaching the ideal case of relaxation oscillations due to synchronization of the dislocations.

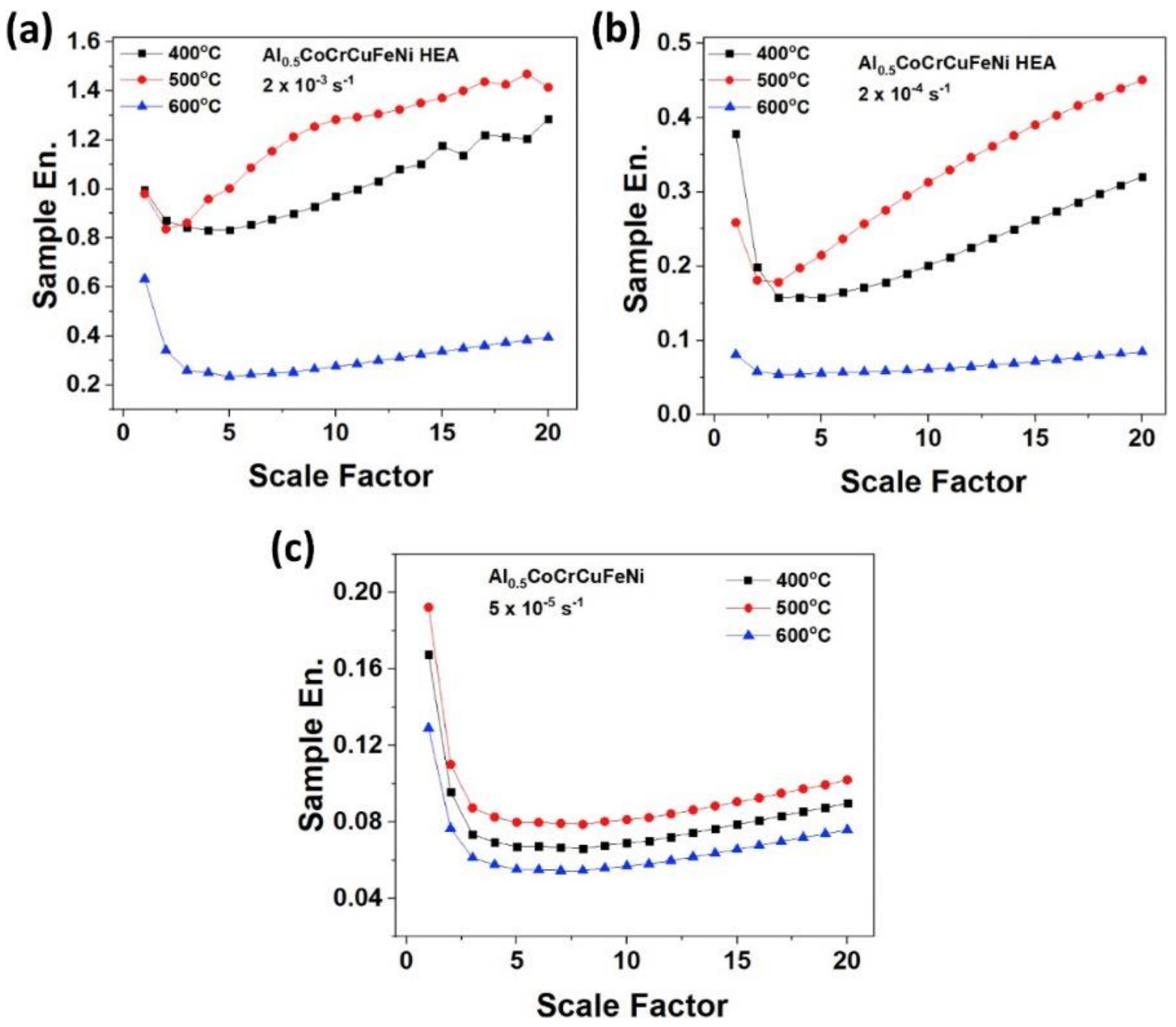

Fig. 28. The sample-entropy vs. scale factor for the $Al_{0.5}CoCrCuFeNi$ HEA specimens compressed at temperatures of 400 - 600 °C with strain rates of (a) 2 × $10^{-3}$ $s^{-1}$, (b) 2 × $10^{-4}$ $s^{-1}$, and (c) 5 × $10^{-5}$ $s^{-1}$. Figures from Brechtl et al. [91].

The accompanying multifractal spectra for the specimens tested at 5 × $10^{-5}$ $s^{-1}$ - 2 × $10^{-3}$ $s^{-1}$ and temperatures of 400 - 600 °C are presented in Figs. 29(a)-(c). The results revealed that all the spectra exhibited the typical inverted parabolic shape with a maximum value of ~ 1. For each strain rate, the spread of the multifractal curve, i.e., the multifractality, was the widest for the specimen tested at 600 ºC. This finding suggests that the Type-C serrations exhibited the greatest multifractality, and thus, the most dynamical heterogeneity, as compared to Type-A and B serrations. For a strain rate of 2 × $10^{-3}$ $s^{-1}$, there was a substantial increase in the multifractality at 600 °C. Such a significant increase may correspond to a transition point between different serration types or forms of dynamical behavior, such as self-organized criticality or chaos [172, 179]. It should be stated, however, that these results suggest that there is more complexity for the Type-C serrations, which contradicts the conclusion drawn from the entropy analysis. This discrepancy may be related to the scale distinction in the serrated flow. For instance, large Type-C serrations tend to periodicity (synchronization toward relaxation oscillations) and carry less complexity, but there are also small serrations such that the analysis of the ensemble gives an apparent complexity [172, 179, 180]. Lastly, the relatively-lower multifractality that was exhibited by the specimens tested at 400 and 500 °C may be attributed to the serrated flow, which exhibits the self-organized criticality [156].

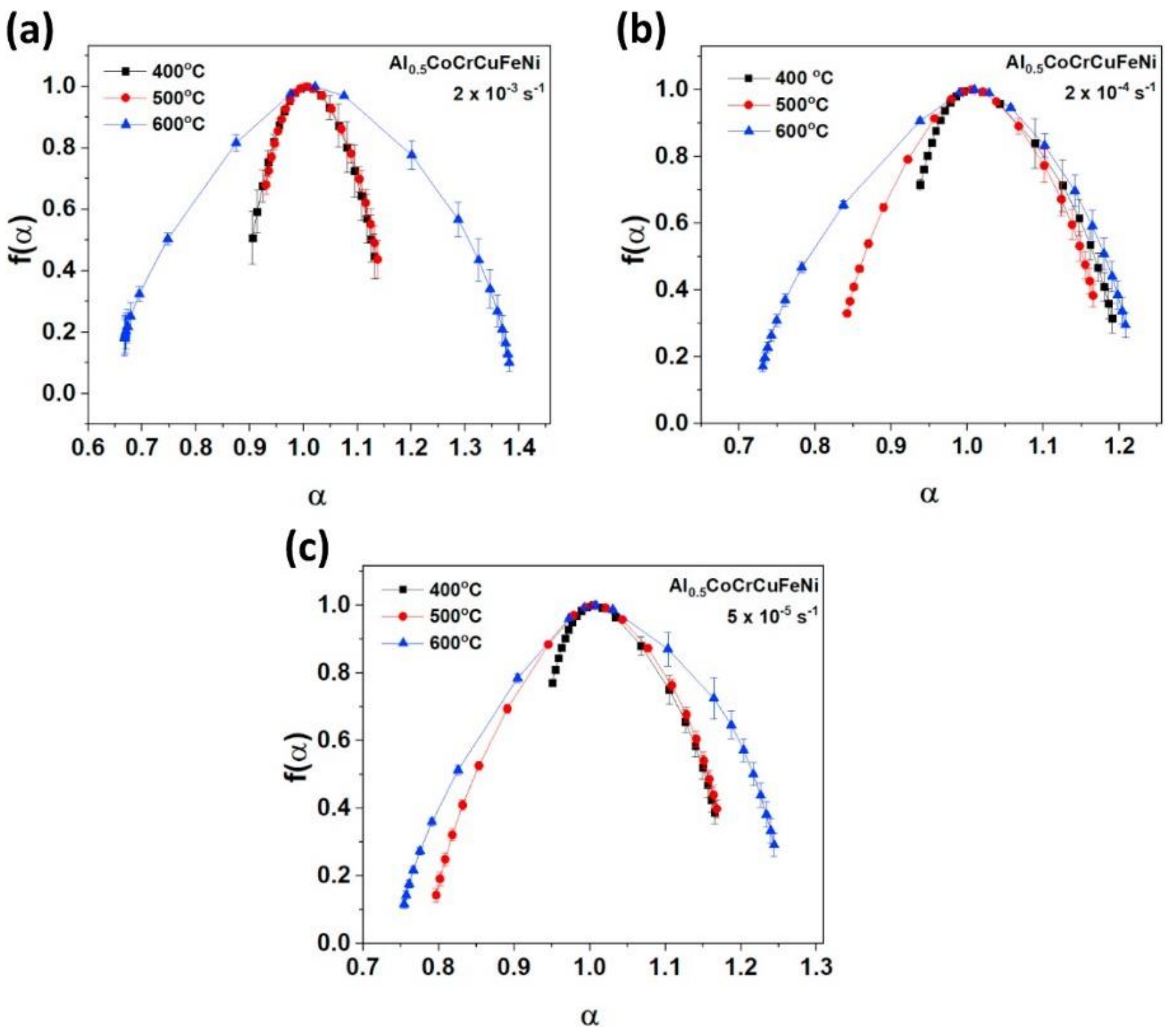

Fig. 29. The multifractal spectra for the $Al_{0.5}$CoCrCuFeNi HEA compressed at temperatures of 400 - 600°C and strain rates of (a) $2 \times 10^{-3}$ s$^{-1}$, (b) $2 \times 10^{-4}$ s$^{-1}$, and (c) $5 \times 10^{-5}$ s$^{-1}$. Figures from Brechtl et al. [91].

In [67], the deformation behavior of the $Al_{0.5}$CoCrCuFeNi HEA that was subjected to compression tests a strain rate of $5 \times 10^{-5}$ s$^{-1}$ and temperatures of RT - 700 °C was examined. The XRD and high-resolution transmission electron microscopy (HRTEM) characterization showed that the HEA contained an FCC phase structure with partially-ordered $L1_2$ particles after testing at 500 °C. Furthermore, it was found that after compression at 600°C (see Fig. 30), BCC and FCC phase structures, dislocations, and fully-ordered $L1_2$ particles were present in the matrix. The results indicated that the $L1_2$ particles can impede dislocation motion, making them susceptible to solute pinning that results in the serrated flow [6]. As for the serrated-flow behavior, the Type-C serrations occurring at 600 °C is possibly attributed to the BCC phase and fully-ordered $L1_2$ nanoparticles that were present in the material. In contrast, the serrated flow that occurred at 500 °C likely resulted from the presence of partially-ordered nanoparticles and a single FCC phase structure. Based on the findings from [91], it can be suggested that the less complex dynamics associated with the Type-C serrations may result from the $L1_2$-dislocation interactions. Whereas the more complex behavior at 500 °C could be attributed to the effects of the FCC phase structure and the partially-ordered $L1_2$ particles on the dislocation dynamics.

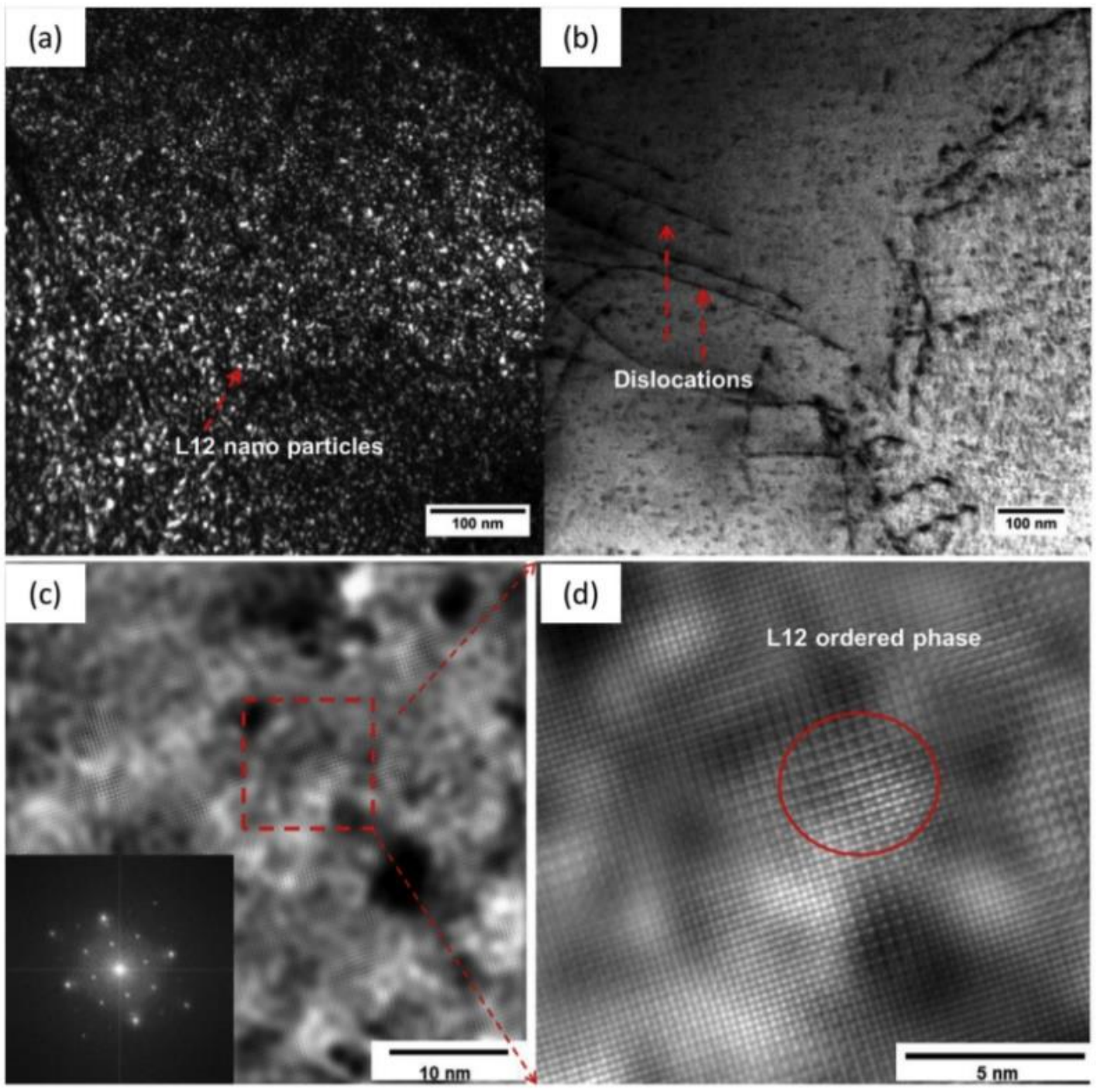

Fig 30. (a,b) The dark- and bright-field (BF) TEM images of the $Al_{0.5}CoCrCuFeNi$ HEA compression tested at a strain rate of $5 \times 10^{-5}$ s$^{-1}$ and temperature of 600 °C and (c,d) The HRTEM images [electron diffraction pattern in the inset of (c)] orientated in the (001) zone axis. Figures from Chen et al. [67].

In [79], the serrated flow of an $Al_{0.5}CoCrCuFeNi$ HEA was analyzed using the MFT. Here, specimens were subjected to uniaxial compression at strain rate of $4 \times 10^{-4}$ s$^{-1}$ and temperatures of 7 - 9 K. The stress vs. strain data, as well as the results of the MFT analysis, are shown in Figs. 31(a)-(b). The results revealed that serrations occurred at all of the testing conditions. Moreover, with an increase in the strain, the magnitude of the serrations significantly increased, which is a trend that has also been observed in bulk metallic glasses during compression [44, 47, 201]. Also, the largest stress-drop magnitude decreased as the temperature increased from 7 to 9 K. Such a trend has also been observed in a previous investigation [202]. Such a trend may be attributed to the twinning phenomenon in the HEA such that a temperature increase reduces the chances that deformation twinning can occur during deformation [6].

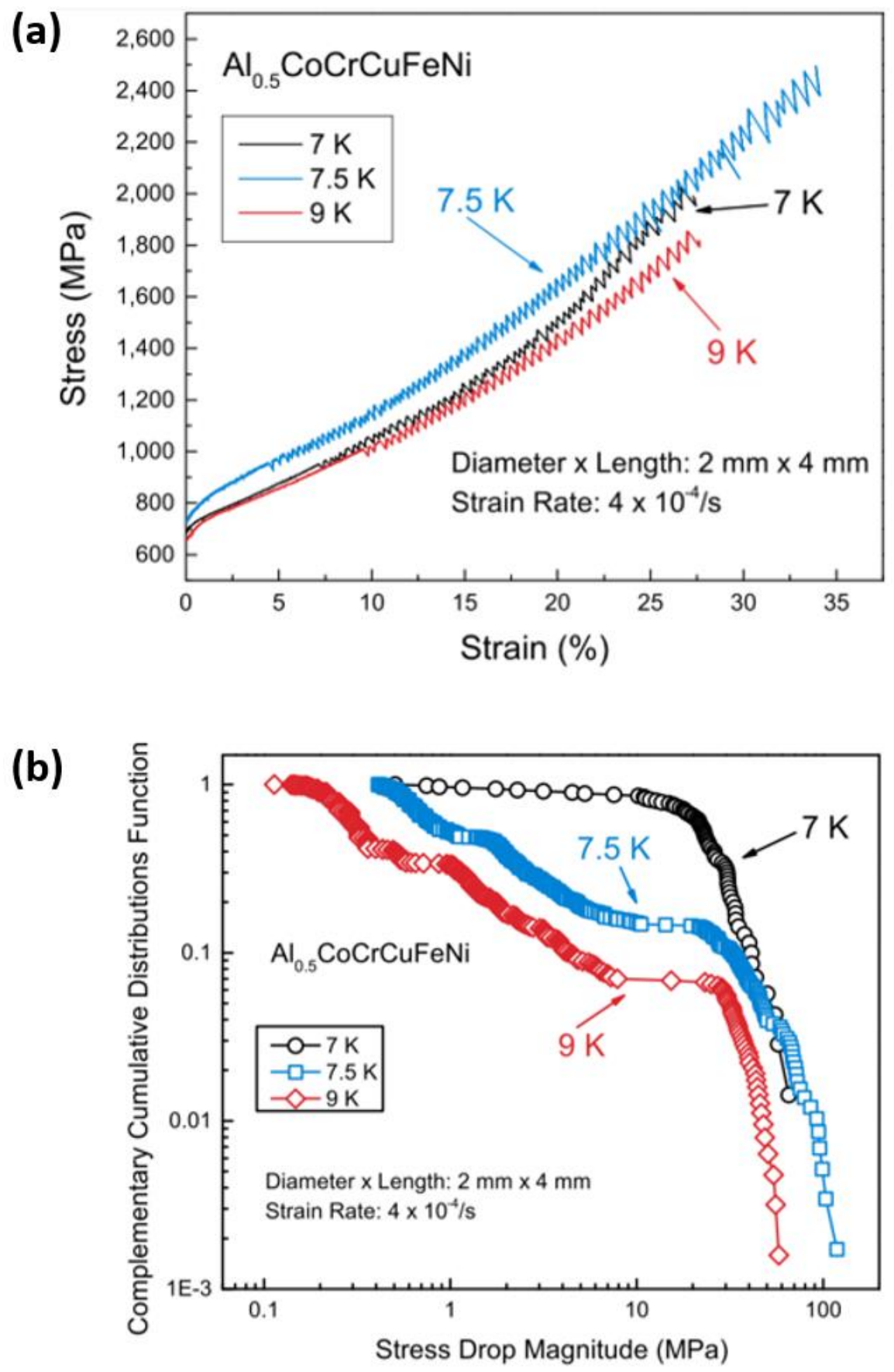

Fig. 31. (a) The stress vs. strain data for the $Al_{0.5}$CoCrCuFeNi HEA compressed at temperatures of 7 - 9 K and a strain rate of $4 \times 10^{-4}$ $s^{-1}$, and (b) the corresponding CCDF. Figures from Antonaglia et al. [79].

Guo et al. [76] study the serrated flow of the $Al_{0.5}$CoCrCuFeN HEA that underwent uniaxial compression tests. For the experiments, cylindrical specimens were tested at a strain rate of $4 \times 10^{-4}$ $s^{-1}$ and temperatures of 4.2 - 9 K. The plastic deformation displayed the serrated flow that were characterized by stair-like fluctuations. Statistical analysis revealed that there was a scaling relationship between the strain jump sizes and the released elastic energy. Further analysis of the serrated flow indicated that the largest Lyapunov exponent was negative for all temperatures, indicating that the serrations did not display chaotic dynamics at any of the conditions. However, it was determined that the exponent increased with a decrease in temperature, suggesting that the serrated flow became more disordered as the temperature was lowered.

### 3.2.5 $Al_x$MoNbTiV HEA

The compressive behavior of an $Al_x$NbTiMoV (x = 0, 0.25, 0.5, 0.75, 1, and 1.5) HEA was investigated in [90]. Here, specimens were subjected to RT-compression testing at strain rates of $5 \times 10^{-5}$ $s^{-1}$ - $5 \times 10^{-2}$ $s^{-1}$. The engineering stress vs. strain data for the specimens subjected to compression at a strain rate of $5 \times 10^{-4}$ $s^{-1}$ is displayed in Fig. 32. The serrated flow could be observed in the specimens with x = 0, 0.25, and 0.5, and was attributed to Cottrell atmospheres that surround moving dislocations and lock them in place [203]. The lack of observable serrations in the other specimens (x = 0.75, 1, and 1.5) may be explained as follows. For x values greater than 0.5 (specimens with no observable serrations), the atomic-size differences exceeded 3.22 % and the lattice parameters were greater than 3.191 Å. This finding suggests that when the lattice volume and atomic-size disparity exceed a threshold value, the drag effect is reduced enough such that no dislocation locking may occur in the HEA.

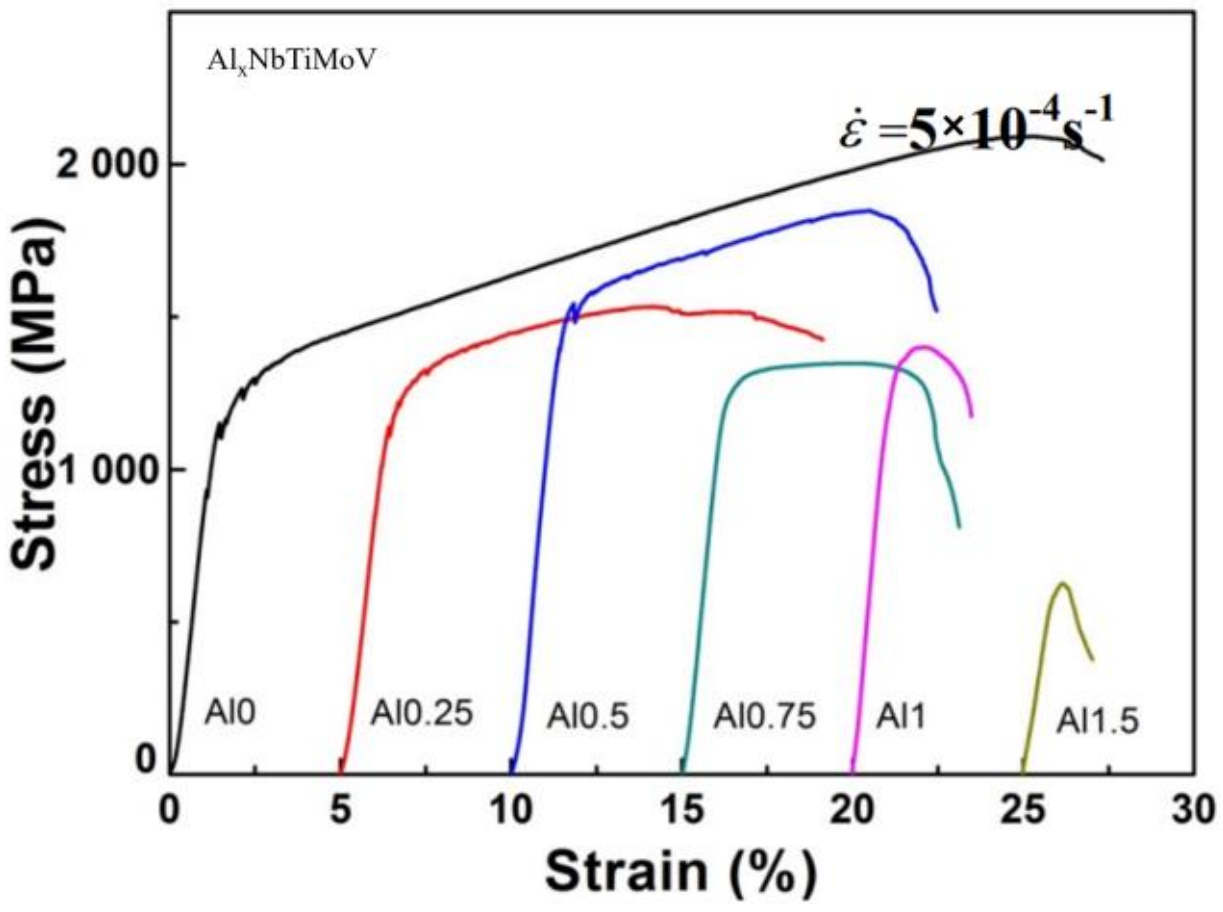

Fig. 32. The engineering stress-strain curves for the $Al_x$NbTiMoV (x = 0, 0.2, 0.5, 0.75, 1, and 1.5) HEA. Figure from Chen et al. [90].

## 3.2.6 $Al_xMoNbTaTiV$ (x = 0, 0.2, 0.4, 0.6, and 1) HEA

Ge et al. [204] investigated the deformation behavior of an $Al_xMoNbTaTiV$ (x = 0, 0.2, 0.4, 0.6, and 1) HEA during compression at a strain rate of $10^{-3}$ $s^{-1}$ and temperatures of RT - 900 $^oC$. Figures 33(a)-(d) present the engineering stress vs. engineering strain graphs for the specimens subjected to compression at the listed conditions, and Table 7 displays the corresponding serration types. It was found that the serrated flow occurred under certain conditions. For example, the specimens with x = 0 and 0.2 displayed strictly Type-A serrations at RT. On the other hand, the specimen with x = 0.4 initially exhibited Type-A serrations that evolved into Type-E serrations towards the end of the curve. At 500 $^oC$, specimens with x = 0.4 – 1 presented the serrated flow, although there were significantly more serrations for x = 0.6 and 1. More specifically, Types-A + B and Type-B serrations were observed for x = 0.6 and 1, respectively. As for 700 $^oC$, the serrated flow was only observed for x = 0.6 and 1. The lack of serrations at 900 $^oC$ was likely a consequence of the relatively-large thermal fluctuations that result in a very low activation energy for dislocation motion.

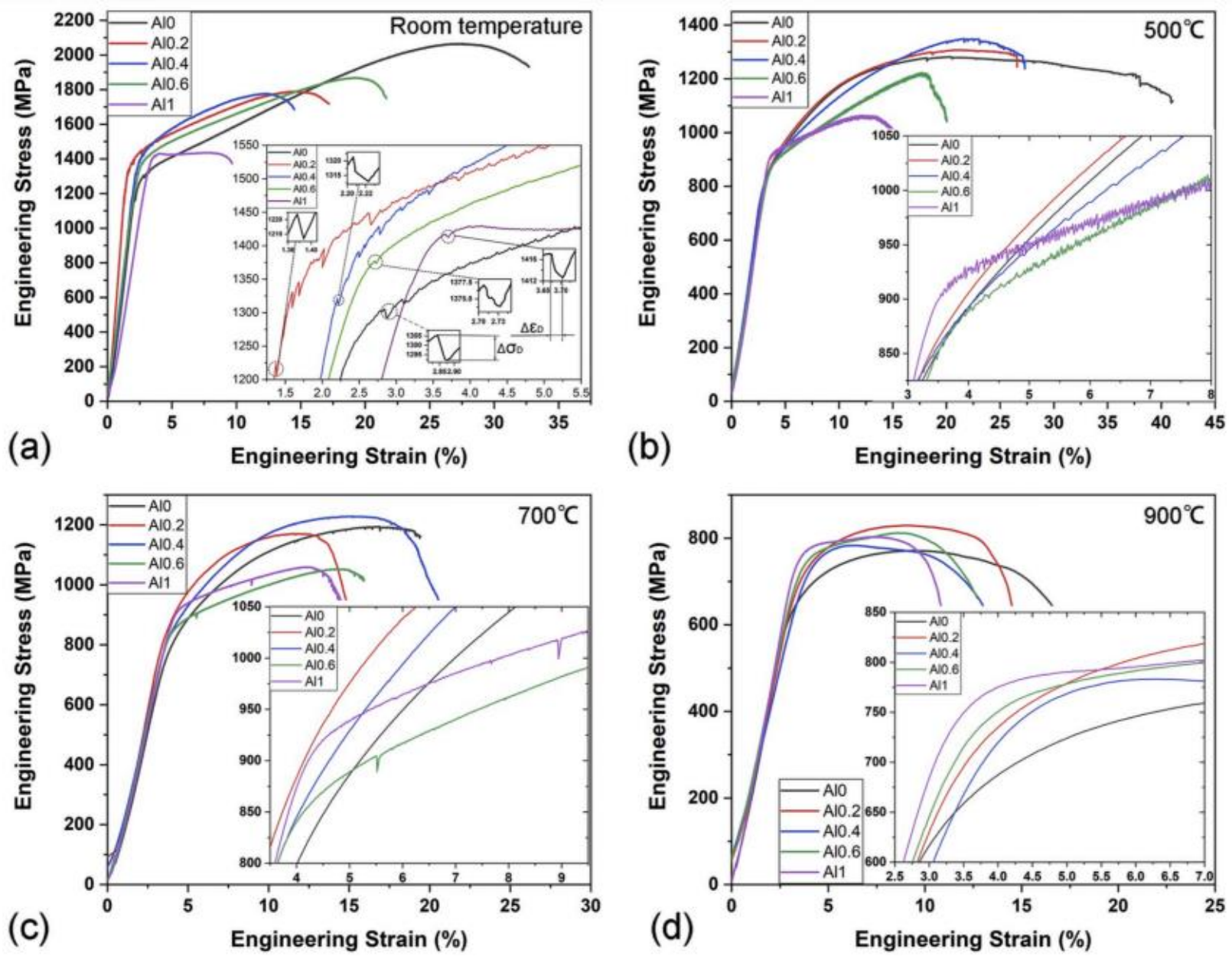

Fig. 33. Engineering stress vs. engineering strain curves for the $Al_xMoNbTaTiV$ HEA that were compressed at a strain rate of $10^{-3}$ $s^{-1}$ and the following temperature conditions: (a) room temperature (b) 500 $^oC$, (c) 700 $^oC$, and (d) 900 $^oC$. Figures from Ge et al. [204].

Table 7. The serration type exhibited by the $Al_xMoNbTaTiV$ (x = 0, 0.2, 0.4, 0.6, and 1) HEA during compression at temperatures of RT - 900 °C and a strain rate of $1 \times 10^{-3}$ $s^{-1}$. From Ge et al. [204].

| Alloy | Temperature (°C) | Serration Type |
|---|---|---|
| MoNbTaTiV | RT | A |
| $Al_{0.2}MoNbTaTiV$ | RT | A |
| $Al_{0.4}MoNbTaTiV$ | RT | A + E |
| $Al_{0.6}MoNbTaTiV$ | 500 | A + B |
| $Al_1MoNbTaTiV$ | 500 | B |

In terms of the serrated flow, it was suggested that the tiny Al addition could form point obstacles in the interdendrites that can repeatedly pin dislocations. Furthermore, as the Al content increases, the Al atoms and their nearest neighbors form large-size clusters, which have weak pinning effects such that the highly-diffused solute atoms can no longer catch and pin dislocations. Finally, additional analysis determined that the critical strain for the onset of serrations was a power-law function in terms of the Al content in the alloy.

### 3.2.7 CoCrFeMnNi HEA (Cantor alloy)

Wang et al. [78] studied the mechanical behavior of an FCC CoCrFeMnNi HEA during RT high-impact compression. Here, specimens were subjected to compression tests at strain rates of $1 \times 10^3$ - $3 \times 10^3$ $s^{-1}$ (1,200 and 2,800 $s^{-1}$) using a split-Hopkinson pressure bar system. The true stress vs. strain data is presented in Fig. 34, and it was found that the serrated flow was affected by changes in the applied-loading rate. It was suggested that as the specimen approached temperatures as high as 1,300 K, local hotspots are created that can weaken intergranular bonding. Also, in such extreme conditions, cracks and microvoids can form in the severe-deformation zone.

The above findings led to the following hypothesis about how the serrated flow occurred during the high impact compression. During the primary stages of deformation, dislocations accumulate along the boundaries of the grains which elongate during the compression. As the stress increases, the grain boundaries continue to elongate such that their width further decreases, making them more susceptible to weakening. Consequently, the thermal hotspots generated under the compressive forces diminish the intergranular bonding, which leads to the eventual collapse of the grain boundary. Subsequently, shear bands develop, which results in the serrated flow during deformation.

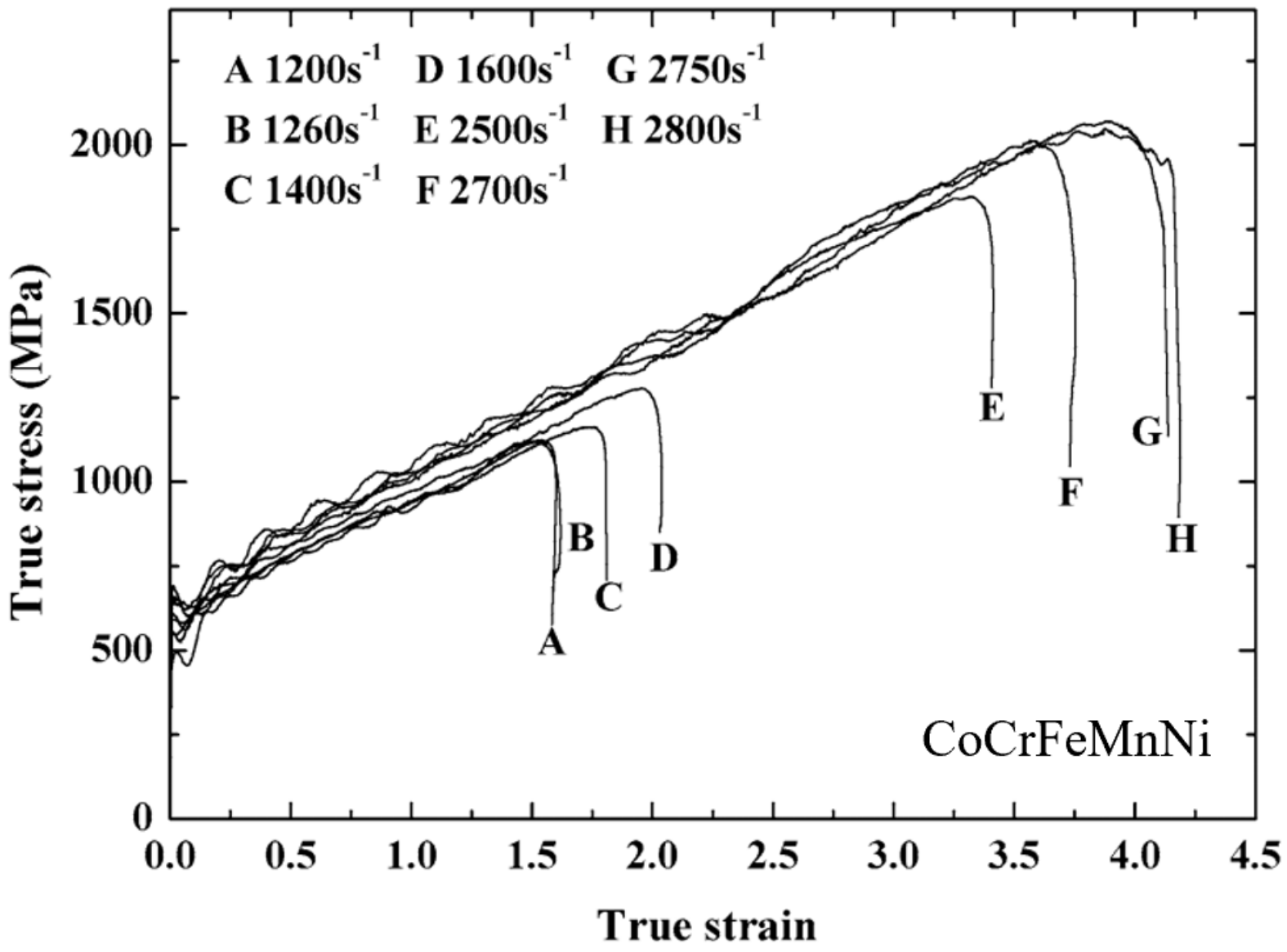

Fig. 34. True stress vs. true strain data for the CoCrFeMnNi HEA compressed at loading rates of 1,200 - 2,800 $s^{-1}$ ($1 \times 10^3$ - $3 \times 10^3$ $s^{-1}$). Figure from Wang et al. [78].

### 3.2.8 $CoCrFeMnNiV_x$ (x = 0, 0.07, 0.3, 0.7, and 1.1) HEA

Fang et al. [205] explored the deformation behavior of a $CrFeMnNiV_x$ (x = 0, 0.07, 0.3, 0.7, and 1.1) HEA thin film via micro-pillar compression experiments. The thin films were deposited on (1 0 0)-Si wafers via radio frequency (RF) magnetron co-sputtering. The V content increased with respect to the RF applied power on the target, where it varied from 0 to 80 W. TEM imaging revealed [see Figs. 35(a)-(e)] that in the specimens with x = 0, 0.07, and 0.3, there were columnar grains with an average width of ~ 45 nm as well as nanotwins. Furthermore, these films consisted of an FCC structure. Interestingly, Fig. 35(e), which represents the image for the HEA with x = 1.1, indicated that the film contained an amorphous structure. Figure 36 shows the engineering stress-strain data for the compressed micro-pillar specimens. It was found that an increase in the V content led to more pronounced serrations in the pillars. Moreover, the results of hardness testing showed that the hardness increased with respect to the amount of V in the specimen.

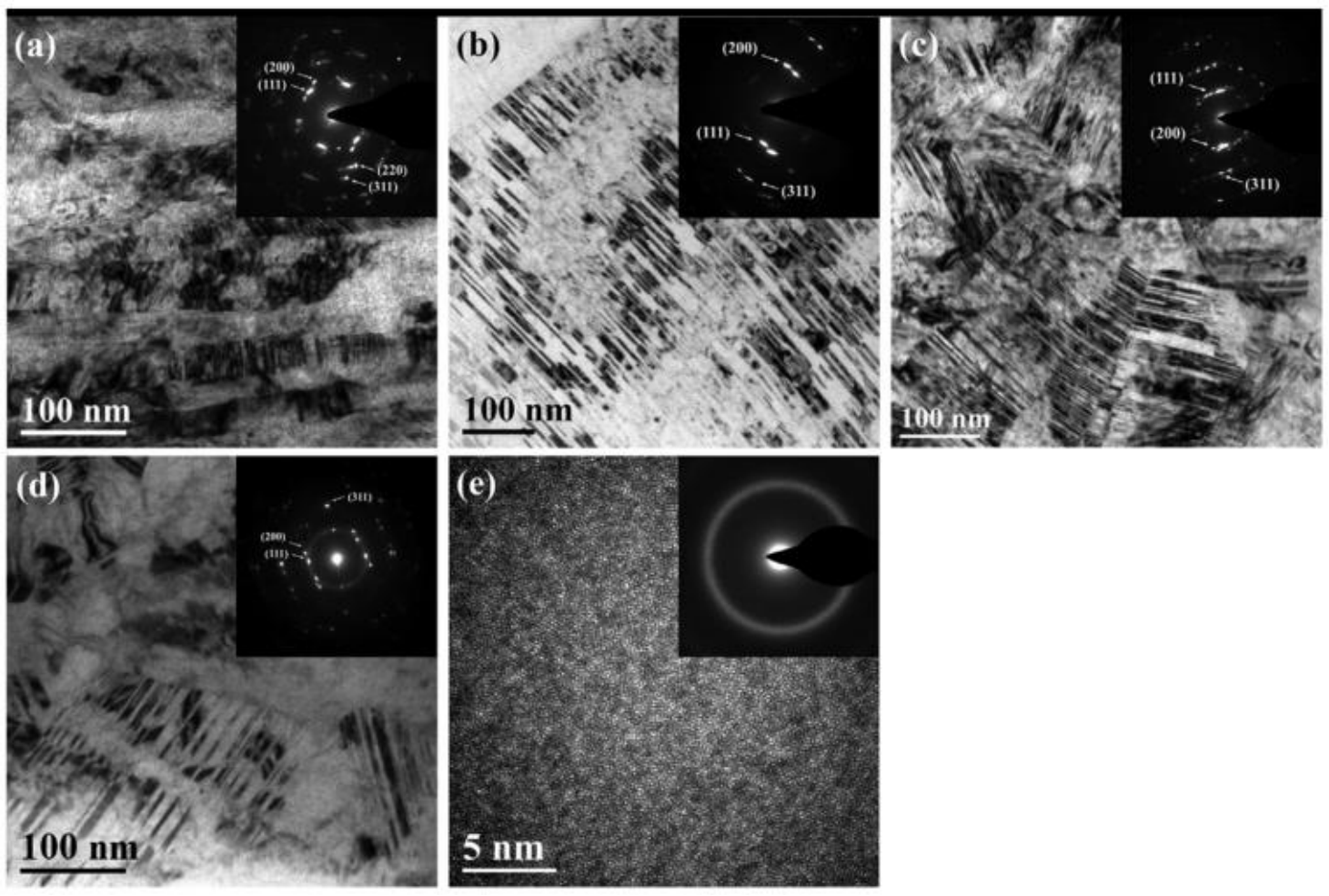

Fig. 35. TEM images of as-received CoCrFeMnNiV$_x$ (x = 0, 0.07, 0.3, 0.7, and 1.1) HEA films with the inset showing the corresponding SAED patterns for (a) x = 0 (b) x = 0.07, (c) x = 0.3, and (d) x = 0.7. (e) The HRTEM image at x = 1.1 with the corresponding SAED pattern, as shown in the inset. Figures from Fang et al. [205].

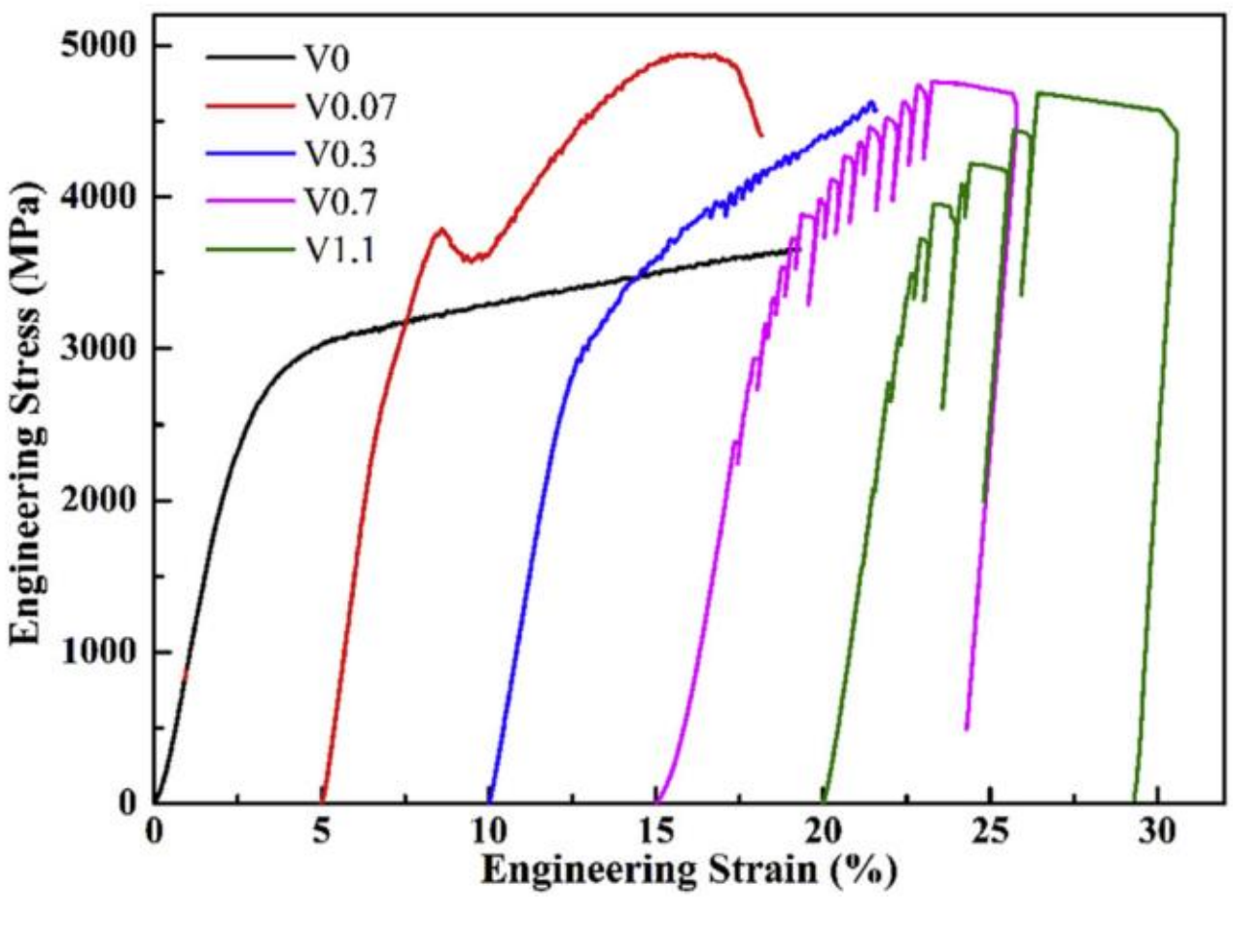

Fig. 36. Engineering stress-strain data of CoCrFeMnNiV$_x$ (x = 0, 0.07, 0.3, 0.7, and 1.1) HEA films where curves are offset horizontally from the origin to improve visibility. Figure from Fang et al. [205].

Figures 37(a)-(e) display the TEM images for the HEA films after micro-pillar compression. The results from Fig. 37(e) show that the amorphous structure remained in the $CrFeMnNiV_{1.1}$ specimen during deformation. It was found that deformation led to a decrease in the amount of nanotwins in the matrix, which suggested that nanotwins were unstable during deformation. Finally, it was concluded that the introduction of nanotwins into the structure via magnetron sputtering with high-energy atomic bombardment and relatively fast quenching is an effective technique to improve the mechanical properties of the HEA.

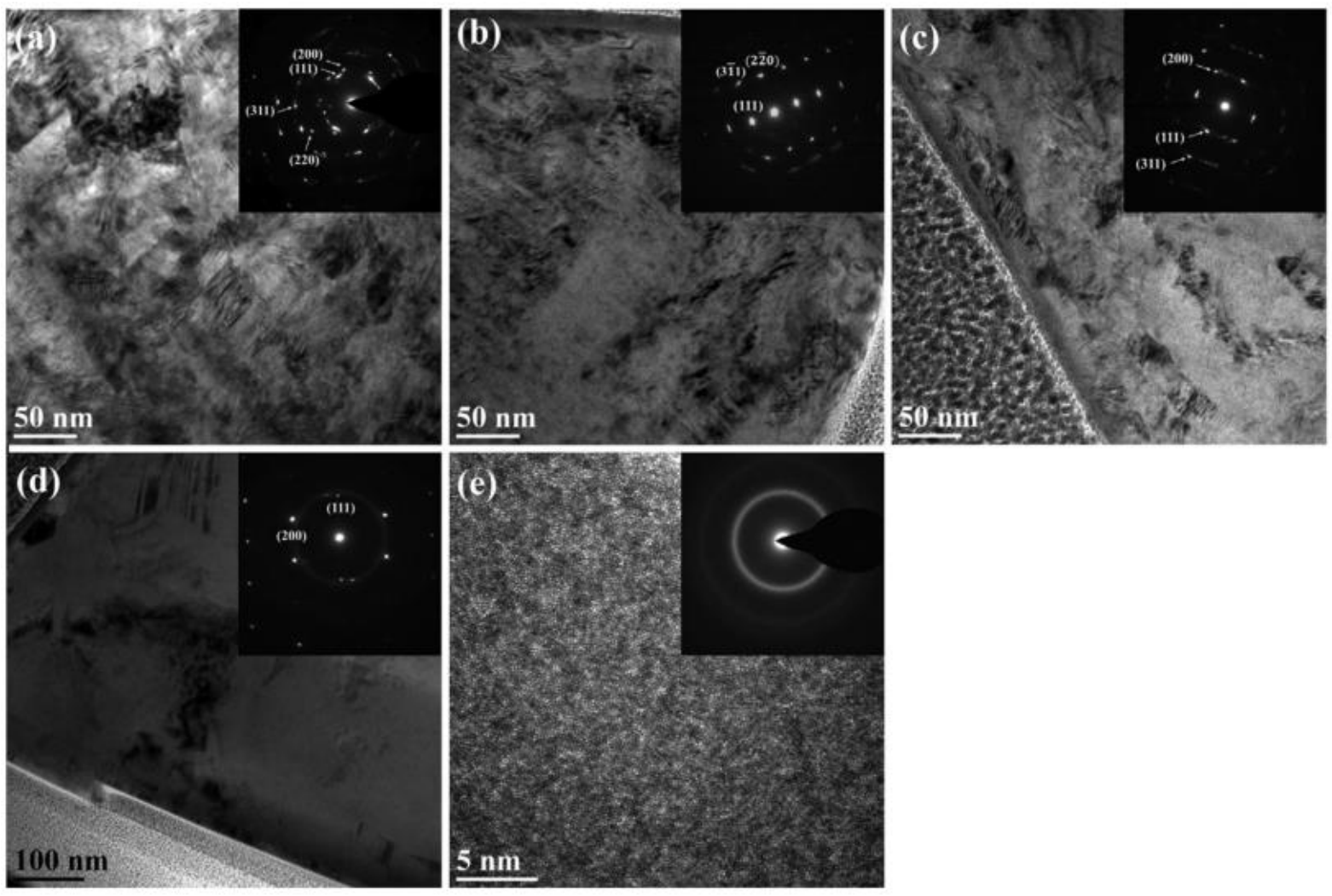

Fig. 37. Magnified images of the deformed zones in the $CoCrFeMnNiV_x$ HEA films after micro-pillar compression tests for (a) x = 0 (b) x = 0.07, (c) x = 0.3, (d) x = 0.7, and (e) x = 1.1. The corresponding SAED pattern is displayed in the inset. Figures from Fang et al. [205].

### 3.2.9 CoCuFeNiTi HEA

In [73], the serrated-flow behavior of a eutectic CoCuFeNiTi HEA during compression was investigated. For the experiment, specimens were subjected to isothermal hot-compression at strain rates of $10^{-3}$ and $10^{-1}$ $s^{-1}$ and temperatures of 1,073 - 1,273 K (800 - 1,000 °C). Differential scanning calorimetry revealed that the alloy underwent no phase transformations up to 1,000 °C, indicating that no phases changes occurred during the deformation at any of the test temperatures. The experimental results can be observed in Figs. 38(a)-(b), which shows the true stress-strain data. For all conditions, the stress initially increased until it reached a maximum, where it subsequently decreased with strain. It was thought that this drop in the stress value may have been due to dynamic recrystallization and softening during hot deformation. It was found that the serrations were more prominent at lower strain rates. Furthermore, the serration amplitude decreased with temperature at both strain rates.

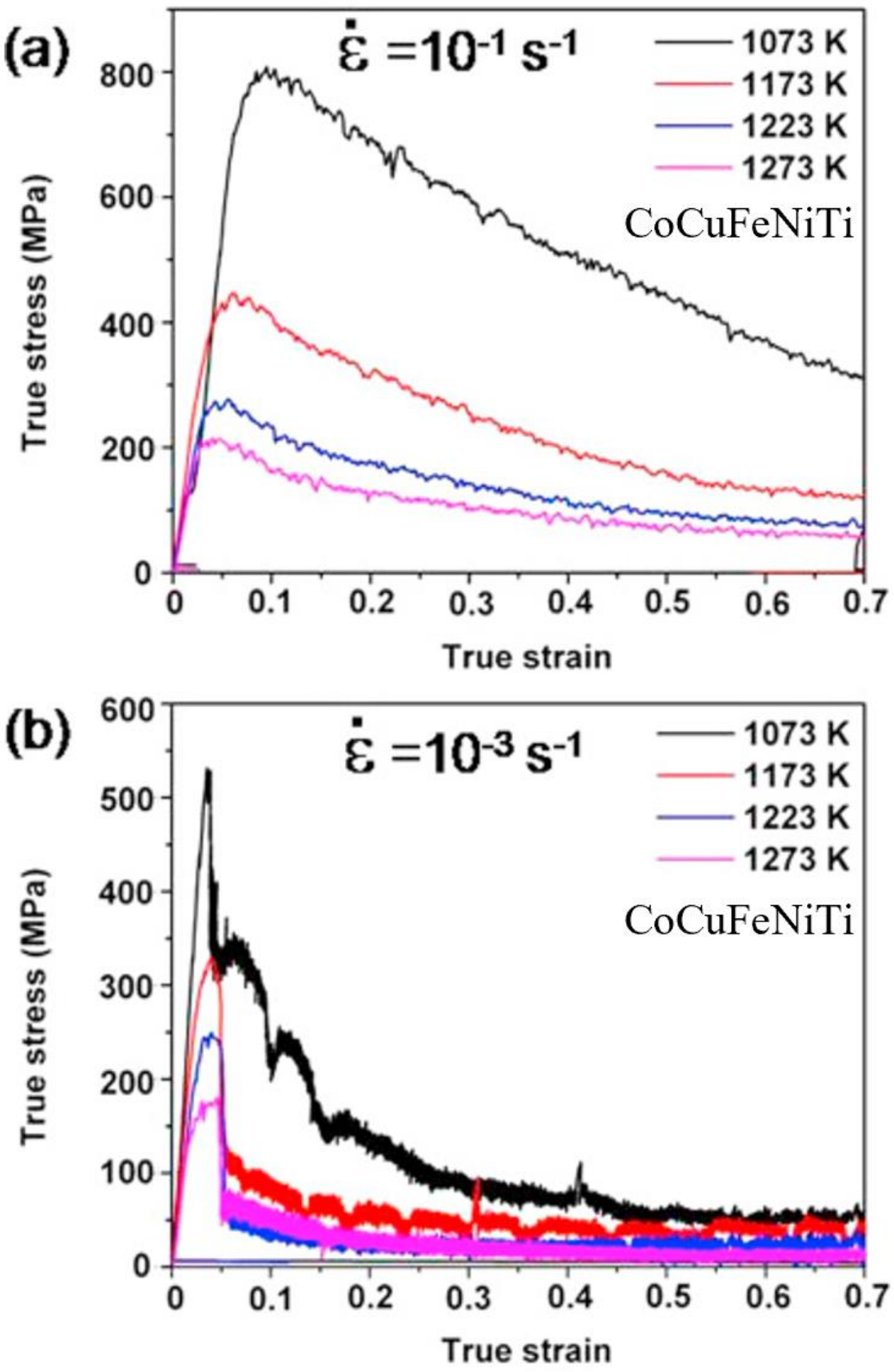

Fig. 38. True stress-true strain plots for the CoCuFeNiTi HEA subjected to compression at temperatures of 1,073 – 1,273 K and strain rates of (a) $10^{-1}$ and (b) $10^{-3}$ $s^{-1}$. Figures from Samal et al. [73].

This decrease in the serration amplitude at higher temperatures was a consequence of the increased thermal-vibration energy necessary to pin dislocations [65, 79]. On the other hand, the serration amplitude increased with a decrease in the strain rate, and was attributed to the increased time that atoms begin to lock dislocations. Table 8 summarizes the observed serration types for the different strain-rate and temperature conditions. Serration Types-A, B, and A + B were observed.

Table 8. The serration types displayed by the CoCuFeNiTi HEA during compression at strain rates of $1 \times 10^{-3}$ $s^{-1}$ and $1 \times 10^{-1}$ $s^{-1}$ and temperatures of 1,073 - 1,273 K (800 °C – 1,000 °C). From Samal et al. [73].

| **Strain Rate ($s^{-1}$)** | **Temperature (°C)** | **Serration Type** |
|---|---|---|
| $1 \times 10^{-3}$ | 800 | A + B |
| | 900 | A + B |
| | 950 | B |
| | 1,000 | B |
| $1 \times 10^{-1}$ | 800 | A |
| | 900 | A + B |
| | 950 | A + B |
| | 1,000 | A + B |

## 3.2.10 MoNbTaW HEA

Antonaglia et al. [79] investigated the serrated flow in a MoNbTaW HEA specimens during compression testing. For the experiment, cylindrical specimens with a diameter of 3.6 mm and length of 5.4 mm were compressed at RT (298 K) and 873 K and a strain rate of $1 \times 10^{-3}$ $s^{-1}$. Figures 39(a) and 39(b) show the plots of the associated stress vs. strain and stress vs. time results, respectively. From the inset of these figures, it was apparent that serrations did occur in both specimens during compression. It should be noted that for the specimen compressed at 873 K, there were some fluctuations in the elastic region of the graph that may have been caused by machine noise during the experiment. It was also surmised that the transition from strain softening at RT to strain hardening at 873 K indicates that different mechanisms were behind the serrated flow behavior at each temperature. At RT, it was believed that plastic instability was the primarily responsible for the serrations. Whereas at 873 K, temperature-induced microstructural evolution was the mechanism responsible for the serrations.

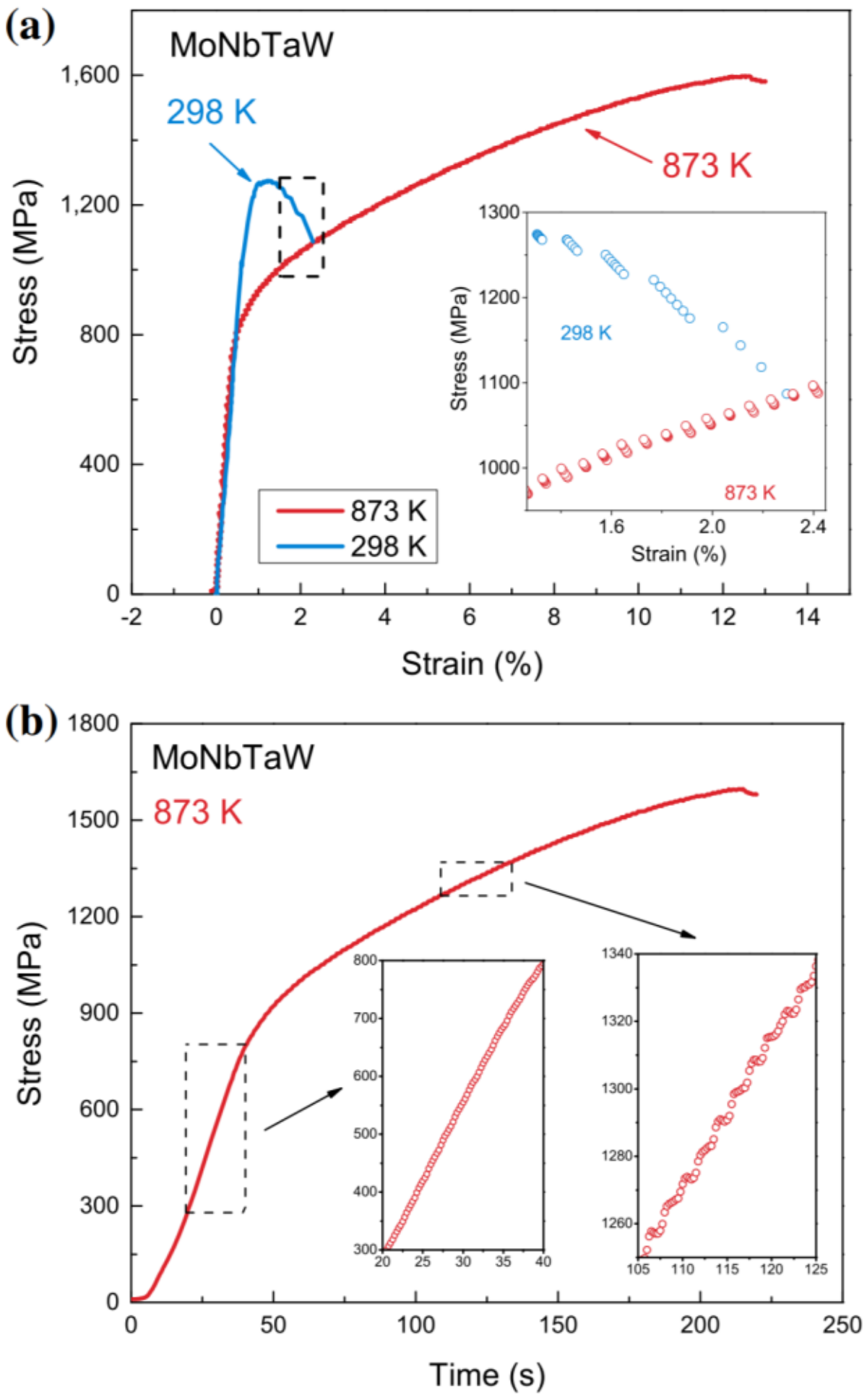

Fig. 39. (a) Engineering stress vs. strain data for the MoNbTaW HEA compressed at temperatures of 298 K and 873 K at a strain rate of $1 \times 10^{-3}$ $s^{-1}$. The inset of the figure displays the magnified region, which is marked by the dashed box. (b) The corresponding stress vs. time curve for the specimen tested at 873 K with the inset that displays the elastic and plastic regimes of the deformation curve. Figures from Antonaglia et al. [79].

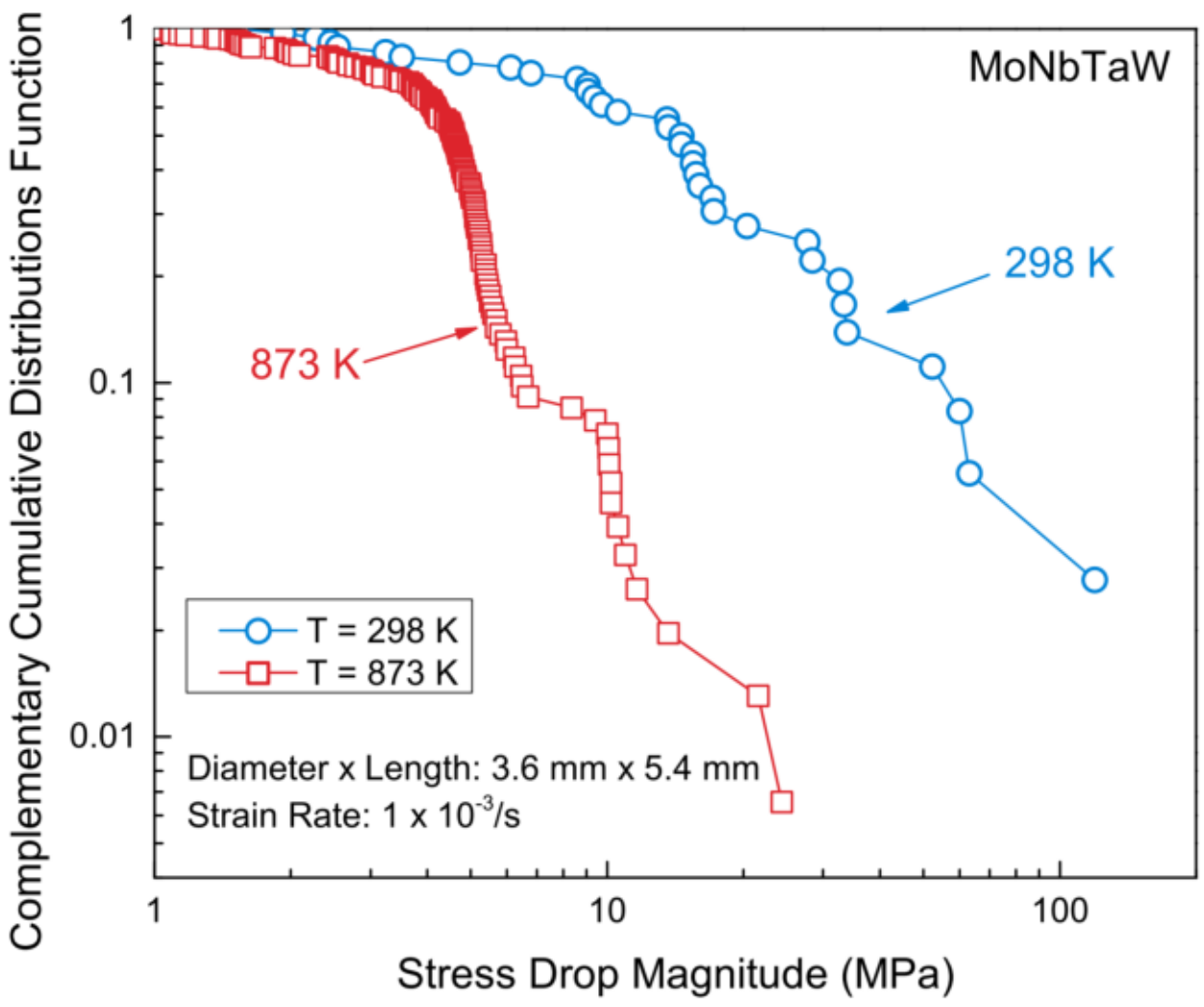

Fig. 40. The results of the CCDF analysis for the MoNbTaW HEA compressed at temperatures of 298 and 873 K and a strain rate of 1 × $10^{-3}$ $s^{-1}$. Figure from Antonaglia et al. [79].

Zou et al. examined the micro- and nano-pillar compression of a MoNbTaW HEA [206]. For the experiments, two sets of pillars with diameters ranging from ~250 nm to ~2 μm were prepared, using a focused ion beam. The first set of pillars consisted of a [0 0 1] orientation, which is a multiple-slip system, whereas the other pillars had the [3 1 6] orientation, which is a single-slip system. Before testing, the specimens were homogenized at 1,800 °C for 7 days. For the compression tests, a strain rate of 2 × $10^{-3}$ $s^{-1}$ was applied, employing a diamond flat-punch tip. Imaging of the compressed pillars was performed, using SEM.

Figures 41(a)-(d) display the SEM images of the compressed MoNbTaW HEA pillars with the [3 1 6] orientation and sizes ranging from ~ 200 nm – 2 μm. From Figs. 41(a)-(b), clear single-slip bands can be observed. In Figs. 41(c)-(d), multiple-slip bands can be seen on the specimen surface, which contribute to strain hardening. It was surmised that the multiple slips that occurred in the specimen may have been due to a slight misalignment between the flat punch and the pillar top or the influence of the tolerance angle. The SEM images for the compressed pillars with the [0 0 1] orientation are displayed in Figs. 42(a)-(d). The specimens exhibited a wavy-slip feature that was thought to be attributed to the cross slip of screw dislocations along the <1 1 1> directions. Such a feature has also been observed in BCC metals [207, 208].

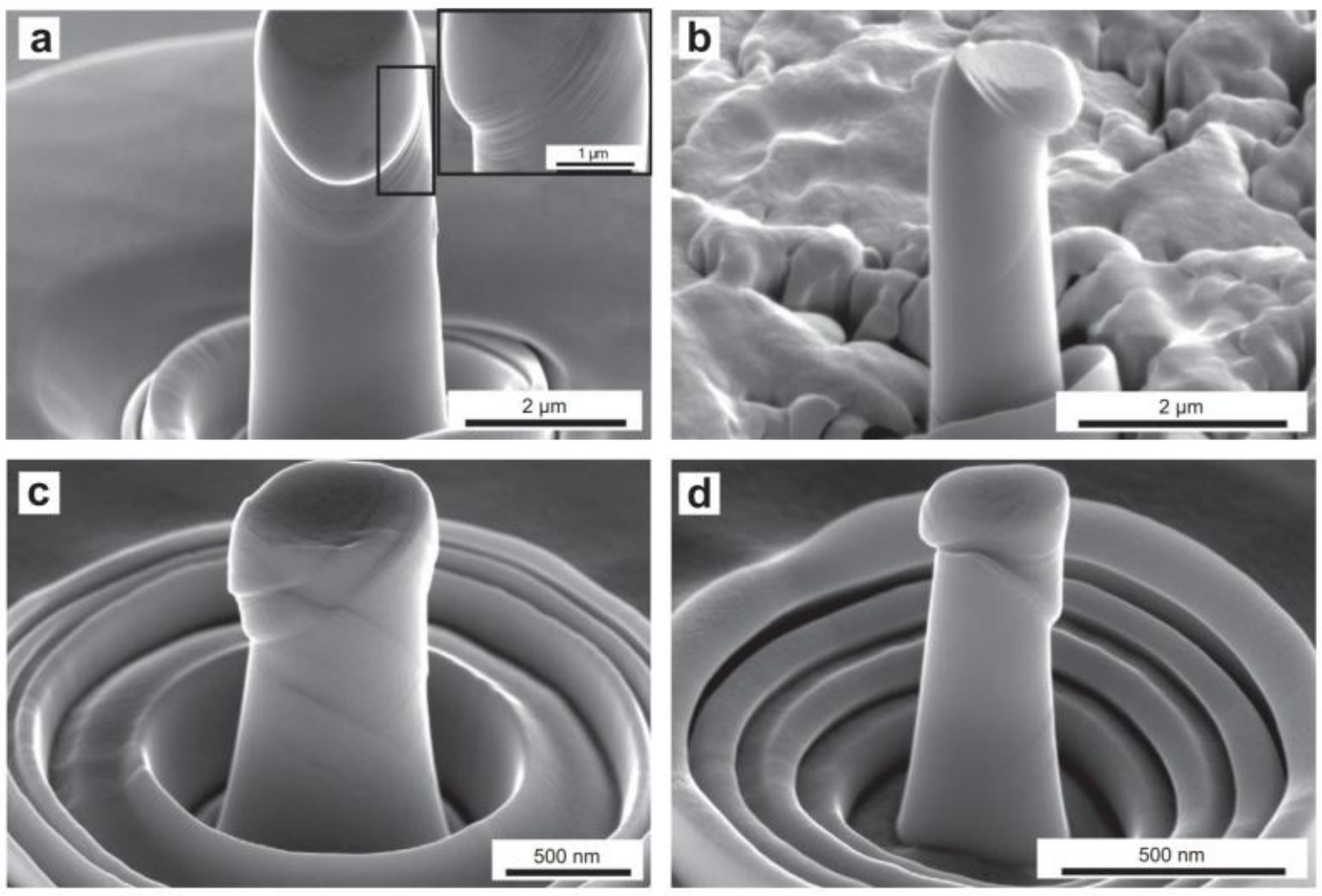

Fig. 41. SEM images of the compressed [3 1 6] oriented MoNbTaW HEA pillars with diameters of approximately (a) 2 μm (the inset features an enlarged image of the sharp slip bands), (b) 1 μm, (c) 500 nm, and (d) 250 nm. Figures from Zou et al. [206].

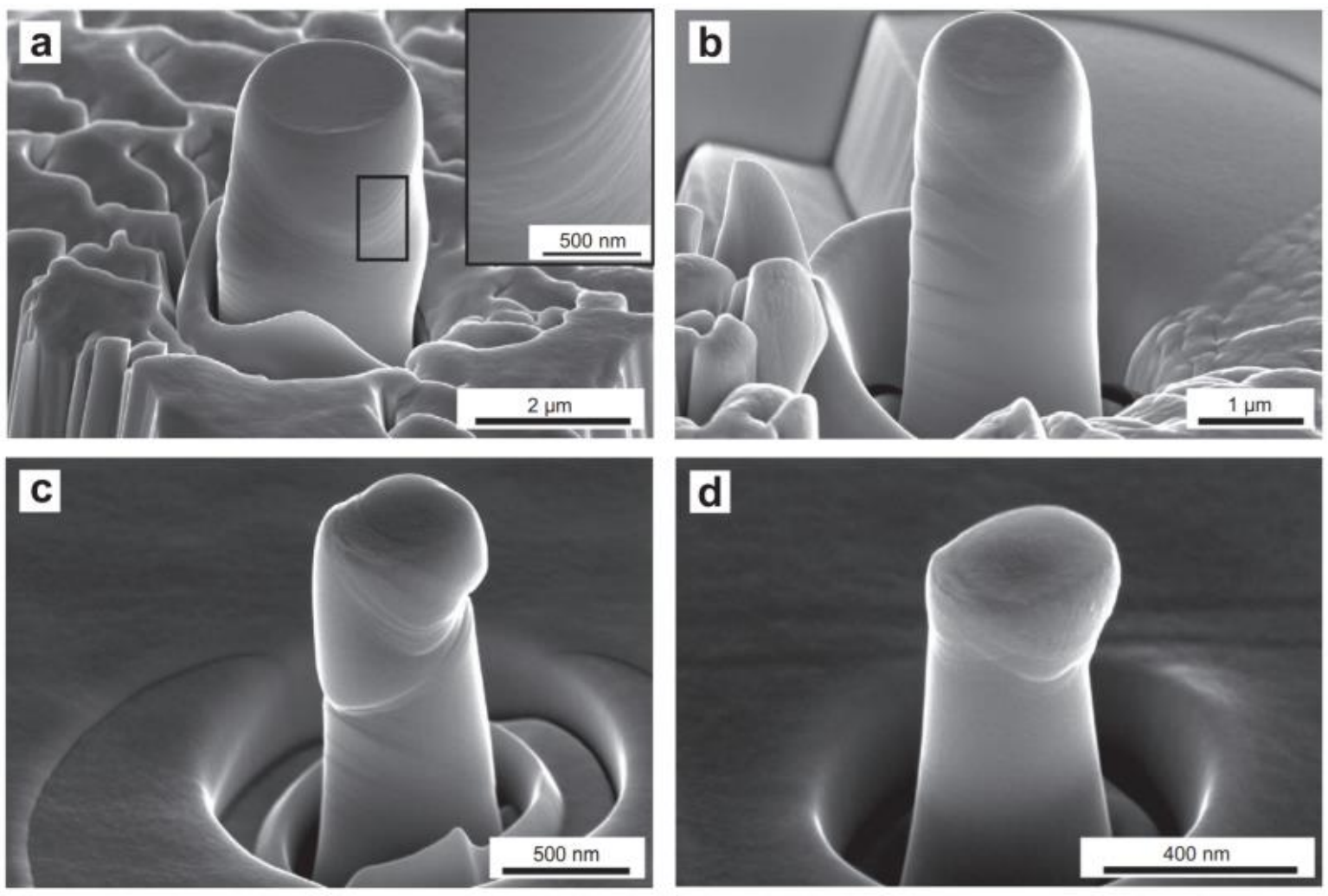

Fig. 42. SEM images of the compressed [0 0 1] oriented MoNbTaW HEA pillars with diameters of approximately (a) 2 μm (inset shows an enlarged wavy surface morphology), (b) 1 μm, (c) 500 nm, and (d) 250 nm. Figures from Zou et al. [206].

Figures 43(a)-(b) present the corresponding engineering stress vs. engineering strain data for the two sets of pillars. It was found that the pillars exhibited similar serration dynamics. For instance, serrations were observed in both sets of pillars and as compared to the larger pillars, the smaller pillars displayed higher flow strengths. Moreover, the smaller pillars exhibited a higher strain-hardening rate, as compared to the larger pillars, which was attributed to the more activated and interacting slip systems in the former. Finally, the results of the study indicated that in the HEA pillars, the yield strength significantly increased and the plastic deformation proceeded as intermittent strain bursts, which is similar to what is observed in conventional alloys [209].

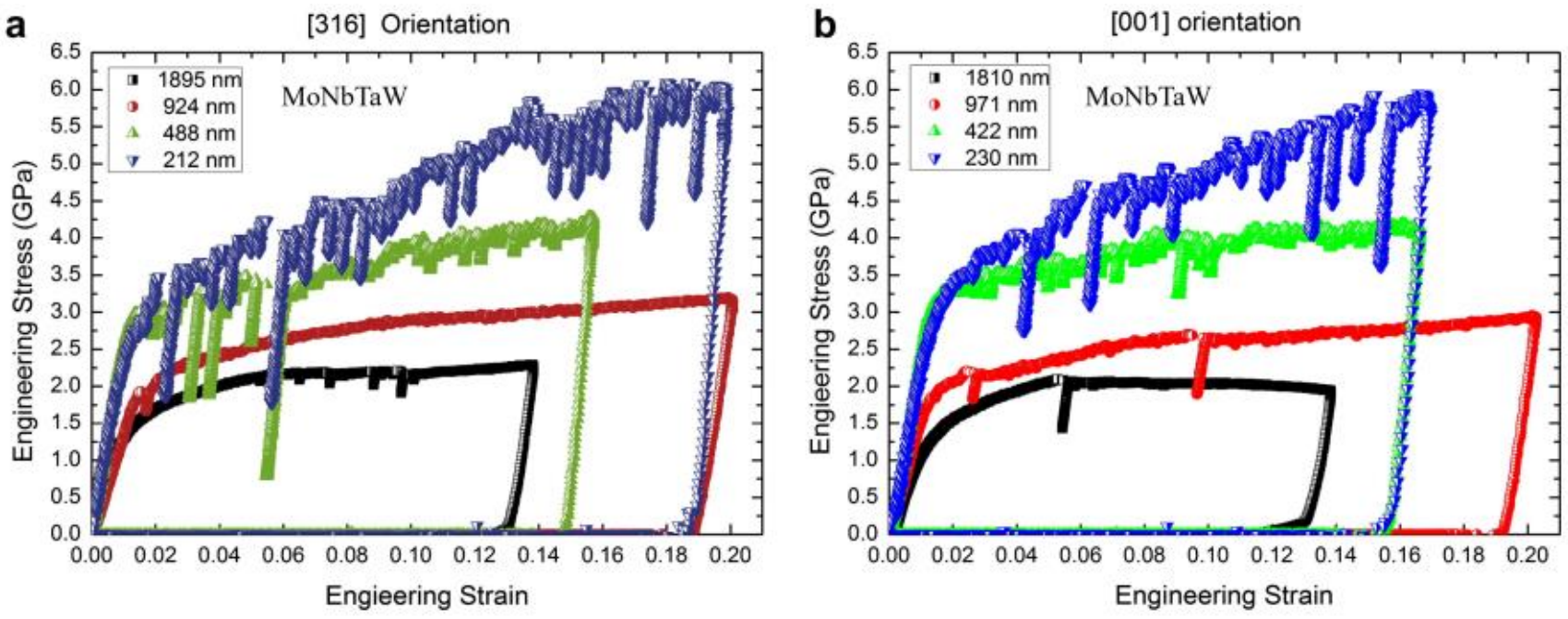

Fig. 43. The engineering stress vs. engineering strain data for the (a) [3 1 6]-oriented and (b) [0 0 1]-oriented single crystalline MoNbTaW HEA pillars with diameters ranging from ~ 200 nm to ~ 2 μm. Figures from Zou et al. [206].

## 3.3 Nanoindentation testing

### 3.3.1 $Al_{0.3}CrCoFeNi$ HEA

Qiang et al. [132] examined the serration behavior of the $Al_{0.3}CrCoFeNi$ HEA during RT nanoindentations. For the experiments, as-cast and torsionally-deformed specimens were indented, using a Hysitron Triboindenter TI950 equipped with a Berkovich tip. The torsional-deformation process consisted of deforming specimens for 1, 3, and 10 rotations under a pressure of 10 GPa. For the nanoindentations, 15 indents using a loading rate of 250 μN/s were performed. TEM, XRD, and Vickers hardness tests were also conducted. According to the XRD results, the specimens are composed of a single-phase FCC structure. The TEM characterization results revealed that after 10 revolutions, the grain size decreased from hundreds of microns (as-cast) to tens of nanometers.

There were some interesting results with regards to the mechanical heavier. For instance, Vickers hardness tests showed that torsion resulted in a substantial increase in the hardness of the HEA. Figure 44(a) displays the nanoindentation-depth-change ($\Delta h$) behavior as a function of the indenter depth. The authors reported that for the deformed specimens, there were no significant displacement bursts characteristic of the serrated flow. On the other hand, the displacement bursts were much larger for the as-cast specimen, indicating that the serrated flow had occurred. The lack of serrations displayed by the deformed specimens may be a consequence of the affluent grain boundaries that can act as extra mediators for plastic deformation as well as obstacles for dislocation motion [6]. Figure 44(b) displays the cumulative-probability distribution (CPD) [$P(> S) = AS^{-\beta} e^{-\frac{S}{S_c}}$] of the displacement-burst size, $S = \Delta h/h$, for the as-cast specimen. The CPD was found to significantly decrease with an increase in the size of the displacement burst. The values, $S_c$ and $\beta$, were determined to be ~ 0.014 ± 0.000 and 0.15 ± 0.02, respectively, indicating that in the as-cast specimen, slip avalanches occurred during nanoindentation.

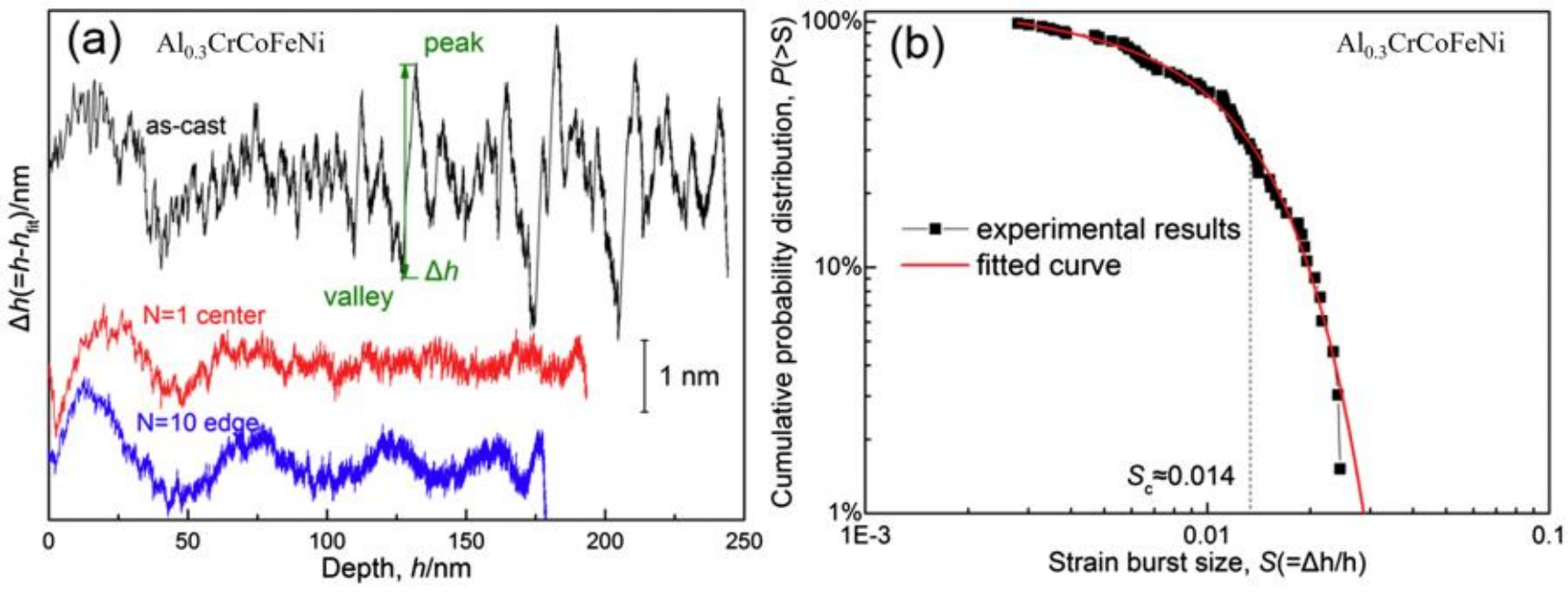

Fig. 44. The nanoindentation data for the $Al_{0.3}$CrCoFeNi HEA (as-cast) with respect to (a) the depth change, $\Delta h$, vs. depth and (b) the CPD of the strain-burst size. Figures from Qiang et al. [132].

### 3.3.2 $Al_{0.7}$CrCoFeNi HEA

In [210], the nanoindentation behavior was examined in a multi-phase $Al_{0.7}$CrCoFeNi HEA. For the experiment, indents were performed inside a BCC containing grain. The BCC phases consisted of the (Fe, Cr)-rich disordered A2 phase and the (Ni, Al)-rich ordered B2 phase, (Fe, Cr)-rich disordered A2 phase, as determined by the energy dispersive spectroscopy (EDS). The corresponding load vs. depth data for four different indents is displayed in Fig. 45(a). A graph, which features a magnification consisting of multiple pop-in events (serrated flow), is shown in Fig. 45(b). The graph also indicates that the pop-in serrations occur after the elastic to plastic transition. Thus, the nanoindentation-serrated flow is associated with non-hertzian behavior [211]. It was thought that the serrations were due to the resistance caused by the interfacial-strengthening mechanisms among the dislocations, the elastically-stiff B2 matrix, and the soft A2 phase.

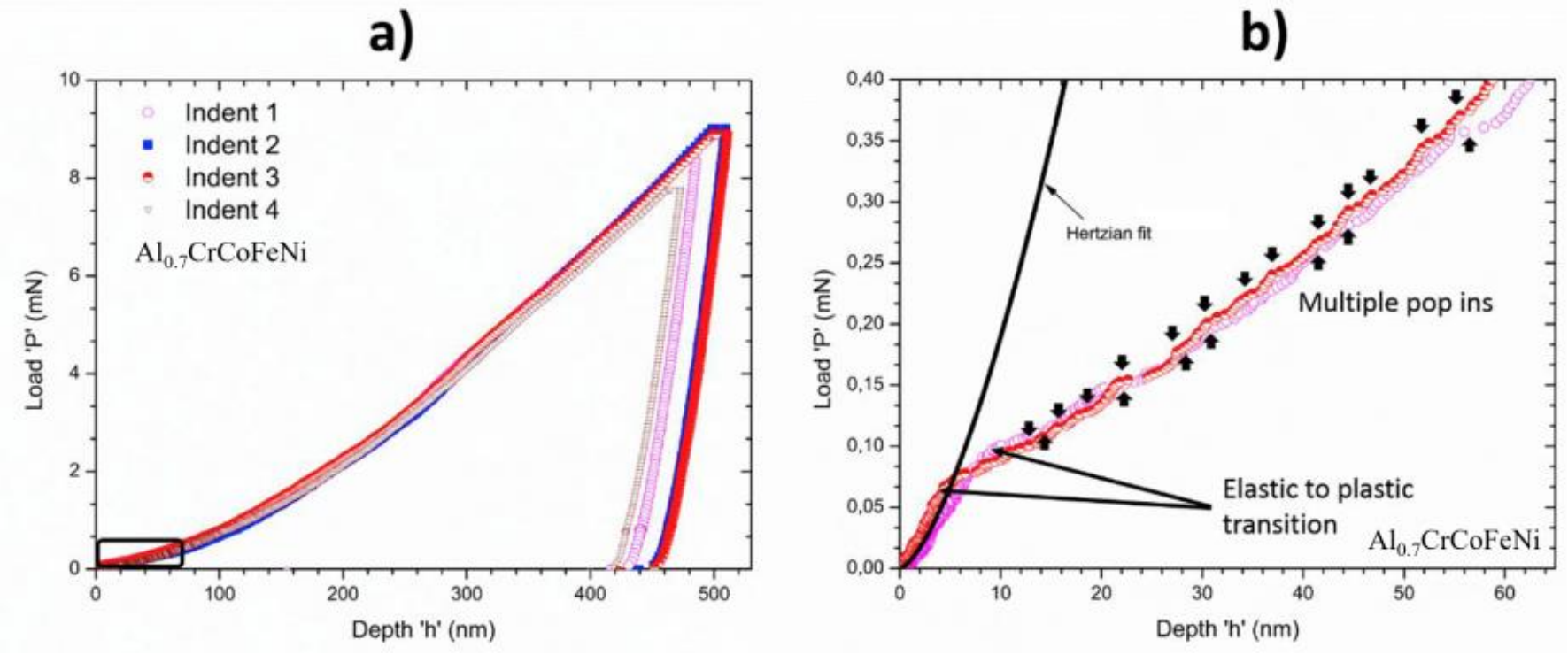

Fig. 45. (a) Load-displacement data for four different indents conducted on a BCC grain in the $Al_{0.7}CrCoFeNi$ HEA for an indentation depth of 500 nm, and (b) the corresponding magnified regions that features the data for two randomly-selected indents. Figures from Basu et al. [210].

### 3.3.3 $Al_{0.5}CoCrCuFeNi$ HEA

In [8], Chen et al. investigated the serrated flow dynamics of $Al_{0.5}CoCrCuFeNi$ HEA during nanoindentation. Nanoindentations were performed on specimens at RT and 200 °C where a maximum load of 100 mN was used. Figure 46 shows the nanoindentation load as a function of the tip displacement. For both temperatures, there were apparent displacement bursts in the data, which correspond to the serrated flow. Various phenomena may have been responsible for the serrated flow, such as the evolution of dislocation cells or tangles, dislocation multiplication, or dislocation breakaway from obstacles such as atoms or precipitates [212]. It was also reported that the displacement bursts were relatively larger for the specimen tested at 200 °C. The larger displacement bursts suggest that at elevated temperatures, there are more dislocations available to participate in the serration process. Furthermore, it was thought that the promoted thermal activation of dislocations during nanoindenation at the higher temperature was responsible for the softening of the material that led to greater tip penetration.

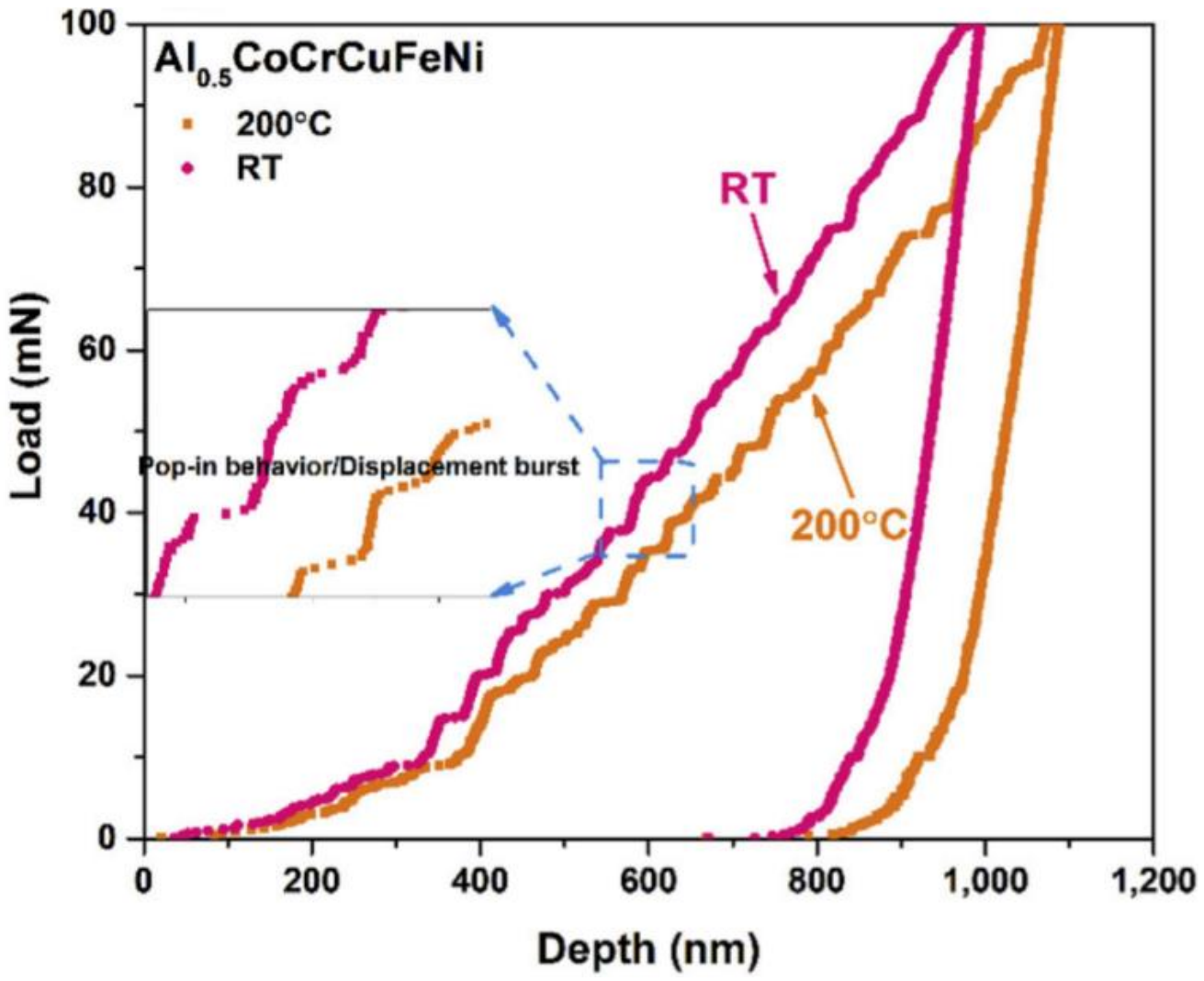

Fig. 46. Load vs. depth curves for the $Al_{0.5}CoCrCuFeNi$ HEA nanoindented at RT and 200 °C. Figure from Chen et al. [8].

Subsequently, Chen et al. [66] studied the nanoindentation induced serrated flow of the $Al_{0.5}CoCrCuFeNi$ HEA. Methods such as the ApEn and chaos formalism were used to analyze the serrations. Similar to [8], indentations were performed at RT and 200 °C to a maximum load of 100 mN. At the maximum load, the load was kept constant for holding times of 5 - 20 s. The corresponding depth vs. time data is presented in Fig. 47. The inset of the figure (RT, 5 s) shows the typical serrated flow that occurred during the constant depth period of the nanoindentation. The associated largest Lyapunov exponent, $\lambda_1$, as well as the ApEn results are shown in Fig. 48 for the prescribed holding time. It was found that the $\lambda_1$ values were positive for all the holding times and temperatures, which indicates that the serrations were associated with chaotic slip-band dynamics. For a given holding time, $\lambda_1$ increased with respect to temperature, indicating that the serrations were more chaotic behavior at 200 °C, and was perhaps a consequence of greater dislocation mobility. The ApEn values followed a similar trend as the largest Lyapunov exponent values at 200 °C, where they exhibited a minimum at a loading time of 10 s, although the reason for this trend is not known. On the other hand, both the $\lambda_1$ and ApEn values reached a maximum (200 °C) for a holding time of 5 s, which was ascribed to the greater amount of interactions among the slip bands. Interestingly, for a holding time of 10 s at RT, the $\lambda_1$ and ApEn values attained a minimum and maximum, respectively. This result suggests that that at this condition, the serrations displayed dynamical behavior that has both a greater degree of freedom as well as a relatively-low sensitivity to initial conditions [66].

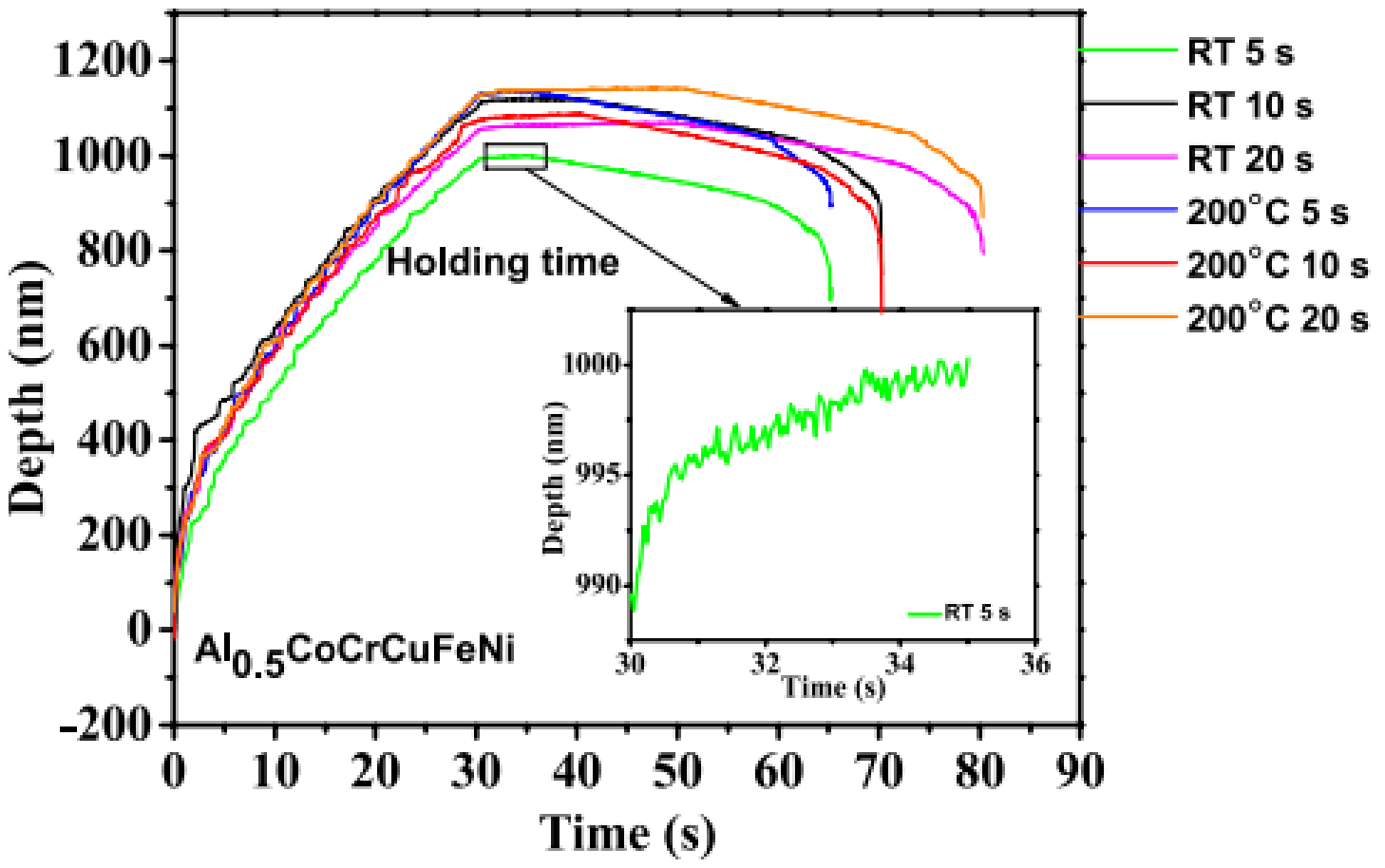

Fig. 47. The depth vs. time data for the $Al_{0.5}CoCrCuFeNi$ HEA nanoindented for holding times of 5 - 20 s at RT and 200 °C. Figure from Chen et al. [66].

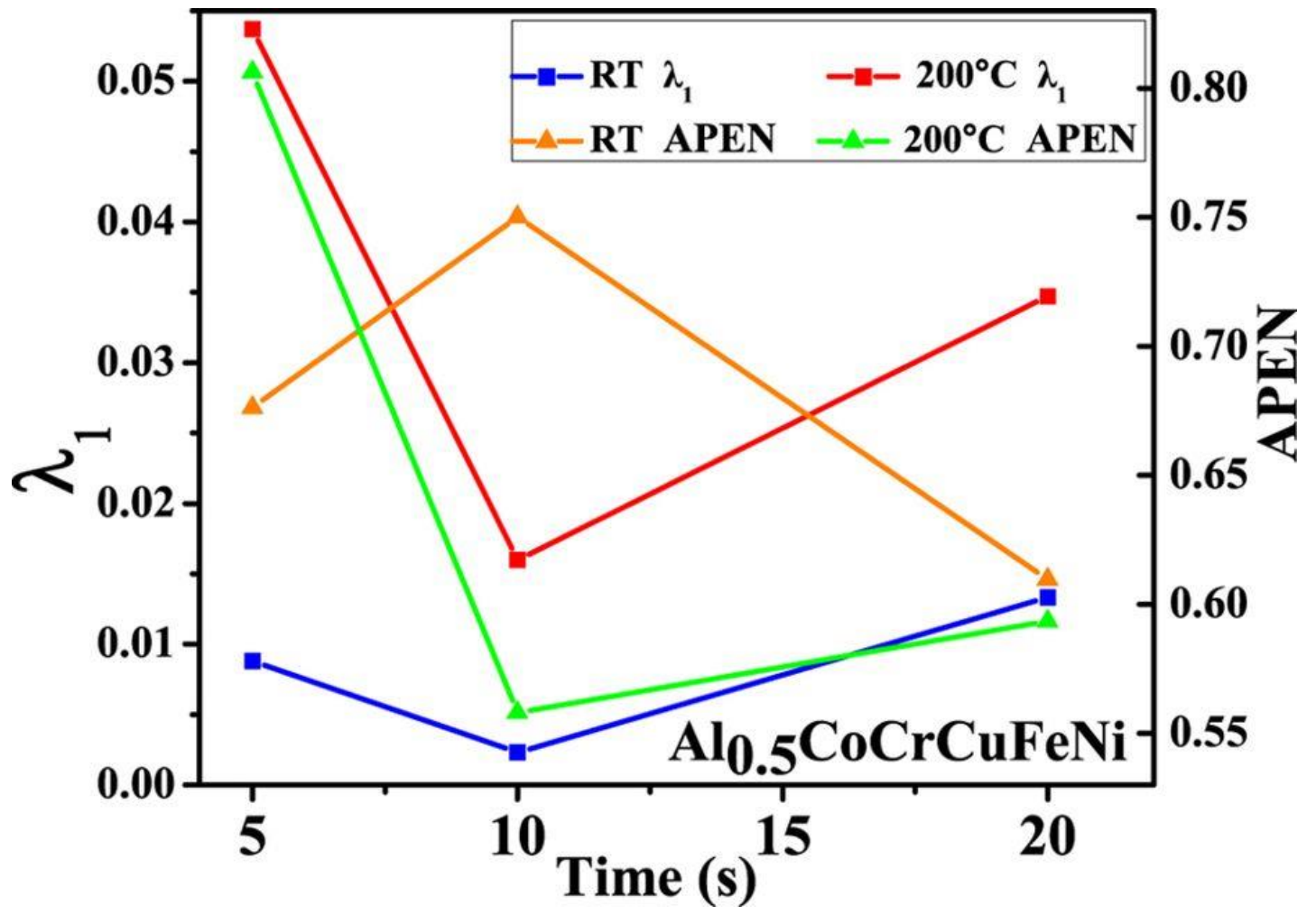

Fig. 48. The $\lambda_1$ (left axis), and ApEn (right axis) values for the data from Fig. 47 for holding times of 5 - 20 s at RT and 200 °C. Figure from Chen et al. [66].

## 4. Summary

Tables 9 and 10, as well as Figs. 49(a)-(e) and 50(a)-(d), offer a survey of the serration types for different kinds of HEAs during tension and compression experiments, as reported in the literature [6]. Table 9 and Figs. 49(a)-(e) provide the serration data for the tension tests. Here, Type-A, Type-B, Type-C, Types-A + B, and Types-B + C serrations were reported. The serration Types-A + E was not featured in the graphs due to their being only one data point mentioned in the chapter. Serrations were reported for temperatures of RT - 600 $^{o}$C and strain rates of 1 × $10^{-4}$ $s^{-1}$ - 1 × $10^{-4}$ $s^{-1}$. Type-A serrations were the most frequent type of serrations observed, while Type-C were the least observed. Furthermore, Type-A serrations were found to occur over the largest range of strain rates and temperatures of 1 × $10^{-4}$ $s^{-1}$ - 1 × $10^{-2}$ $s^{-1}$ and RT - 600 $^{o}$C, respectively. Whereas Type-C serrations were observed over the smallest range of values, i.e., strain rates of $10^{-4}$ $s^{-1}$ - 3 × $10^{-4}$ $s^{-1}$ and temperatures of 500 - 600 $^{o}$C.

Table 10 and Figs. 50(a)-(d) feature the reported results for serrations that occurred during the compression tests. The specimens were reported to exhibit serration Type-A, Type-B, Type-C, Types-A + B, Types-B + C, and Types-A + E. As compared to the tension testing, serrations exhibited a wider range of reported temperatures (RT – 1,100 $^{o}$C) and strain rates (1 × $10^{-4}$ $s^{-1}$ – 1 $s^{-1}$) for the compression tests. It is important to note that in contrast to the other types of serrations, serration Types-A + B occurred at relatively-higher temperatures. Finally, Types-A and C serrations were the most commonly reported serration types, while B + C and A + E were the least reported.

A qualitative plot featuring the relationship between the extent of serrations and temperature in HEAs, is presented in Fig. 51(a) [6]. In the graph, there are two primary temperature ranges where serrations can occur. The lower temperature region (4.2 – 77 K) corresponds to the serrated flow that occurs primarily due to the twinning phenomenon. A previous study suggested that the extent of the serrations may increase as the temperature is decreased [78]. In this scenario, the decrease in temperature is accompanied by an increase in the twinning activity that is accompanied by an increased serrated flow. For temperatures between 77 K and RT, there is neither significant twinning activity nor dislocation pinning by solute atoms, and therefore, no serrated flow occurs. Once the temperature exceeds RT, serrations can be observed again. The mechanisms for the serrated flow in this temperature range is due to the solute pinning of dislocations. Initially, the extent of the serrations increases with temperature due to the increased solute mobility and consequent increase in the rate of dislocation pinning. As the temperature exceeds a certain value, however, the extent of the serrations decreases, as the thermal vibrations of the solutes reduce their ability to pin dislocations [6]. As the temperature rises above a certain value (1,100 $^{o}$C for the reported HEAs), the thermal vibrations prevent solutes from pinning dislocations, and serrations no longer occur.

Table 9. The reported serration type for Types-A, B, C, A + B, and B + C for a range of HEAs tested at different strain rates and temperatures during tension testing. RT: room temperature. Adapted from Brechtl et al. [6].

| Alloy | Strain rate ($s^{-1}$) | Temperature (°C) | Serration Type | Source |
|---|---|---|---|---|
| $Al_{0.5}CoCrFeNi$ | $1 \times 10^{-4}$ | 200 | A | [5] |
| | | 300 | A + B | |
| | | 400 | B + C | |
| | $5 \times 10^{-4}$ | 300 | A | |
| | | 400 | A + B | |
| | $1 \times 10^{-3}$ | 300 | A | |
| | | 400 | A + B | |
| | | 500 | B + C | |
| $Al_{0.4}CrMnFeCoNi$ | $3 \times 10^{-4}$ | 300 | A + B | [183] |
| | | 400 | B | |
| | | 500 | B + C | |
| | | 600 | C | |
| $Al_{0.5}CrMnFeCoNi$ | $3 \times 10^{-4}$ | 300 | A + B | [183] |
| | | 400 | B | |
| | | 500 | B + C | |
| $Al_{0.6}CrMnFeCoNi$ | $3 \times 10^{-4}$ | 300 | A + B | [183] |
| | | 400 | B | |
| | | 500 | C | |
| CoCrFeMnNi | $1 \times 10^{-4}$ | 300 | A | [65] |
| | | 400 | B | |
| | | 500 | B | |
| | | 600 | C | |
| | $1 \times 10^{-3}$ | 300 | A | |
| | | 400 | A | |
| | | 500 | B | |
| | | 600 | B | |
| | $1 \times 10^{-2}$ | 400 | A | |
| | | 500 | A | |
| | | 600 | A | |
| CoCrFeMnNi | $1 \times 10^{-5}$ | 500 | B + C | [74] |
| | $1 \times 10^{-4}$ | 500 | B | |
| | $3 \times 10^{-4}$ | 300 | A | |
| | | 400 | A + B | |
| | | 500 | B | |
| | | 550 | B + C | |
| | | 600 | C | |
| | $1 \times 10^{-3}$ | 500 | A + B | |
| | $5 \times 10^{-3}$ | 500 | A | |
| CoCrFeMnNi (C ~ 0.9 at.%) | $1.6 \times 10^{-3}$ | RT | A | [129] |

Table 10. The reported serration type for Types-A, B, C, A + B, B + C, and A + E for a range of HEAs tested at different strain rates and temperatures during compression testing. RT: room temperature. Adapted from Brechtl et al. [6].

| Alloy | Strain rate ($s^{-1}$) | Temperature (°C) | Serration Type | Source |
|---|---|---|---|---|
| $Al_{0.3}CoCrFeNi$ | $1 \times 10^{-3}$ | 400 | C | [81] |
| | | 500 | C | |
| | | 600 | C | |
| | | 700 | B + C | |
| | | 800 | B | |
| AlCoCrFeNi | $1 \times 10^{-3}$ | 1,100 | C | [213] |
| | 1 | | B | |
| $Al_{0.5}CoCrCuFeNi$ | $5 \times 10^{-5}$ | 400 | A | [91] |
| | | 500 | B | |
| | | 600 | C | |
| | $2 \times 10^{-4}$ | 400 | A | |
| | | 500 | B | |
| | | 600 | C | |
| | $2 \times 10^{-3}$ | 400 | A | |
| | | 500 | A | |
| | | 600 | C | |
| $Al_{0.2}MoNbTaTiV$ | $1 \times 10^{-3}$ | RT | A | [204] |
| $Al_{0.4}MoNbTaTiV$ | $1 \times 10^{-3}$ | RT | A + E | [204] |
| $Al_{0.6}MoNbTaTiV$ | $1 \times 10^{-3}$ | 500 | A + B | [204] |
| $Al_{1}MoNbTaTiV$ | $1 \times 10^{-3}$ | 500 | B | [204] |
| CoCuFeNiTi | $1 \times 10^{-3}$ | 800 | A + B | [73] |
| | | 900 | A + B | |
| | | 950 | B | |
| | | 1,000 | B | |
| | $1 \times 10^{-1}$ | 800 | A | |
| | | 900 | A + B | |
| | | 950 | A + B | |
| | | 1,000 | A + B | |
| MoNbTaTiV | $1 \times 10^{-3}$ | RT | A | [204] |

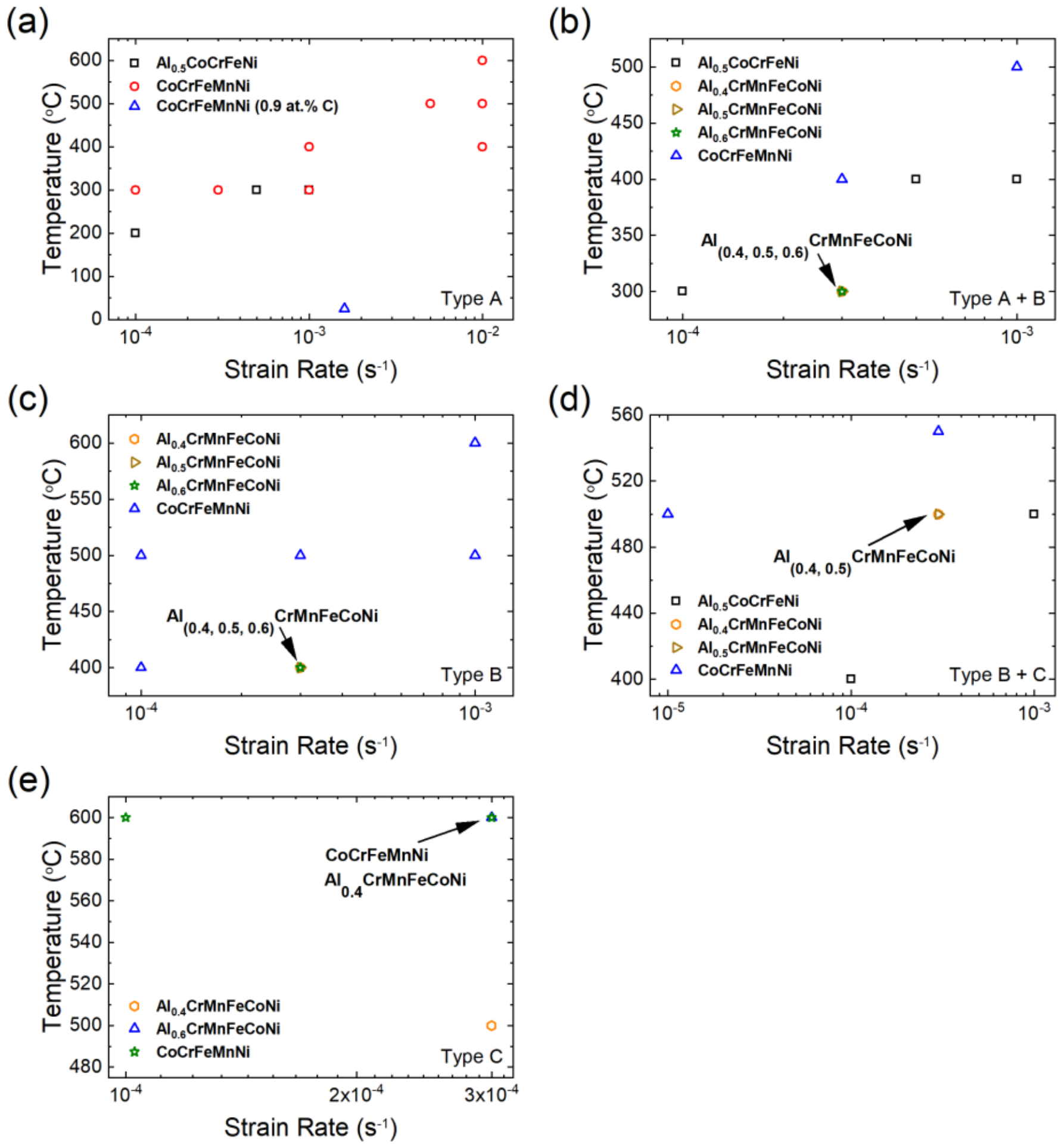

Fig. 49. Graphs for the reported results from Tables 9 and 10 for the serration types (a) A, (b) A + B, (c) B, (d) B + C, and (e) C. Adapted from Brechtl et al. [6].

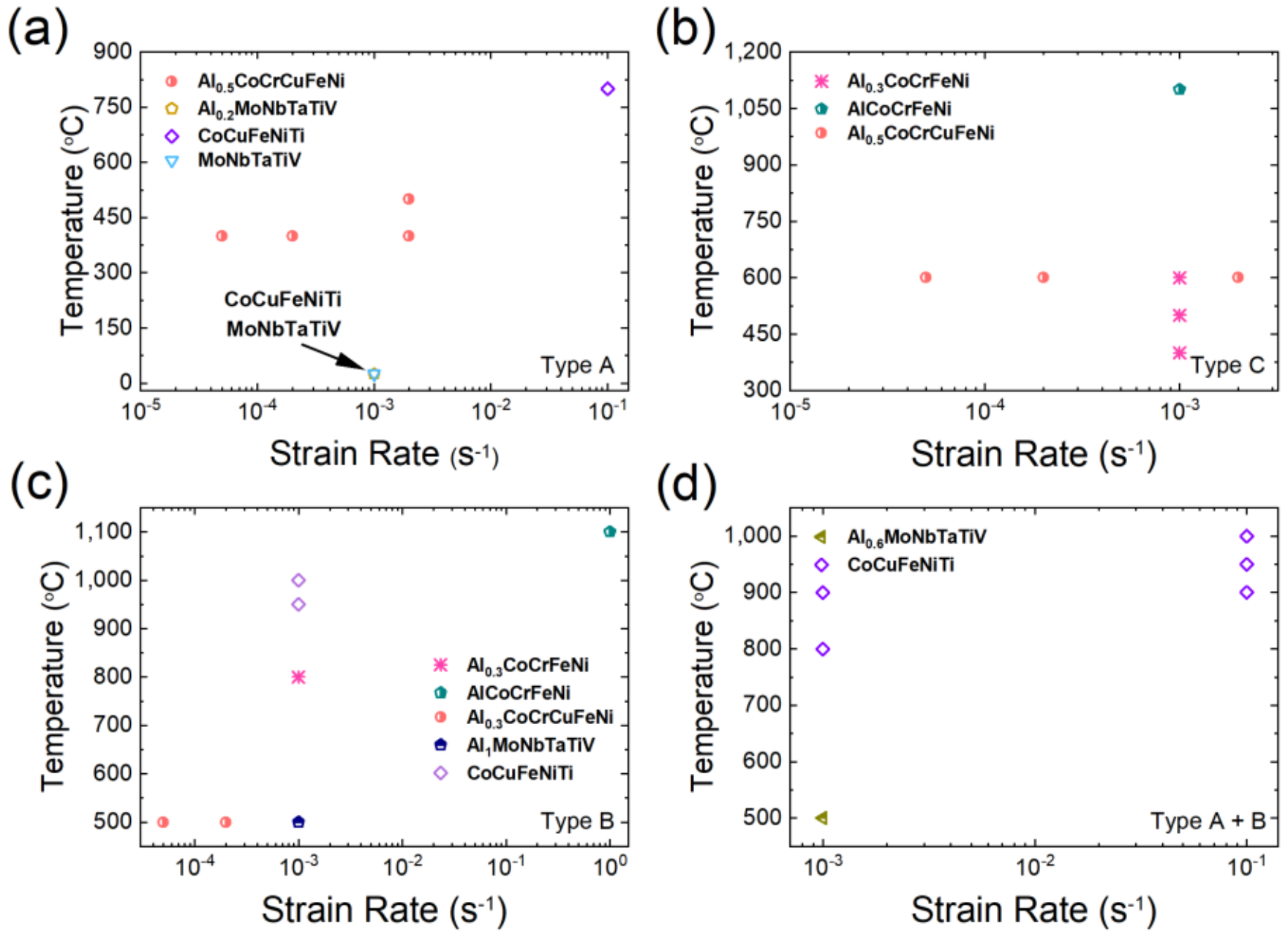

Fig. 50. Graphs for the reported results from Tables 9 and 10 for the serration types (a) A, (b) C, (c) B, and (d) A + B. Adapted from Brechtl et al. [6].

In a similar fashion to Fig. 51(a), Fig. 51(b) displays a qualitative graph depicting the relationship between the extent of serrations and the strain rate [6]. For strain rates of $10^{-5}$ – 1 $s^{-1}$, dislocation pinning is the mechanism that is primarily responsible for the serrated flow [6]. As the strain rate increases in this range, the speed of dislocations increases such that mobile solute atoms are less able to catch and pin them. As the strain rate surpasses 1 $s^{-1}$, the dislocations outpace the solute atoms, resulting in no serrations. For strain rates beyond $10^3$ $s^{-1}$, the serrated flow is likely caused by twinning mechanisms. As the strain rate increases in this range, the twinning activity increases, which results in an increase in the extent of the serrations. As for strain rates of 1 – $10^3$ $s^{-1}$, no serrated flow occurs due to the absence of both dislocation pinning and twinning. Lastly, no serrations have been reported for strain rates either below $10^{-5}$ $s^{-1}$ or above $10^3$ $s^{-1}$. Therefore, future studies should involve testing in these strain rate regimes.

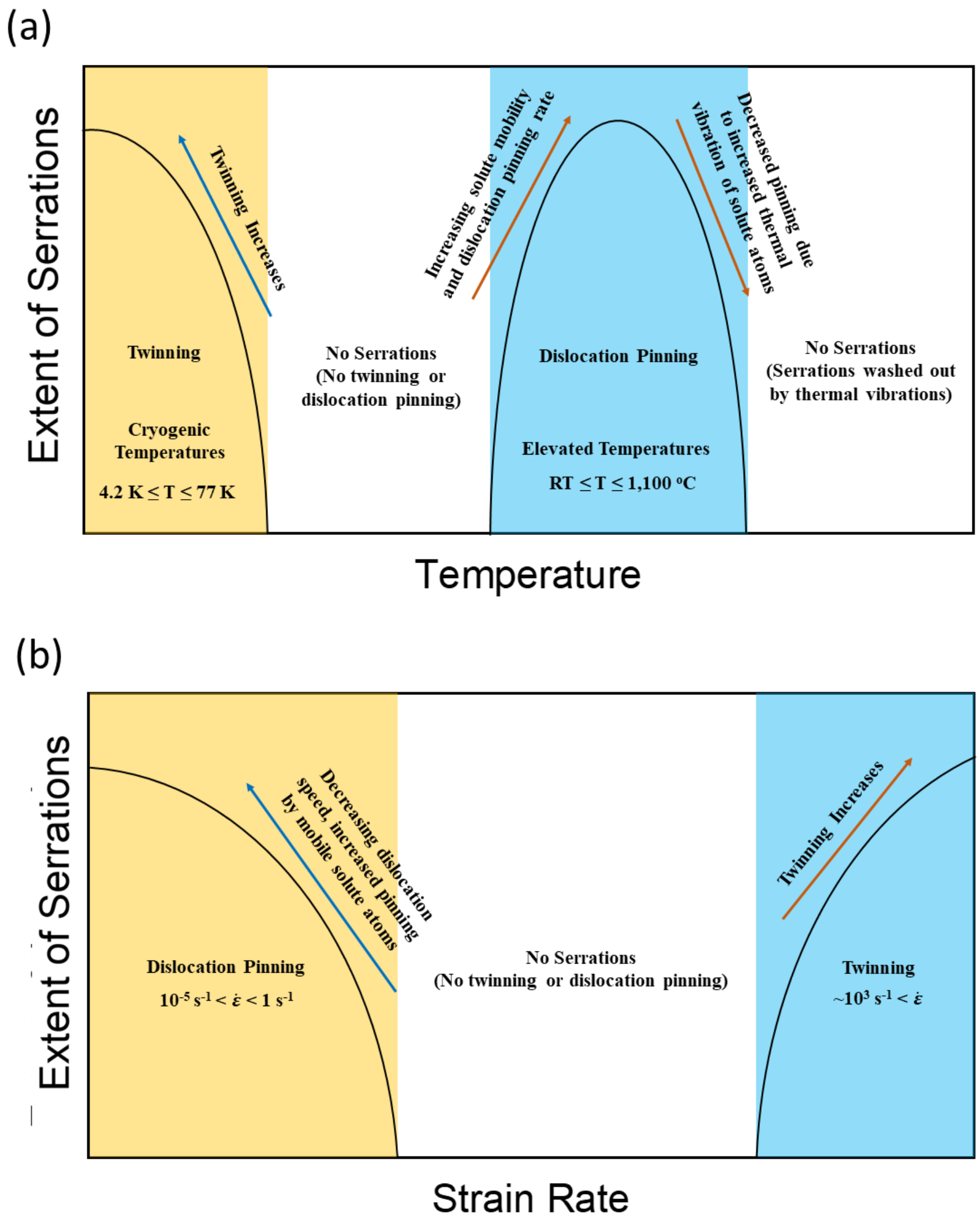

Fig. 51. A qualitative plot featuring the relationship between the extent of serrations and the (a) temperature and (b) strain rate. Adapted from Brechtl et al. [6].

Below are some key points on the serrated flow phenomenon in HEAs, as determined from the literature that was reviewed for the present chapter [6]:

- The reported serration types that have been observed are Types-A, B, C, A + B, B + C, and A + E. On the other hand, serration Types-D, E, and A + C have not been observed.

- The serration type is affected by the temperature. For instance, Type-A serrations are typically observed at relatively lower temperatures as compared to other serration types.
- At cryogenic temperatures, i.e., 4.2 K – 77 K, serrations are caused primarily by twinning mechanisms since twin boundaries can hinder dislocation motion.
- For temperatures of 77 K – RT, serrations have not been reported. The lack of serrated flow for this temperature range is attributed to the reduced solute mobility that renders them unable to catch and pin moving dislocations.
- For temperatures of RT – 1,100 $^{o}$C, dislocation pinning by solute atoms is the primary mechanism responsible for the serrated flow. In this temperature range, the extent of the serrations initially increases (due to the increased pinning rate), but then decreases (thermal vibrations decrease pinning) with temperature.
- The strain rate also influences the serration behavior. Types-A and B have been reported at relatively-higher strain rates, whereas Type-C serrations are usually observed at lower strain rates.
- When strain rates are below 1 $s^{-1}$, the extent of serrations increases with a decrease in the strain rate and is attributed to the reduced dislocation mobility that make them more susceptible to solute pinning.
- For strain rates exceeding $10^3$ $s^{-1}$, an increase in the strain rate results in increased twinning activity that enhances the extent of serrations.
- Certain solutes, such as C impurities and Al, as well as nanoparticles and phase structures, can play an important role in the serration behavior of HEAs.
- Various analytical techniques, including the mean-field theory analysis, chaos analysis, complexity analysis, and multifractal analysis, have been implemented to analyze and model the serration behavior in HEAs.

## 5. Future directions

Since HEAs are a relatively-new class of materials, much work needs to be done to elucidate a more fundamental understanding of the serrated-flow phenomenon in this alloy system. One such a phenomenon that should be investigated in the future involves the twinning-induced serrated flow in HEAs. For the experiments, HEA specimens could be mechanically tested at both cryogenic temperatures and strain rates exceeding $10^3$ $s^{-1}$. The microstructure could be characterized, using *ex situ* and *in situ* techniques, including XRD, EBSD, and TEM. To attain a more synergistic understanding of twinning on the serrated flow, the microstructural analysis could be paired with different analytical techniques that would be used to analyze and model the serration dynamics. It is expected that the work, as discussed above, would elucidate a more complete picture regarding the link between the serrated flow and the twinning phenomenon in HEAs.

Additionally, studies that investigate how irradiation dose affects the serrated flow in HEAs would be fundamentally important. Moreover, such studies could examine how ion (or neutron) irradiation at different temperatures affect the serrated flow in HEAs. Importantly, such studies would examine the effects of irradiation-induced dislocation loops, voids, or precipitates on the serrated flow. Other studies could involve how implanted ions, such as carbon, affect the serration dynamics in HEAs. Due to the typically-low penetration depth of ions, which is usually on the order of a few microns, mechanical testing would involve smaller-scale mechanical testing experiments, such as micro/nano pillar compression. Such studies, as those listed above, would elucidate how irradiation-displacement damage affects the serrated flow of HEAs.

On a final note, future investigations should involve the modeling and analysis of the serration behavior of HEAs, using a combination of different techniques. The methods could include the MFT analysis, complexity-analysis, multifractal-analysis, and the chaos-analysis techniques. The results of the analysis and modeling would be combined with those from different advanced microstructural-characterization techniques, such as *in situ* XRD, neutron diffraction, and TEM. Such combinations of these techniques would, in theory, provide a more holistic picture of the dynamical behavior that underlies the serrated flow in HEAs.

## 6. Conclusions

This chapter has provided a comprehensive review of the serrated-flow phenomenon for different HEAs. This review covered many important topics, such as how temperature, strain rate, and impurities affect this type of plastic deformation. Furthermore, the chapter examined the effects of the serrated flow on the microstructures under different experimental conditions. The results of several studies revealed that at cryogenic temperatures, twinning is a primary mechanism behind the serrated flow in HEAs. On the other hand, dislocation pinning is a major contributing factor for serrations at temperatures of RT - 1,100 $^{o}$C. In terms of the dynamical behavior, the serration Types-A, B, C, A + B and B + C were primarily observed in HEAs. The serrated flow in HEAs can be successfully modeled and analyzed, using multiple analytical methods, such as the mean-field theory formalism, chaos formalism, and complexity analysis. Despite the recent advances in our understanding regarding the serration behavior in HEAs, further exploratory work is needed. Examples of future phenomena to be studied include twinning, irradiation effects, influence of C impurities and nanoparticles, and the use of advanced characterization techniques. It is anticipated that such endeavors would lead us to the path to gaining a more fundamental understanding of the serrated-flow behavior in HEAs.